\documentclass[prd,aps,twocolumn,a4paper,showkeys,nofootinbib,10pt]{revtex4-1}

\usepackage{amsmath}
\usepackage{amsfonts}
\usepackage{amssymb}	
\usepackage{graphicx}
\usepackage{bm}
\usepackage{color}
\usepackage{commath}
\usepackage{hyperref}
\usepackage[T1]{fontenc}
\usepackage{xcolor}
\usepackage{verbatim}
\usepackage{comment}
\usepackage{float}
\usepackage{amsbsy}
\usepackage{makecell}
\usepackage{multirow}
\usepackage{tikz}
\usepackage{tikz-feynman}
\tikzfeynmanset{compat=1.1.0}

\usepackage[normalem]{ulem}
\allowdisplaybreaks

\newcommand{\mpeg}[1]{{\textcolor{teal}{\texttt{MP: #1}} }}

\newcommand{\A}{\ensuremath{\mathrm{A}}}
\newcommand{\diff}{\ensuremath{\mathrm{d}}}

\tikzfeynmanset{
	worldline/.style={
		black,very thick},
	potentialsigma/.style={
		draw=green!90!black,densely dotted,thick },       
	potentialphi/.style={
		draw=blue!90!black, dashed }
}
\tikzset{
	worldlinevertex/.style={
		font = {\tiny},  circle, inner sep=1.25pt,draw=black,thick, minimum size=1.5mm},
}

\newcommand{\orcid}[1]{\href{https://orcid.org/#1}{
\includegraphics[width=10pt]{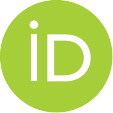}
}}

\begin{document}
\title{Charged Black-Hole Binary Evolution at Second Post-Newtonian Order}

\author{Andrea Placidi${}^{1}$\orcid{0000-0001-8032-4416}}
\author{Elisa Grilli${}^{1,2}$\orcid{0009-0007-7529-4962}}
\author{Marta Orselli${}^{1,2}$\orcid{0000-0003-3563-8576}}


\affiliation{${}^1$Dipartimento di Fisica e Geologia, Universit\`a di Perugia, I.N.F.N. Sezione di Perugia, \\ Via Pascoli, I-06123 Perugia, Italy}
 \affiliation{${}^2$ Center of Gravity, Niels Bohr Institute, Copenhagen University,\\ Blegdamsvej 17, DK-2100 Copenhagen \O{}, Denmark}

\author{Matteo Pegorin${}^{3,4,5}$\orcid{0009-0003-1248-871X}}
\author{Nicola Bartolo${}^{3,4,6}$\orcid{0000-0001-8584-6037}}
\author{Pierpaolo Mastrolia${}^{3,4}$\orcid{0000-0001-9711-7798}}
\affiliation{${}^3$ Dipartimento di Fisica e Astronomia ``Galileo Galilei'', Universit{\`a} degli Studi di Padova, via Marzolo 8, I-35131 Padova, Italy}
\affiliation{${}^4$INFN, Sezione di Padova, Via Marzolo 8, I-35131 Padova, Italy}
\affiliation{${}^5$Max Planck Institute for Gravitational Physics (Albert Einstein Institute), Am M{\"u}hlenberg 1, Potsdam 14476, Germany}
\affiliation{${}^6$INAF - Osservatorio Astronomico di Padova, Vicolo dell’Osservatorio 5, I-35122 Padova, Italy}


\email{andrea.placidi@unipg.it \\
elisa.grilli@nbi.ku.dk  \\
orselli@nbi.ku.dk \\
matteo.pegorin.1@phd.unipd.it \\
nicola.bartolo@unipd.it \\
pierpaolo.mastrolia@unipd.it 
}

\begin{abstract}
We study the dynamics of electrically charged black-hole binaries and their gravitational-wave emission during the inspiral phase. Within the post-Newtonian framework, we derive the conservative and dissipative dynamics up to second order (2PN), combining Effective Field Theory and classical methods. We compute the next-to-next-to-leading-order conservative Lagrangian, leading-order dissipative effects in harmonic and Lorenz gauges, and provide the equations of motion, center-of-mass transformations, and the Lagrangian/Hamiltonian in Arnowitt–Deser–Misner-type coordinates. We also obtain gauge-invariant expressions for the binding energy, periastron advance in quasi-circular orbits, and the scattering angle in unbound orbits. Our results extend previous analyses and are fully consistent with recent post-Minkowskian findings. 

\end{abstract}
\maketitle
\newpage

\section{Introduction}
\label{sec:intro}   

In conventional astrophysical contexts, black holes are expected to be electrically neutral to a good approximation. Even in the presence of external magnetic fields, selective charge accretion is strongly suppressed and various physical processes, such as vacuum breakdown, plasma screening, and pair production, act to rapidly neutralize any significant charge imbalance~\cite{Gibbons, Znajek, PhysRevD.10.1680, Palenzuela:2011es}. This expected neutrality also underpins the standard modeling of gravitational waves (GWs) from binary black hole (BBH) mergers, as implemented in the LIGO and Virgo data analysis pipelines~\cite{AdvLIGO, AdvVirgo, O3a_catalog, Abbott:2020gyp}.

Nevertheless, studying compact binaries of electrically charged black holes remains of prime importance for GW astronomy, as it enables the extraction of direct observational constraints on black hole charge from GW analyses. This is an especially relevant task if one considers scenarios involving hidden-sector physics. In these cases, the ``charge'' appearing in the Reissner-Nordström (or Kerr-Newman) metric may correspond to more exotic quantities: magnetic monopole charge~\cite{Preskill_monopoles, Bozzola:2020mjx, Liu:2020cds}, vector charges in modified gravity~\cite{Bozzola:2020mjx}, or minicharges associated with dark matter~\cite{Cardoso:2016}\footnote{Recent studies have also investigated how tidal effects can substantially modify the dynamics of charged BBHs, particularly in the extreme–mass–ratio limit~\cite{Grilli:2024fds}, as well as how electric charge influences the evolution of the event horizon in such systems~\cite{Pina:2022dye}.}. Crucially, such charges may evade the discharge mechanisms that occur in standard electromagnetism, allowing them to persist through the coalescence process of BBHs, leaving a potentially measurable imprint on the emitted gravitational radiation.

While several tests of general relativity (GR), sensible also to the eventual presence of charge, have been applied to GW signals~\cite{Abbott:2018lct,TGR-LVC2016, LIGOScientific:2019fpa, Ghonge:2020suv,Abbot:O3a_TGR,Brito:2018rfr, ghosh:2021c, Carullo:2019a, isi:2019, Isi:2021iql,Ghosh:2016a, Ghosh:2018}, these analyses typically follow an agnostic phenomenological approach, where the looked-for GR deviations have no aprioristic modeling. While this model-independent approach has the clear advantage of being basically valid for any possible alternative theory to GR, it also implies a relevant loss in sensitivity to specific beyond-GR signatures compared with what can be achieved while leaning on more precise theoretical predictions.

At the level of the post-merger phase, these motivations have led to charge-targeted ringdown analyses such as Ref.~\cite{Carullo:2021oxn}. On parallel, concerning the inspiral phase, several works have been dedicated to computing the motion and radiation of charged BBHs within the Post-Newtonian (PN) perturbative framework~\cite{Blanchet:2013haa,Damour:2014jta,Bernard:2015njp,Foffa:2019yfl,Blanchet:2023sbv,Blanchet:2023bwj,Ghosh:2025ban}, a weak-field and slow-velocity expansion of Einstein's equations organized in power of $1/c$ (the $n$PN order corresponding to $1/c^{2n}$ corrections), where $c$ is the velocity of light. More in details, the 1PN Lagrangian for a charged BBH has been independently obtained in Refs.~\cite{Julie:2017rpw,Khalil:2018aaj,Patil:2020dme,Julie:2018lfp}~\footnote{We point out that Refs.~\cite{Julie:2017rpw,Khalil:2018aaj,Julie:2018lfp} study the more general case of BBHs in Einstein-Maxwell-dilaton theories, that reduce to charged BBHs when the dilaton coupling constant is set to 0. 
Moreover, Ref.~\cite{Khalil:2018aaj} also provides the 1PN energy flux for this class of BBHs, and then uses it, together with the results for the binding energy, to derive a Fourier-domain waveform model in the stationary-phase approximation.}, and later extended to 2PN, using effective field theory (EFT) techniques, in Ref.~\cite{Gupta:2022spq}, result that however presents some possible inconsistencies that will be more thoroughly discussed below, in Sec.~\ref{subsec:chi}.

With the present work we wish to contribute in this endeavor by offering a thorough 2PN-accurate description of the dynamics of charged BBHs, comprehensive of conservative and dissipative contributions (which enter at 1.5PN due to charge-related dipolar emissions). In particular we provide: (i) a 2PN-accurate Lagrangian in harmonic coordinates, compatible with known PN results for neutral BBHs~\cite{Blanchet:2013haa} and with the post-Minkowskian results of~\cite{Wilson-Gerow:2023syq,Alonzo-Artiles:2026wbe};
(ii) the corresponding conservative part of the 2PN equations of motion (EoMs);
(iii) the associated two-body Hamiltonian in an Arnowitt–Deser–Misner– (ADM-) type coordinate system designed to give back the known ADM results in the neutral limit; (iv) the 2PN transformations to the center of mass (CoM) frame; (v) the 2PN expansion of three gauge invariant quantities, i.e.~binding energy and periastron advance, for quasi-circular orbits, and the scattering angle, for generic orbits; (vi) the 1.5PN contributions, due to dipolar-radiation dissipations, to EoMs and transformations to the CoM frame. 

In this paper, we focus on a binary system of two Reissner–Nordstr\"om black holes, which are non-rotating but electrically charged. However, our results apply more generally to any binary system comprising spherically symmetric, electrically charged and spinless compact objects. This is because for compact objects with vanishing permanent multipole moments, finite size effects arise only at higher PN orders. Moreover, the results remain valid if the electromagnetic interaction is replaced by any other massless minimally-coupled $U(1)$ gauge interaction, with the compact objects coupled through the corresponding electric-type monopole charge.

We anticipate that this work will be followed by a companion paper~\cite{Placidi:future} that will focus on the derivation of the 2PN energy flux radiated by a charged BBH, on the corresponding corrections to the radiative spherical modes of the waveform at infinity, and, additionally, on the analysis of the extremal limit.

The paper is organized as follows. Sec.~\ref{sec:notation} clarifies our notation for the parameters and the dynamical variables used in the rest of the paper. In Sec.~\ref{sec:EFT_conservative} we present the fundamental action for Einstein-Maxwell theory coupled to charged black holes. Afterwards, we set up the EFT approach to PN theory, which is used to derive the expression for the conservative harmonic and Lorenz gauge effective Lagrangian up to 2PN, also providing the double zero terms to recast it in classical form. The expanded action relevant for deriving the Feynman rules used in the EFT computations is explicitly given in App.~\ref{app:Feynman_rules}. In Sec.~\ref{sec:eom and Hamiltonian} we expand on the conservative dynamics of the previous section deriving the associated EoMs, the conserved total linear momentum, and the two-body Hamiltonian in ADM-type coordinates (whose expression can be found in App.~\ref{app: center-of-mass frame}). In Sec.~\ref{sec:CoM_frame} we obtain the CoM frame transformations. These are then used to derive explicit results for the 2PN dynamics in this frame, which are collected in App.~\ref{app: center-of-mass frame}. Sec.~\ref{sec:guage_inv_quantities} is dedicated to computing the 2PN expansion of three gauge-invariant quantities: the binding energy and the periastron advance, for quasi-circular orbits, and the scattering angle along unbound orbits. Finally, to complete the description of the dynamics at 2PN order, in Sec.~\ref{sec:dissipations} we consider the effect of the dipolar-radiation dissipations and we compute the 1.5PN dissipative corrections to the EoMs and the CoM frame transformations; further computational details can be found in App.~\ref{app: explicit calculation dissipative contributions}.

The explicit results for Lagrangian, Hamiltonian, EoMs, and CoM transformations are also provided in electronic form in the Supplementary Material~\cite{SupplementalMaterial} accompanying this paper.
\section{Notation}
\label{sec:notation}

We denote by  $y_A^i(t)$ the instantaneous position, $v^i_A= dy_A^i(t)/dt$ coordinate velocities, and $a^i_A= dv_A^i(t)/dt$ coordinate accelerations, {where $A=1,2$ labels the two bodies in the system and $i=1,2,3$ is the index of the space components. The harmonic-coordinate distance between the two bodies is defined by $r=|\mathbf{y}_1(t)-\mathbf{y}_2(t)|$. The relative binary's separation is $x^i= y_1^i-y_2^i$, with $r=|\mathbf{x}|$ and $n^i=x^i/r$. The relative velocity and acceleration are respectively defined as $v^i= dx^i/dt= v_1^i\, -v_2^i$, and $a^i=dv^i/dt\, =a_1^i-a_2^i$.  All expressions in the CoM frame are parametrized by the total mass $M=m_1\, +m_2$, the reduced mass $\mu= m_1 m_2/M$, the mass ratio $\nu=\, \mu/M$, and the dimensionless charge-to-mass ratios $\eta_1= q_1/\sqrt{G}m_1$ and $\eta_2= q_2/\sqrt{G}m_2$. To further simplify the CoM frame expressions we also introduce the parameter $X_{12} \equiv (m_1-m_2)/M \equiv \sqrt{1-4\nu}$, which is assumed to be non-negative by choosing $m_1>m_2$.

In this work we use Gaussian units, in which $c=1$, and consequently $\mu_0=\, (\varepsilon_0)^{-1}=4\pi $. Occasionally we retain  explicit factors of $\epsilon_0$, to make the power counting transparent, and of $c$, to clearly indicate the PN order of different terms.

We adopt the mostly minus metric, with signature $(+, -, - ,-)$. The sign convention we adopt for the Riemann tensor, in a coordinate basis, reads
\begin{equation}
	R^{\mu}{}_{\nu\rho\sigma} = \partial_\rho \Gamma^{\mu}_{\nu \sigma} - \partial_\sigma \Gamma^{\mu}_{\nu \rho} + \Gamma^{\mu}_{\alpha \rho} \Gamma^{\alpha}_{\nu \sigma} - \Gamma^{\mu}_{\alpha \sigma} \Gamma^{\alpha}_{\nu \rho}\ .
\end{equation}

\section{Harmonic-coordinate 2PN Lagrangian within the Effective Field Theory approach}
\label{sec:lagrangian} 
\label{sec:EFT_conservative} 

To obtain the Lagrangian describing the conservative dynamics of the system we employ the EFT approach to PN theory which was first introduced in Ref.~\cite{Goldberger:2004jt} (see~\cite{Porto:2016pyg, Levi:2018nxp} for reviews).

The EFT approach has proven to be a powerful framework for deriving high-precision results in the study of compact binary dynamics. Building on~\cite{Goldberger:2004jt,Gilmore:2008gq}, the application of modern Feynman integrals techniques~\cite{Foffa:2011ub, Kol:2013ega,Foffa:2016rgu} enabled breakthrough calculations at higher PN orders~\cite{Foffa:2011ub, Foffa:2016rgu,Foffa:2019rdf, Foffa:2019yfl,Blumlein:2020pog, Foffa:2019hrb,Blumlein:2019zku,Blumlein:2021txe, Blumlein:2020znm}.
The EFT approach is well suited also to incorporate spin and finite size effects of the compact objects in the system~\cite{Porto:2005ac,Levi:2015msa,Kim:2021rfj,Kim:2022pou,Mandal:2022nty,Kim:2022bwv,Mandal:2022ufb,Levi:2022rrq,Mandal:2023lgy,Mandal:2023hqa}, to evaluate dissipative effects and gravitational radiation from compact binary systems~\cite{Goldberger:2004jt,Galley:2009px,Galley:2010es,Galley:2012qs,Goldberger:2020wbx,Cho:2022syn,Amalberti:2023ohj,Amalberti:2024jaa,Mandal:2024iug}, as well as hereditary effects~\cite{Goldberger:2009qd,Galley:2015kus,Almeida:2021xwn,Almeida:2022jrv,Brunello:2022zui, Almeida:2023yia,Porto:2024cwd,Almeida:2024lbv}, also in theories beyond vacuum GR~\cite{Sanctuary:2010dao,Endlich:2017tqa, Huang:2018pbu, Kuntz:2019zef,Brax:2019tcy, Wong:2019yoc, Riva:2019xjs, Bhattacharyya:2023kbh, Bernard:2023eul, Diedrichs:2023foj, Wu:2023jwd, Almeida:2024uph, Almeida:2024cqz, Bernard:2025dyh}, such as Einstein-Maxwell theory~\cite{Patil:2020dme,Gupta:2022spq,Martinez:2021mkl,Martinez:2022vnx}.

A crucial observation for setting up the formalism is that, in the non-relativistic two body problem, several scales come into play: the characteristic length scale $R_s$ of each compact object, the orbital separation $r$ between them, and the wavelength $\lambda$ of the gravitational waves emitted by the system.
In particular, under the assumption of bound systems and slow velocities, a separation of scales becomes apparent
\begin{equation}
\label{eq:separation_of_scales}
    R_s \ll r \ll \lambda\ .
\end{equation}
Consequently, we can identify three corresponding regions in our problem: the internal zone, with length scales comparable to $R_s$, the near zone, comprising longer length scales, up about to $r$, and the far zone, comprising length scales comparable to $\lambda$.

This hierarchy makes the EFT framework well suited for application to the PN formalism since it allows for the construction of a separate effective theory for each of the different zones, iteratively integrating out the short scale modes of the relevant degrees of freedom. In this way, the shorter scales are systematically removed, while their effect on the dynamics of the coarse-grained system is fully accounted for in an effective way.

The general idea then is to start from the fundamental action of our theory, implement the suitable PN expansion and power counting, and integrate out the gravitational and electromagnetic degrees of freedom, while assuming the worldlines to be non-propagating; that is, for the compact objects to be non-dynamical background sources~\cite{Goldberger:2004jt,Porto:2016pyg,Levi:2018nxp}. This will result in an effective action in which only the worldlines degrees of freedom will appear. More specifically, we will obtain the 2PN Lagrangian governing the motion of the binary system.

\subsection{Fundamental action}
Our starting fundamental action, which governs the dynamics of the system, is the Einstein-Maxwell action, which is given by the combination of Einstein's theory of GR and Maxwell's theory of electromagnetism, coupled to matter
%
\begin{equation}
    \label{eq: total E-M action}
    S\, =S_g\, +S_{\rm EM}\, +S_m, 
\end{equation}
where $S_g$ is the Einstein-Hilbert gravitational action, $S_{\rm EM}$ is the electromagnetic action and $S_{m}$ is the matter action.

Explicitly, the Einstein-Hilbert action reads
\begin{equation}
    \label{eq:S_EH}
    S_g= - 2 \Lambda^2 c^4 \int \diff t \diff^3 \mathbf{x}\sqrt{-g}\, R, 
\end{equation}
where $R$ is the Ricci scalar and $\Lambda \equiv (32 \pi G)^{-\frac{1}{2}}$. In the following, we adopt the harmonic gauge and accordingly supplement the action with the corresponding harmonic gauge-fixing term
\begin{equation}
    \label{eq:S_EH_gf}
    S_{g,{\rm gf}}= \Lambda^2  c^4 \int \diff t \diff^3 \mathbf{x}\sqrt{-g}\, g_{\mu \nu} \Gamma^{\mu} \Gamma^{\nu} ,
\end{equation}
where $\Gamma^{\mu} \equiv \Gamma^\mu {}_{\nu\rho} g^{\nu\rho}$.
Employing the relation $\mu_0 \epsilon_0 c^2 = 1$, the electromagnetic action $S_{\rm EM}$ can be written as~\footnote{In the electromagnetic action we choose $\epsilon_0$ as the coupling constant, instead of $\mu_0$, since this choice makes our $\frac{1}{c}$ PN power counting scheme manifest, as detailed in Sec.~\ref{sec:power_counting}. We also explicitly show the contraction of Lorentz indices using the metric $g_{\mu \nu}$, necessary to have a diffeomorphism-invariant action, to highlight the contributions to the non-linear bulk interactions vertices due to the metric.}
\begin{equation}
    \label{eq:S_EM}
    S_{\rm EM }=\, - \frac{\Lambda_{\text{EM}}^2 c^2}{4} \int \diff t \diff^3\mathbf{x}\sqrt{-g}\, g^{\mu\rho} g^{\nu\sigma} F_{\mu \nu}F_{\rho \sigma},
\end{equation}
with $F_{\mu \nu}= \partial_\mu A_\nu - \partial_\mu A_\nu$ where $A_{\mu}$ is the covariant vector potential, and $\Lambda_{\text{EM}} \equiv \sqrt{\epsilon_0}$.

To impose the Lorenz gauge on the vector potential, we additionally introduce the generally covariant gauge-fixing term
\begin{equation}
    \label{eq:S_EM_gf}
    S_{\rm EM, gf }=\, - \frac{\Lambda_{\text{EM}}^2 c^2}{2} \int \diff t \diff^3\mathbf{x}\sqrt{-g}\, (g^{\mu\nu} \nabla_\mu A_\nu)^2  .
\end{equation}

Concerning the matter action $S_{m}$, we can resort to the principles of EFTs to obtain an effective description of the internal, short scale, dynamics taking place inside each compact object. In particular we can employ a bottom-up approach, writing the most generic possible action, consistent with the symmetries of the problem, and concerning only the long-wavelength degrees of freedom.
Assuming spinless, spherically symmetric compact objects, and a target accuracy of 2PN, the resulting effective matter action is simply the action for charged point-particle worldlines, given by~\cite{Julie:2017rpw,Khalil:2018aaj, Goldberger:2004jt, Porto:2016pyg, Levi:2018nxp, Patil:2020dme, Martinez:2021mkl, Martinez:2022vnx,Gupta:2022spq}~\footnote{
For compact objects, at higher PN order new operators in the worldline action would become relevant. They would encode finite size effects, such as electromagnetic polarizability or also acceleration-induced dipole moment~\cite{Patil:2020dme, Goldberger:2005cd, Galley:2010es, Martinez:2021mkl, Martinez:2022vnx}. Alternatively, for non-spherically symmetric compact objects, Refs.~\cite{Henry:2023guc,Henry:2023len} employed a multipole expansion of the four-current $j_A^\mu$, as presented in~\cite{Ross:2012fc, Amalberti:2023ohj}, and a matching procedure, to implement magnetic- and electric-type moments. Extending the analysis to spinning compact objects would introduce new operators in the worldline action as well~\cite{Porto:2005ac,Levi:2015msa}.}
\begin{align}
    \label{eq:S_m}
    \notag S_m=&\, -\sum_{A}\int dt \Bigl[ m_A\,  c^2\sqrt{g_{\mu \nu} v_A^\mu v_A^\nu/c^2}\, \\ 
    &- c\,  q_A A_\mu (v^\mu_A / c )\Bigr].
\end{align}

In the computations that follow, to systematically deal with divergent integrals, we employ the dimensional regularization scheme. 
To do so we promote the number of spatial dimensions from 3 to a generic $d$, working hence on a $d+1$ dimensional manifold, finally taking the limit $d \rightarrow 3$ only at the end.
To account for the change in the mass dimension of Newton's constant $G$, we promote the gravitational coupling $\Lambda$ as~\cite{Mandal:2022nty,Mandal:2022ufb,Mandal:2023lgy,Mandal:2023hqa}
\begin{equation}
    \label{eq:Lambda_def}
    \Lambda \rightarrow \Lambda\, (\sqrt{4 \pi e^{\gamma_E}} R_0)^{-\frac{(d-3)}{2}}\ ,
\end{equation}
where $\gamma_E$ is the Euler-Mascheroni constant and $R_0$ is an arbitrary length scale. Similarly, for the electromagnetic action, we promote 
\begin{equation}
    \label{eq:Lambda_EM_def}
    \Lambda_{\text{EM}} \rightarrow \Lambda_{\text{EM}}\, (\sqrt{4 \pi e^{\gamma_E}} R_{0,\mathrm{EM}})^{-\frac{(d-3)}{2}}\ ,
\end{equation}
with $R_{0,\mathrm{EM}}$ standing for a new independent arbitrary length scale.
With these definitions, we consistently generalize the fundamental actions given in Eqs.~\eqref{eq:S_EH} through~\eqref{eq:S_EM_gf} to generic $d$-spatial dimensions.

\subsection{Metric and EM potential decomposition}
Under the PN assumptions of slow velocity, by choosing an appropriate coordinate frame, the spatial components of the
metric are suppressed with respect to the temporal $g_{00}$ component. 
We exploit this fact by performing a temporal Kaluza-Klein (KK) decomposition of the metric $g_{\mu\nu}$ and the electromagnetic potential $A_{\mu}$, equivalent to a 
threading decomposition of spacetime. This allows us to simplify several calculations, avoiding the evaluation of some diagrams altogether, at the expense of having more degrees of freedom to track~\cite{Kol:2007bc, Kol:2010si}.

Specifically we employ the parametrization given in terms of the Kol-Smolkin variables~\cite{Kol:2007bc, Kol:2010si} to decompose the metric, by introducing the scalar field $\phi$, the $d$-dimensional vector field $\A_i$, and the $d$-dimensional symmetric rank-2 tensor field $\sigma_{ij}$. We have
%
\begin{subequations}
    \label{eq:metric_decomposition}
    \begin{equation}
        g_{00} = e^{2 \frac{\phi}{\Lambda c^2}}\ ,
    \end{equation}
    \begin{equation}
        g_{0i} = - e^{2 \frac{\phi}{\Lambda c^2}} \frac{\A_i}{\Lambda c^2}\ ,
    \end{equation}
    \begin{equation}
        g_{ij} = e^{2 \frac{\phi}{\Lambda c^2}} \left( \frac{\A_i}{\Lambda c^2} \frac{\A_j}{\Lambda c^2} - e^{-c_d \frac{\phi}{\Lambda c^2}} \left(  \delta_{ij} + \frac{\sigma_{ij}}{\Lambda c^2} \right) \right)\ ,
    \end{equation}
\end{subequations}
where we have normalized the fields by $\Lambda\, c^2$, as defined in Eq.~\eqref{eq:Lambda_def}, to obtain canonically normalized kinetic terms for the fields (up to dimensionless variables), and we have used the definition
\begin{equation}
    c_d \equiv 2 \frac{(d-1)}{(d-2)} \xrightarrow{d \rightarrow 3} 4 \ .
\end{equation}
Let us notice that in the limit of vanishing gravitational fields, the metric parametrization in Eq.~\eqref{eq:metric_decomposition} correctly reduces to the Minkowski metric $\eta_{\mu\nu}$.

We extend this decomposition also to the electromagnetic potential, as
\begin{subequations}
    \label{eq:potential_decomposition}
    \begin{equation}
        A_0 = \frac{\phi_{\rm EM}}{\Lambda_{\text{EM}}\, c}\ ,
    \end{equation}
    \begin{equation}
        A_i = \frac{\A_{{\rm EM},i}}{\Lambda_{\text{EM}}\, c}\ ,
    \end{equation}
\end{subequations}
where we have similarly normalized the electromagnetic KK fields by $\Lambda_{\text{EM}}\, c$, defined in Eq.~\eqref{eq:Lambda_EM_def}.

Upon implementing the above decomposition in the fundamental action, we are able to perturbatively expand the latter under the PN assumption of weak field, to obtain a polynomial expression in the fields that specifies the interaction terms for the perturbative calculations. We will elaborate on this in Sec.~\ref{sec:power_counting}.

To recover a well-defined scaling in the PN parameter $v/c$, recalling the relation in Eq.~\eqref{eq:separation_of_scales} for the separation of scales, we further split the five fields $\hat{W} = (\phi, \phi_{\rm EM}, \A_i, \A_{{\rm EM},i}, \sigma_{ij})$ in potential ($W$) and radiation $(\bar{W})$ modes
\begin{equation}
    \hat{W} = W + \overline{W}\ ,
    \label{eq:potential_radiation_decomposition}
\end{equation}
respectively with wavelengths $R_s < k^{-1} < r$, associated to the near zone, and $k^{-1} > r$, associated to the far zone. Doing so the momenta of the potential fields scale as $(k^0, \mathbf{k}) \sim (\frac{v}{c\,r}, \frac{1}{r})$, whereas the radiation fields, which are allowed to be on shell, have momenta that scale as $(k^0, \mathbf{k}) \sim (\frac{v}{c\,r}, \frac{v}{c\,r})$~\cite{Goldberger:2004jt}.

In this work we are interested only in the dynamics of the binary system, to NNLO (next-to-next-to-leading order) beyond the leading Newtonian and Coulomb one. Hence, apart from the radiation reaction due to electromagnetic radiation which first contributes at 1.5PN order (and is evaluated in Sec.~\ref{sec:dissipations}), the relevant contributions come solely from the conservative near zone EFT~\footnote{Further hereditary contributions to the conservative sector due to the radiation EFT, such as the tail effect resulting from electromagnetic waves back-scattering off the gravitational potential, are expected to first contribute at 3PN order~\cite{Wilson-Gerow:2023syq,Blanchet:1987wq,Blanchet:1992br,Damour:2014jta,Galley:2015kus,Brunello:2022zui,Porto:2024cwd}. Therefore we expect no divergence in the conservative instantaneous Lagrangian at the order currently considered.}. 
Therefore in the remainder of this section only potential fields in the near zone are considered, and it suffices to set $\overline{W} = 0$.

\subsection{Power counting}
\label{sec:power_counting}

To obtain the relevant action terms, specifying the interactions of the near zone EFT we are constructing, we insert the above decomposition, given by Eqs.~\eqref{eq:metric_decomposition}~-~ \ref{eq:potential_radiation_decomposition}, into the full 
action, given 
in \eqref{eq: total E-M action}, and expand it to obtain a series expansion in the PN parameter $v/c$ and polynomial expression in the Kol-Smolkin fields.
Such expansion contains an infinite number of terms, therefore it is necessary to introduce a power counting scheme. When computing quantities to a given PN order, this set of rules allows to select only a finite subset of interaction terms relevant to the computation, yielding a finite number of diagrams to evaluate.

To introduce the PN expansion power counting in Einstein-Maxwell theory, we consider the most generic scenario of astrophysical interest, comprising bound binary systems where the magnitude of the leading order Coulomb force is comparable to the Newtonian gravitational force, i.e.
\begin{equation}
    \frac{G m_1 m_2}{r^2} \sim \frac{1}{\epsilon_0} \frac{q_1 q_2}{r^2}\ .
    \label{eq:similar_Coulomb_Newton_forces}
\end{equation}
Further assuming a bound binary system with comparable masses of order $m$, and using Eq.~\eqref{eq:similar_Coulomb_Newton_forces}, from the virial theorem it follows that the typical velocities $v$ will be of order
\begin{equation}
   \overline{\epsilon} \sim \frac{v^2}{c^2} \sim \frac{G m}{r c^2} \sim \frac{1}{\epsilon_0} \frac{q_1 q_2}{m r c^2}\ .
\end{equation}
We recall that the PN expansion is organized as a series in powers of $\overline{\epsilon}$. We can see then each of the above three dimensionless terms increases the PN count by one order, and all of them carry a factor of $1/c^2$ by dimensional analysis. Moreover, the PN suppression due to time derivatives $\partial_0 = \frac{1}{c} \partial_t$ and three velocities $v^\mu/c = (1, \mathbf{v}/c)$ is also correctly accounted for by the factors of $c$. Furthermore one can see that, when employing $\epsilon_0$ as the electromagnetic coupling constant, the leading-order effective Lagrangian for the Newton and Coulomb potential have no factor of $c$. 
Therefore tracking the powers of $1/c^2$ in the effective action, as dictated by dimensional analysis, will correctly indicate the PN order of any given term.

Additionally, given our choice of field normalization performed in Eqs.~\eqref{eq:metric_decomposition} and~\eqref{eq:potential_decomposition}, the kinetic terms of the expanded action, and therefore the field propagators, do not introduce any factor of $c$; nor do intermediate computations yielding the effective action from the diagrams.
This implies that to determine the PN order to which a diagram will contribute it suffices to multiply the factors of $1/c^2$ present in the Feynman rules for the bulk and worldline vertices that enter any diagram: that is, a diagram with a total factor $1/c^{2n}$ will contribute at order $n$-PN. This procedure provides the power counting rule for the EFT, which allows us to enumerate and consider only a finite amount of Feynman rules and diagrams at any given PN order.

Let us notice that the assumption of comparable electromagnetic and gravitational forces we made in Eq.~\eqref{eq:similar_Coulomb_Newton_forces} implies that the PN expansion we introduced holds even for the case of extremal Reissner-Nordstr\"om black holes. 
Conversely, when the electromagnetic force is weaker than gravity, the corresponding electromagnetic contributions are suppressed. In this regime, the PN expansion of the electromagnetic terms attains an accuracy that exceeds the estimate suggested by the PN counting alone.

Given the power counting rules presented above, we proceed with the PN expansion of the full action. 
As a first step, we implement the KK decomposition, given in Eqs.~\eqref{eq:metric_decomposition} and~\eqref{eq:potential_decomposition}, in the action~\eqref{eq: total E-M action}. 
Applying the PN power counting, we expand the action to obtain a polynomial expression in the fields, which we truncate at 2PN order, i.e.~$1/c^4$. 
The resulting expression specifies the interactions present in our theory at the desired PN accuracy.
These steps were carried out with a \texttt{Mathematica} code based on the \texttt{EFTofPNG} package~\cite{Levi:2017kzq}. 
In App.~\ref{app:Feynman_rules}, we elaborate on this procedure, recalling the presence of retardation corrections in the PN formalism~\cite{Goldberger:2004jt}, and we present in Eqs.~\eqref{eq:expanded_fundamental_action} the explicit action terms required to compute the conservative point-particle Lagrangian at 2PN order.

\subsection{Evaluation of conservative corrections}

\begin{figure*}
    \includegraphics[width=0.8436\textwidth]{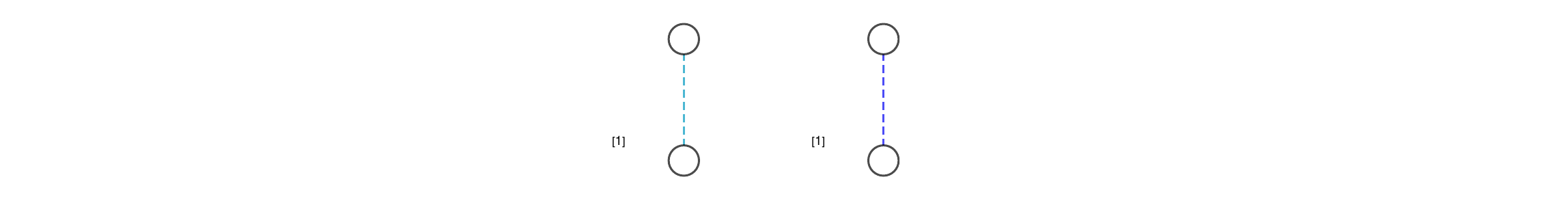}
    \includegraphics[width=0.8436\textwidth]{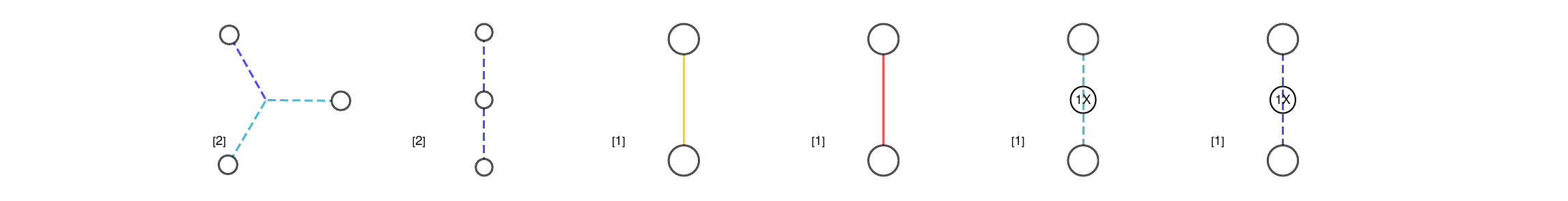}
    \includegraphics[width=0.95\textwidth]{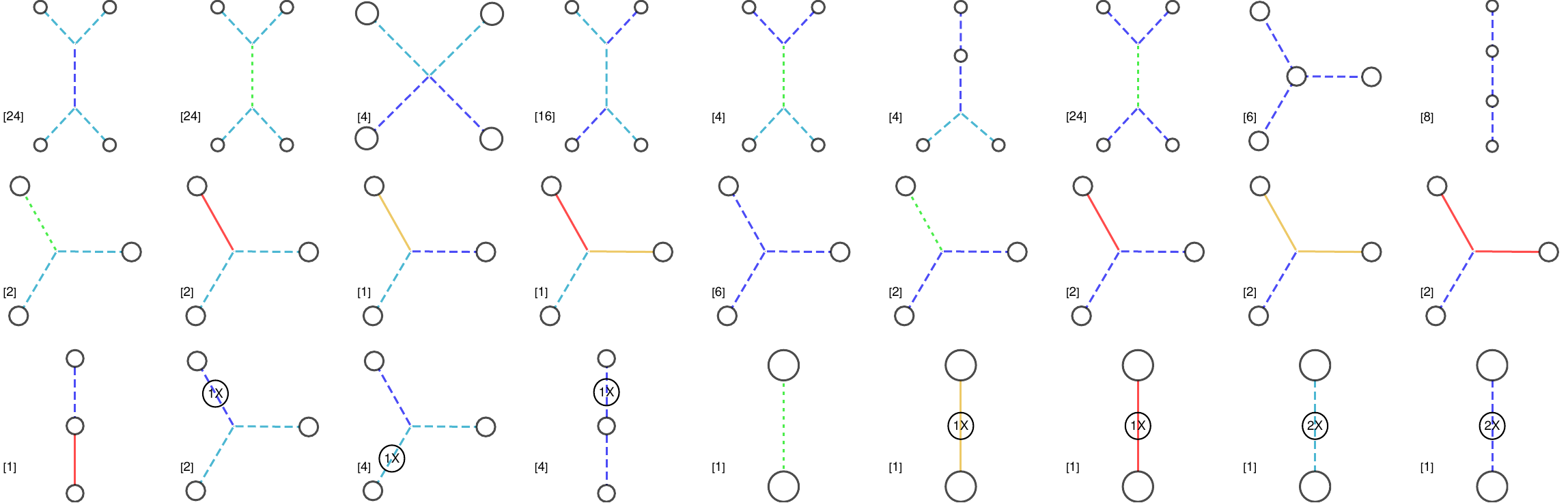}
    \caption{The 35 Feynman diagrams contributing up to second PN order (2PN) in the conservative sector. \\
    The first row shows the 2 diagrams first contributing at 0PN order, the second row represents the 6 diagrams first contributing at 1PN order, whereas the last three rows represent the 27 diagrams first contributing at 2PN order. In this figure we employ the generic worldline diagram representation: its connection to the explicit worldline diagram representation is reported in Fig.~\ref{fig:diagrams_equivalence}. 
    The worldline and bulk interaction vertices represented in each diagram are understood to include also higher PN order terms, whenever such terms are present in the action: therefore, we show each diagram only at the first PN order to which it contributes, although it may contribute at higher orders as well. Propagators of scalar fields are represented with dashed lines, in blue for $\phi$ and light blue for $\phi_{\rm EM}$, propagators of vector fields are represented with solid lines, in red for $\A_i$ and in yellow for $\A_{{\rm EM}, i}$, propagators of the $\sigma_{ij}$ tensor field are represented with green dotted lines. The annotation $n{\rm X}$ over a propagator indicates it carries $n$ retardation propagator insertions. The number in square brackets on the lower left of each diagram indicates the positive multiplicity of the corresponding diagram. This number must be divided by the inverse multiplicity to obtain the total symmetry factor. The inverse multiplicity is given by the product of the number of permutations of identical legs in each vertex and the number of permutations of identical vertices.}
    \label{fig:diagrams_PN}
\end{figure*}

To obtain the PN corrections to the conservative effective Lagrangian governing the motion of the binary system, we proceed by integrating out the potential fields $W = (\phi, \phi_{\rm EM}, \A_i, \A_{{\rm EM},i}, \sigma_{ij})$ from the fundamental action, reported in Eq.~\eqref{eq:expanded_fundamental_action}, while regarding the worldlines degrees of freedom $\{x_A^\mu\}$ as external sources; that is~\cite{Goldberger:2004jt,Porto:2016pyg,Levi:2018nxp}
\begin{equation}
    \exp\left(i \int \mathrm{d}t\, L_{\rm eff}[\{x_A^\mu\}] \right) = \int \mathrm{D}[W] \exp \left(i \, S[\{x_A^\mu\}, W]\right)\ .
\end{equation}  
We do so in a perturbative way, by summing over the relevant Feynman diagrams contributing up to 2PN order, which are constructed from the Feynman rules prescribed by the fundamental action in Eqs.~\eqref{eq:expanded_fundamental_action}. Being interested only in the classical contributions to the Newtonian and Coulomb potentials, we discard diagrams with graviton or photon loops, since those would result in quantum corrections.
The relevant diagrams to be computed are reported in Fig.~\ref{fig:diagrams_PN}.

To evaluate the relevant diagrams we recognize their equivalence to multi-loop diagrams, as first shown in Refs.~\cite{Foffa:2011ub,Kol:2013ega,Foffa:2016rgu}. Specifically each diagram can be mapped to the Fourier transform of Feynman integrals arising in a $d$-dimensional massless Euclidean quantum field theory, as depicted diagrammatically in Fig.~\ref{fig:diagrams_equivalence}.
This fact allows us to employ multi-loop Feynman integral techniques, first developed in particle physics, to streamline the computations. 

\begin{figure*}
    \centering

{
\begin{equation*}
\begin{aligned}
\renewcommand{\arraystretch}{2}
\begin{array}{ @{\hspace{0.25cm}} c @{\hspace{0.25cm}} c @{\hspace{0.25cm}} c @{\hspace{0.5cm}} | @{\hspace{0.5cm}} c @{\hspace{0.25cm}} c @{\hspace{0.25cm}} c @{\hspace{0.25cm}} }
\hline \hline
\makecell{\mathrm{Generic\ worldline}\\\mathrm{diagrams}} &  & \makecell{\mathrm{Explicit\ worldline}\\\mathrm{diagrams}} & \makecell{\mathrm{Post\!-\!Newtonian}\\\mathrm{diagrams}} &  & \makecell{\mathrm{Multi\!-\!loop}\\\mathrm{diagrams}} \\
\hline
\begin{gathered}
	\begin{tikzpicture}
	\begin{feynman}
	\def\l{1}
	\def\h{1}
	\def\dotsize{1mm}
	\def\oneoversqrtthree{0.57735026919}
	\def\apothem{2*\h/3}
	\node[worldlinevertex] (a) at (-\h/2,0);
	\node[worldlinevertex] (b) at (\h/2,0);
	\node[worldlinevertex] (c) at (-\h/2,-\h);
	\node[worldlinevertex] (d) at (\h/2,-\h);
	\vertex (ia) at (-\l/2,0);
	\vertex (fa) at (\l/2,0);
	\vertex (ib) at (-\l/2,-\h);
	\vertex (fb) at (\l/2,-\h);
	\vertex (mupper) at (0,-\h/3);
	\vertex (mlower) at (0,-2*\h/3);
	
        \diagram* {		
		(ia) --[draw=none]  (fa),
		(ib) --[draw=none] (fb),
		(a) -- [potentialphi] (mupper),
		(b) -- [potentialphi] (mupper),
		(c) -- [potentialphi] (mlower),
		(d) -- [potentialphi] (mlower),
		(mupper) -- [potentialsigma] (mlower)
	};
	
	\end{feynman}
	\end{tikzpicture}
\end{gathered}
&
= 
&
\begin{gathered}
	\begin{tikzpicture}
	\begin{feynman}
	\def\l{1.6}
	\def\h{1}
	\def\dotsize{1mm}
	\def\apothem{2*\h/3}
	\def\oneoversqrtthree{0.57735026919}
	\vertex (a) at (-\h/2,0) ;
	\vertex (b) at (\h/2,0) ;
	\vertex (c) at (-\h/2,-\h);
	\vertex (d) at (\h/2,-\h) ;
	\vertex (ia) at (-\l/2,0);
	\vertex (fa) at (\l/2,0);
	\vertex (ib) at (-\l/2,-\h);
	\vertex (fb) at (\l/2,-\h);
	\vertex (mupper) at (0,-\h/3);
	\vertex (mlower) at (0,-2*\h/3);
	
	\diagram* {		
		(ia) --[worldline]  (fa),
		(ib) --[worldline] (fb),
		(a) -- [potentialphi] (mupper),
		(b) -- [potentialphi] (mupper),
		(c) -- [potentialphi] (mlower),
		(d) -- [potentialphi] (mlower),
		(mupper) -- [potentialsigma] (mlower)
	};
	
	\end{feynman}
	\end{tikzpicture}
\end{gathered} 
+
\begin{gathered}
	\begin{tikzpicture}
	\begin{feynman}
	\def\l{1.6}
	\def\h{1}
	\def\dotsize{1mm}
	\def\apothem{2*\h/3}
	\def\oneoversqrtthree{0.57735026919}
	\vertex (a) at (-\h/2,0) ;
	\vertex (b) at (\h/2,0) ;
	\vertex (c) at (-\h/2,-\h);
	\vertex (d) at (\h/2,-\h) ;
	\vertex (e) at (-\h/2,-\h/2);
	\vertex (f) at (\h/2,-\h/2) ;
	\vertex (ia) at (-\l/2,0);
	\vertex (fa) at (\l/2,0);
	\vertex (ib) at (-\l/2,-\h);
	\vertex (fb) at (\l/2,-\h);
	\vertex (mupper) at (0,-\h/3);
	\vertex (mlower) at (0,-2*\h/3);
	
	\diagram* {		
		(ia) --[worldline]  (fa),
		(ib) --[worldline] (fb),
		(a) -- [potentialphi] (c),
		(b) -- [potentialphi] (d),
		(e) -- [potentialsigma] (f)
	};
	
	\end{feynman}
	\end{tikzpicture}
\end{gathered}
+
\begin{gathered}
	\begin{tikzpicture}
	\begin{feynman}
	\def\l{1.6}
	\def\h{1}
	\def\dotsize{1mm}
	\def\apothem{2*\h/3}
	\def\oneoversqrtthree{0.57735026919}
	\vertex (a) at (-\h/2,0) ;
	\vertex (b) at (\h/2,0) ;
	\vertex (c) at (-\h/2,-\h);
	\vertex (d) at (\h/2,-\h) ;
	\vertex (e) at (-\h/2,-\h/2);
	\vertex (f) at (\h/2,-\h/2) ;
	\vertex (h) at (0,-\h) ;
	\vertex (ia) at (-\l/2,0);
	\vertex (fa) at (\l/2,0);
	\vertex (ib) at (-\l/2,-\h);
	\vertex (fb) at (\l/2,-\h);
	\vertex (mupper) at (0,-\h/3);
	\vertex (mlower) at (0,-2*\h/3);
	
	\diagram* {		
		(ia) --[worldline]  (fa),
		(ib) --[worldline] (fb),
		(a) -- [potentialphi] (mupper),
		(d) -- [potentialphi] (mupper),
		(h) -- [potentialphi] (mlower),
		(c) -- [potentialphi] (mlower),
		(mupper) -- [potentialsigma] (mlower)
	};
	
	\end{feynman}
	\end{tikzpicture}
\end{gathered}
+ ( 1 \leftrightarrow 2)	&
\begin{gathered}
	\begin{tikzpicture}
	\begin{feynman}
	\def\l{1.8}
	\def\h{1}
	\def\dotsize{1mm}
	\def\apothem{2*\h/3}
	\def\oneoversqrtthree{0.57735026919}
	\vertex (a) at (-\h/2,0) ;
	\vertex (b) at (\h/2,0) ;
	\vertex (c) at (-\h/2,-\h);
	\vertex (d) at (\h/2,-\h) ;
	\vertex (e) at (-\h/2,-\h/2);
	\vertex (f) at (\h/2,-\h/2) ;
	\vertex (h) at (0,-\h) ;
	\vertex (ia) at (-\l/2,0);
	\vertex (fa) at (\l/2,0);
	\vertex (ib) at (-\l/2,-\h);
	\vertex (fb) at (\l/2,-\h);
	\vertex (mupper) at (0,-\h/3);
	\vertex (mlower) at (0,-2*\h/3);
    
        \fill[gray!30] (a) -- (b) -- (d) -- (c) -- cycle;
    
	\diagram* {		
		(ia) --[worldline]  (fa),
		(ib) --[worldline] (fb)
	};
	
	\end{feynman}
	\end{tikzpicture}
\end{gathered}
& 
\simeq
&
\begin{gathered}
\begin{tikzpicture}
\begin{feynman}
    \def\l{1.8}
    \def\h{1}
    \def\dotsize{1mm}
    \def\apothem{2*\h/3}
    \def\oneoversqrtthree{0.57735026919}
    
    \vertex (a) at (-\h/2,0);
    \vertex (b) at (\h/2,0);
    \vertex (c) at (-\h/2,-\h);
    \vertex (d) at (\h/2,-\h);
    \vertex (e) at (-\h/2,-\h/2);
    \vertex (f) at (\h/2,-\h/2);
    \vertex (h) at (0,-\h);
    \vertex (ia) at (-\l/2,0);
    \vertex (fa) at (\l/2,0);
    \vertex (ib) at (-\l/2,-\h);
    \vertex (fb) at (\l/2,-\h);
    \vertex (mupper) at (0,-\h/3);
    \vertex (mlower) at (0,-2*\h/3);
    \vertex (mtop) at (0,0);
    \vertex (mbottom) at (0,-\h);
    \vertex (midcenter) at ($ (mtop)!0.5!(mbottom) $);
    \vertex (mtopupper) at (0,\h/3);
    \vertex (mbottomlower) at (0,-4*\h/3);

    \diagram* {		
        (mtopupper) --[worldline]  (mbottomlower)
    };

    \fill[gray!30] (midcenter) circle (\h/2);
	
\end{feynman}
\end{tikzpicture}
\end{gathered} \\
\hline \hline
\end{array}
\end{aligned} 
\end{equation*}
}

    \caption{Relation between different representations of PN diagrams (left side), and their subsequent connection to multi-loop diagrams (right side). 
    Focusing first on the left side of the figure, the left-hand side of the equality shows the generic worldline diagram representation (employed in Fig.~\ref{fig:diagrams_PN}), where the worldline vertices are represented as empty circles, and their Feynman rules understand the summation over the worldine indices $\sum_{A=1,2}$. The right-hand side of the equality shows the explicit worldline diagram representation, with worldlines already specified to either $A = 1$ or $A = 2$, and depicted explicitly as horizontal thick lines, even though they are not propagating. 
    As shown in the figure, these two representations are equivalent. More specifically, each generic worldline diagram represents an equivalence class of possibly several explicit worldline diagrams, obtained by fixing the worldline labels. Incidentally, the symmetry factors reported in Fig.~\ref{fig:diagrams_PN} refer to the generic worldline diagrams, whereas the explicit worldline diagrams belonging to the same equivalence class may carry different symmetry factors each.
    Focusing now on the right part of the figure (adapted from~\cite{Foffa:2016rgu, Mandal:2022nty, Mandal:2022ufb, Mandal:2023hqa, Mandal:2023lgy}), we depict the connection between each PN diagram, here with explicit worldlines, and multi-loop diagrams. In particular, each of the PN diagrams, after solving the worldline algebra, can be mapped to (the Fourier transform of) multi-loop diagrams, in particular exactly to two-point loop diagrams arising in a $d$-dimensional Euclidean quantum field theory with a single massless scalar field. The gray-colored area represents arbitrarily complicated loop structures.
    }
    \label{fig:diagrams_equivalence}

\end{figure*}
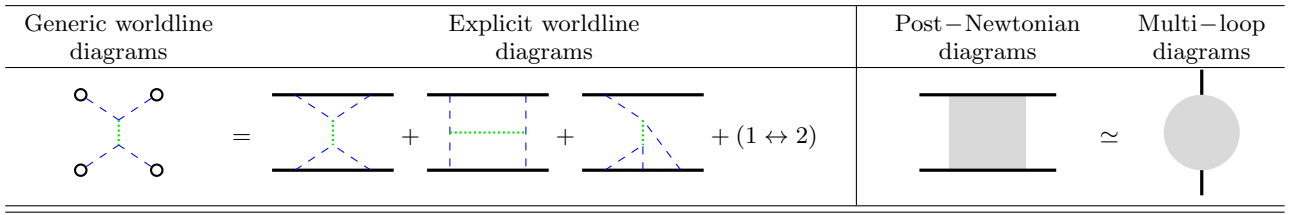

Specifically we employ dimensional regularization, to regularize divergent integrals, tensor decomposition to extract scalar Feynman integral from the intermediate tensorial expressions, and integration-by-parts identities~\cite{Chetyrkin:1981qh,Tkachov:1981wb,Laporta:2000dsw}, to reduce via algebraic relations the evaluation of the many scalar integrals to the computation of a minimal set of master integrals. 

The master integrals needed to evaluate the conservative Lagrangian at NNLO in Einstein-Maxwell theory, which have up to two loops, are the same master integrals appearing in the corresponding vacuum GR limit. In particular the two two-loop master integrals are either nested, or factorizable into, one loop integrals. The single one loop master integral necessary to perform all the computations, as well as the Fourier transform integral, are reported in Refs.~\cite{Gilmore:2008gq,Foffa:2011ub}.

The algorithmic procedure used to generate and evaluate the conservative diagrams has been implemented in a in-house software package developed in the \texttt{Mathematica} computer algebra system, and will be explained in more detail in a future publication~\cite{Bartolo:2025preparation}. This package interfaces with the \texttt{EFTofPNG}  package~\cite{Levi:2017kzq} for KK field decomposition in the fundamental action, \texttt{xAct/xTensor} package~\cite{xAct} for tensor algebra, the \texttt{FiniteFlow} package \cite{Peraro:2019svx} for linear algebra evaluation over finite fields, and the \texttt{LiteRed} and \texttt{Mint} packages~\cite{Lee:2012cn,Lee:2013mka,Lee:2013hzt} interfaced with \texttt{Fermat} CAS~\cite{Fermat} for Feynman integral topology mapping and reduction through integration-by-parts identities.

\subsection{The 2PN Lagrangian in Einstein-Maxwell theory}
We report here the expression for the 2PN Lagrangian that governs the conservative dynamics of a charged binary system, in the harmonic gauge for the metric and in the Lorenz gauge for the electromagnetic field. 
The Lagrangian is organized into distinct contributions: the kinetic terms $L_{\mathrm{kin}}$, the 2PN contributions from pure General Relativity $L_{\mathrm{2PN}}$, the 2PC (post-Coulombian) contributions from electromagnetism $L_{\mathrm{2PC}}$, and the mixed terms arising in the full Einstein-Maxwell theory at order $\frac{1}{c^4}$, denoted $L_{\mathrm{mixed}}$.
The full Lagrangian $L$ is then given by~\cite{SupplementalMaterial}
\begin{widetext}
    \begin{subequations}
    \label{eq:Conservative_Lagrangian_EFT}
    \begin{equation}
        L = L_{\mathrm{kin}} + L_{2\mathrm{PN}} + L_{2\mathrm{PC}} + L_{\mathrm{mixed}}  \ ,
    \end{equation}
    \begin{equation}
        L_{\mathrm{kin}} = \frac{m_1 v_1^2}{2} + \frac{1}{c^2} \left\{ \frac{ m_1 v_1^4}{8} \right\} + \frac{1}{c^4} \left\{ \frac{m_1 v_1^6}{16}  \right\} +\big( 1 \leftrightarrow 2\big) \ ,
    \end{equation}
\begin{align} \notag
            L_{2\mathrm{PN}} =& \frac{G m_1 m_2}{2 r} \\ \notag
            &+ \frac{1}{c^2} \left\{ G \left(\frac{m_1 m_2}{r} \left(\frac{1}{4} \left(-\left(\mathbf{v_1}\cdot\mathbf{n}\right) \left(\mathbf{v_2}\cdot\mathbf{n}\right)-7 \left(\mathbf{v_1}\cdot\mathbf{v_2}\right)\right)+\frac{3 v_1^2}{2}\right)\right)  - G^2 \left( \frac{ m_1^2 m_2}{2 r^2} \right) \right\} \\ \notag
            & + \frac{1}{c^4} \left\{ G \left( m_1 m_2 \left(\frac{1}{16} r \left(15 \left(\mathbf{a_1}\cdot\mathbf{a_2}\right)-\left(\mathbf{a_1}\cdot\mathbf{n}\right) \left(\mathbf{a_2}\cdot\mathbf{n}\right)\right) \right.  \right. \right. \\ \notag
            & \hspace{1.5cm} \left. \left. \left. +\frac{1}{8} \left(v_2^2 \left(\mathbf{a_1}\cdot\mathbf{n}\right)+\left(\mathbf{a_2}\cdot\mathbf{n}\right) \left(\mathbf{v_1}\cdot\mathbf{n}\right){}^2+14 \left(\mathbf{a_2}\cdot\mathbf{v_1}\right) \left(\mathbf{v_1}\cdot\mathbf{n}\right)-12 \left(\mathbf{a_2}\cdot\mathbf{v_2}\right) \left(\mathbf{v_1}\cdot\mathbf{n}\right)\right) \right. \right. \right. \\ \notag
            & \hspace{1.5cm} \left. \left. \left.  +\frac{1}{16 r}\left( -12 v_1^2 \left(\mathbf{v_1}\cdot\mathbf{n}\right) \left(\mathbf{v_2}\cdot\mathbf{n}\right)-2 v_2^2 \left(\mathbf{v_1}\cdot\mathbf{n}\right){}^2+3 \left(\mathbf{v_1}\cdot\mathbf{n}\right){}^2 \left(\mathbf{v_2}\cdot\mathbf{n}\right){}^2 \right. \right. \right. \right. \\ \notag
            & \hspace{1.5cm} \left. \left. \left. \left. +12 \left(\mathbf{v_1}\cdot\mathbf{v_2}\right) \left(\mathbf{v_1}\cdot\mathbf{n}\right) \left(\mathbf{v_2}\cdot\mathbf{n}\right) -20 v_1^2 \left(\mathbf{v_1}\cdot\mathbf{v_2}\right)+2 \left(\mathbf{v_1}\cdot\mathbf{v_2}\right){}^2+14 v_1^4+3 v_2^2 v_1^2 \right)\right) \right)  \right.   \\ \notag
            & \hspace{1.15cm}+G^2 \left(\frac{m_2^2 m_1}{4 r^2} \left(2 \left(\mathbf{v_1}\cdot\mathbf{n}\right){}^2+7 v_1^2\right)+\frac{m_2 m_1^2}{2 r^2} \left(4 v_1^2-7 \left(\mathbf{v_1}\cdot\mathbf{v_2}\right)\right)\right) \\ 
            & \hspace{1.15cm} \left. +G^3 \left(\frac{m_2 m_1^3}{2 r^3}+\frac{3 m_2^2 m_1^2}{2 r^3}\right) \right\} +\big( 1 \leftrightarrow 2\big) \ ,
  \end{align}
        \begin{align}\notag
            L_{2\mathrm{PC}} =& -\frac{q_1 q_2}{2 r} \\ \notag
            &+ \frac{1}{c^2} \left\{ q_1 q_2 \left( \frac{1}{4 r} \left(\left(\mathbf{v_1}\cdot\mathbf{n}\right) \left(\mathbf{v_2}\cdot\mathbf{n}\right)+\mathbf{v_1}\cdot\mathbf{v_2}\right)  \right) \right\} \\ \notag
            & + \frac{1}{c^4} \left\{ q_1 q_2 \left(\frac{1}{16} r \left(\left(\mathbf{a_1}\cdot\mathbf{n}\right) \left(\mathbf{a_2}\cdot\mathbf{n}\right)-3 \left(\mathbf{a_1}\cdot\mathbf{a_2}\right)\right) \right. \right. \\ \notag
            & \hspace{1.5cm} +\frac{1}{8} \left(v_2^2 \left(-\left(\mathbf{a_1}\cdot\mathbf{n}\right)\right)-\left(\mathbf{a_2}\cdot\mathbf{n}\right) \left(\mathbf{v_1}\cdot\mathbf{n}\right){}^2-2 \left(\mathbf{a_2}\cdot\mathbf{v_1}\right) \left(\mathbf{v_1}\cdot\mathbf{n}\right)\right)  \\ 
            &  \hspace{1.5cm} \left. \left. +\frac{1}{16 r}\left( 2 v_2^2 \left(\mathbf{v_1}\cdot\mathbf{n}\right){}^2-3 \left(\mathbf{v_1}\cdot\mathbf{n}\right){}^2 \left(\mathbf{v_2}\cdot\mathbf{n}\right){}^2+2 \left(\mathbf{v_1}\cdot\mathbf{v_2}\right){}^2-v_1^2 v_2^2 \right)\right) \right\} +\big( 1 \leftrightarrow 2\big) \ ,
        \end{align}
        \begin{align} \notag
        \label{eq:Lmixed}
            L_{\mathrm{mixed}} =& \frac{1}{c^2} \left\{ G \left(\frac{1}{r^2}\left(m_1 q_1 q_2-\frac{1}{2} m_2 q_1^2\right) \right)\right\} \\ \notag
            & + \frac{1}{c^4} \left\{  G \left( \frac{3 q_2^2 q_1^2}{2 r^3} +\frac{m_2 q_1^2}{2 r^2} \left(2 \left(\mathbf{v_1}\cdot\mathbf{n}\right){}^2+3 \left(\mathbf{v_1}\cdot\mathbf{v_2}\right)-2 v_1^2\right) \right.  \right.   \\ \notag
        &  \hspace{1.5cm} \left. \left. + \frac{m_1 q_2}{2 r^2} \left(\frac{q_2}{2} \left(-2 \left(\mathbf{v_1}\cdot\mathbf{n}\right){}^2-v_1^2\right) +q_1 \left(-2 \left(\mathbf{v_1}\cdot\mathbf{n}\right){}^2-2 \left(\mathbf{v_1}\cdot\mathbf{v_2}\right)+v_1^2\right) \right)\right)  \right. \\ 
            & \hspace{1.15cm} \left. +G^2 \left(-\frac{m_1^2 q_1 q_2}{r^3}+ \frac{m_2 m_1 q_1}{2 r^3} \left(q_1 - 8 q_2 \right)+\frac{m_2^2 q_1^2}{r^3}\right) \right\} +\big( 1 \leftrightarrow 2\big) \ .
        \end{align}
    \end{subequations}
\end{widetext}

This result is free of poles in $\epsilon = (d - 3) \to 0$, as expected. Furthermore, in the neutral limit of vanishing electric charges $q_A \rightarrow 0$, with $A = 1,2$, it correctly reduces to the 2PN point-particle Lagrangian of Ref.~\cite{Gilmore:2008gq}, which was analogously obtained with the EFT approach in General Relavity. Moreover, in the limit $G \rightarrow 0$ (which interestingly is different from the limit of vanishing masses, as discussed below), it correctly reduces to the conservative Lagrangian at second order in the post-Coulombian expansion (2PC) computed in Refs.~\cite{Golubenkov:1956, Damour:1990jh, Buonanno:2000qq, 2001JNS....11..321K}.

Regarding the results computed in the full Einstein-Maxwell theory, we find perfect agreement with the 1PN Lagrangian computed with the EFT approach of Ref.~\cite{Patil:2020dme}\footnote{Up to a typo in Eq.~(30) therein: the signs of the last two terms in the final line should be reversed. We thank the author for confirming this to us.}. Results to 2PN order in Einstein-Maxwell theory have been computed in Ref.~\cite{Gupta:2022spq} using the EFT approach, but the Lagrangian reported therein does not include the neutral sector: if we use the neutral 2PN Lagrangian of Ref.~\cite{Gilmore:2008gq} to complete the charged 2PN Lagrangian of Ref.~\cite{Gupta:2022spq}, we find a result that is in disagreement with Eq.~\ref{eq:Conservative_Lagrangian_EFT}, specifically at the $1/c^4$ order in the charged sector. We note however that, as discussed in Sec.~\ref{subsec:DZ_terms}, starting at 2PN order, double zero terms added to the neutral sector can alter the charged sector contributions. Therefore, we cannot rule out the possibility that the Lagrangian of Ref.~\cite{Gupta:2022spq} is physically equivalent to ours once supplemented with the right neutral part.  We will come back to this in Sec.~\ref{subsec:chi}, where we provide additional checks in the full Einstein-Maxwell theory by comparing the 2PN Lagrangians with the independent calculation made at the third order in the post-Minkowskian (PM) expansion in Refs.~\cite{Wilson-Gerow:2023syq,Alonzo-Artiles:2026wbe}. 
We notice moreover that the mixed Lagrangian in Eq.~\eqref{eq:Lmixed} presents the 2PN contribution $\frac{G}{c^4}\frac{3 q_1^2 q_2^2}{r^3}$ which does not depend on the masses of the compact objects and hence, as anticipated, it is present even in the limit of vanishing masses, while still depending on the gravitational constant $G$. This term corresponds in fact to an interaction between the gravitational field and the electromagnetic field, with the compact objects sourcing only the latter, and corresponds to the first two diagrams in the third row of Fig.~\ref{fig:diagrams_PN}. We may interpret this term as the contribution involving a gravitational field sourced solely by the energy carried by the electromagnetic waves, via the two-photons one-graviton bulk vertex.
From another perspective, coupling electromagnetism to GR induces nonlinear interactions in the electromagnetic field once the gravitational field is integrated out. In particular, the coupling between the electromagnetic field and the metric generates an effective four-point (and higher) bulk vertex for the electromagnetic field. Consequently, the term discussed above can also be interpreted as an irreducible general-relativistic classical correction to the Coulomb potential, since the coupling between the electromagnetic and gravitational fields, as specified by the fundamental action in Eq.~\eqref{eq:S_EM}, cannot be eliminated by any experiment or observation.

Conversely, within the Einstein-Maxwell theory, a contribution to the binding energy arising from a component of the electromagnetic field sourced solely by the gravitational field, i.e.~a diagram where the electromagnetic field does not couple to any worldline, is forbidden at the classical level. Therefore, in the limit of vanishing electric charges of the compact objects, no electromagnetic contribution can appear in the effective Lagrangian of the system. This restriction follows from the fact that the electromagnetic field appears at least (and actually only) quadratically in the bulk action in Eqs. \eqref{eq:S_EM} and \eqref{eq:S_EM_gf}, with no terms linear in the electromagnetic field. However, such contribution is present at the quantum level via diagrams with at least one quantum loop, for example from diagrams which include quantum loop corrections to the graviton propagator \cite{Capper:1973pv,tHooft:1974toh, Donoghue:1993eb,Donoghue:1994dn,Muzinich:1995uj, Hamber:1995cq, Donoghue:1996mt, Donoghue:2001qc, Bjerrum-Bohr:2002fji, Burgess:2003jk,Brunello:2022zui}.

\subsection{Double Zero Terms}
\label{subsec:DZ_terms}
Based on the calculations carried out so far, we have derived an expression for the Lagrangian describing a binary system of two charged black holes. 

It is important to point out that, in the neutral limit, the Lagrangian given in Eq.~\eqref{eq:Conservative_Lagrangian_EFT} does not match the form of the Lagrangian presented in Ref.~\cite{Blanchet:2013haa}. However, the neutral limit of Eq.~\eqref{eq:Conservative_Lagrangian_EFT} can be related to the well-known Lagrangian in Ref.~\cite{Blanchet:2013haa} by adding a total derivative term and two double zero terms at 2PN order, as previously shown in Ref.~\cite{Gilmore:2008gq}. Inserting these terms modifies the form of the Lagrangian but leaves unchanged the EoMs, because they vanish at 2PN when the lower order of EoMs are used. Indeed, this also means that we remain in the harmonic gauge. 

This procedure can be generalized to the charged case by identifying a new total derivative term and two new double-zero terms, all of which depend on the charges. The explicit expressions for the additional terms are

\begin{widetext}
\begin{subequations}
\label{eq: doubleZeroTerms}
    \begin{align}
        &\delta L_1=\, \frac{1}{c^4}\frac{G m_1 m_2 r}{8}\left( \,  \mathbf{n}\cdot \mathbf{a}_1\, + \frac{G m_2}{r^2}- \frac{q_1 q_2}{m_1 r^2}\right)\left( \,  \mathbf{n}\cdot \mathbf{a}_2\, - \frac{G m_1}{r^2}+ \frac{q_1 q_2}{m_2 r^2}\right), \\ 
        &\delta L_2=\, - \frac{1}{c^4}\frac{15G m_1 m_2 r}{8} \left(\mathbf{a}_1+ \frac{G m_2}{r^2}\mathbf{n}- \frac{q_1 q_2}{m_1 r^2}\mathbf{n}\right)\cdot \left(\mathbf{a}_2- \frac{G m_1}{r^2}\mathbf{n}+ \frac{q_1 q_2}{m_2 r^2}\mathbf{n}\right), \\ 
        &\delta L_3= \frac{1}{c^4}\frac{d}{dt}\left[ \frac{7 G^2 m_1 m_2}{4 r}\Big(m_2 (\mathbf{n}\cdot \mathbf{v}_2)- m_1 (\mathbf{n}\cdot \mathbf{v}_1)\Big)+ \frac{3 G m_1 m_2}{4}\Big( v_2^2( \mathbf{n}\cdot \mathbf{v}_1)- v_1^2 (\mathbf{n}\cdot \mathbf{v}_2) \Big) \right].
        \end{align}
\end{subequations}
\end{widetext}
The Lagrangian obtained through this procedure is physically equivalent to the original one, as both yield the same EoMs. 

Before presenting the result, we apply an additional modification to the Lagrangian, which does not alter the gauge condition and the EoMs. Specifically, we introduce other double-zero terms that allow us to eliminate any quadratic dependence on the accelerations. This procedure enables us to obtain a Lagrangian that is linear in the accelerations, as required to apply the method employed in Ref.~\cite{deAndrade:2000gf}, which we will follow in Sec.~\ref{subsec:adm_coord} to obtain the Lagrangian in ADM-type coordinates. In this case the idea is to reformulate quadratic-acceleration terms as a combination of a linear-in-acceleration term and a double zero term. This can be also generalized to terms involving higher powers of the accelerations. In our specific case we can rewrite the two terms of the Lagrangian quadratic in the accelerations as follows
\begin{widetext}
\begin{subequations}
\begin{align}
    &(\mathbf{n}\cdot \mathbf{a}_1) (\mathbf{n}\cdot \mathbf{a}_2) \to \frac{1}{r^4}\left( G^2 m_1 m_2 - 2 G q_1 q_2+\frac{q_1^2 q_2^2}{m_1 m_2}\right) + (\mathbf{n}\cdot \mathbf{a}_1) \left(\frac{G m_1}{r^2}- \frac{q_1 q_2}{m_2 r^2}\right)-(\mathbf{n}\cdot \mathbf{a}_2) \left(\frac{G m_2}{r^2}- \frac{q_1 q_2}{m_1 r^2}\right), \\ 
    &\mathbf{a}_1 \cdot \mathbf{a}_2\to \frac{1}{r^4}\left( G^2 m_1 m_2 - 2 G q_1 q_2+\frac{q_1^2 q_2^2}{m_1 m_2}\right) + (\mathbf{n}\cdot \mathbf{a}_1) \left(\frac{G m_1}{r^2}- \frac{q_1 q_2}{m_2 r^2}\right)-(\mathbf{n}\cdot \mathbf{a}_2) \left(\frac{G m_2}{r^2}- \frac{q_1 q_2}{m_1 r^2}\right).
\end{align}
\end{subequations}
Considering the same separation used in Eq.\eqref{eq:Conservative_Lagrangian_EFT}, the final Lagrangian reads~\cite{SupplementalMaterial}
    \begin{subequations}
    \label{eq:: LagrangianHarmonicCoordinates12}
    \begin{equation}
        L_{\mathrm{kin}} = \frac{m_1 v_1^2}{2} + \frac{1}{c^2} \left\{ \frac{ m_1 v_1^4}{8} \right\} + \frac{1}{c^4} \left\{ \frac{m_1 v_1^6}{16}  \right\} +\big( 1 \leftrightarrow 2\big) \ ,
    \end{equation}
        \begin{align} \notag
            L_{2\mathrm{PN}} =& \frac{G m_1 m_2}{2 r} \\ \notag
            &+\frac{1}{c^2} \Biggl\{\frac{G m_1 m_2}{2r}\left(3 v_1^2-\frac{1}{2}(\mathbf{n}\cdot \mathbf{v}_1)(\mathbf{n}\cdot \mathbf{v}_2)-\frac{7}{2}\mathbf{v}_1 \cdot \mathbf{v}_2\right)-\frac{G^2 m_1^2m_2}{2 r^2} \Biggr\} \\ \notag
            &+\frac{1}{c^4}\Biggl\{\frac{G m_1 m_2}{8}\left(14 (\mathbf{a}_2\cdot \mathbf{v}_1)(\mathbf{n}\cdot \mathbf{v}_1)+ (\mathbf{n}\cdot \mathbf{a}_2)(\mathbf{n}\cdot \mathbf{v}_1)^2+7 (\mathbf{n}\cdot \mathbf{a}_1) v_2^2 \right) +\frac{G^3 m_1^2 m_2}{2 r^3}\left(m_1 +\frac{19}{4}  m_2\right) \\ \notag
            &\hspace{1 cm}+ \frac{G^2 m_1^2 m_2}{4 r^2}\left( 14(\mathbf{n}\cdot \mathbf{v}_1)^2-14(\mathbf{n}\cdot \mathbf{v}_1)(\mathbf{n}\cdot \mathbf{v}_2)+2 (\mathbf{n}\cdot \mathbf{v}_2)^2+ v_1^2-7 \mathbf{v}_1 \cdot \mathbf{v}_2 +7 v_2^2 \right) \\ \notag
            &\hspace{1 cm} +\frac{G m_1 m_2}{16 r}\Bigl(3 (\mathbf{n}\cdot \mathbf{v}_1)^2(\mathbf{n}\cdot \mathbf{v}_2)^2+14 v_1^4+12 (\mathbf{n}\cdot \mathbf{v}_1)(\mathbf{n}\cdot \mathbf{v}_2)(\mathbf{v}_1\cdot \mathbf{v}_2)-32 v_1^2 (\mathbf{v}_1\cdot \mathbf{v}_2) \\ 
            & \hspace{1.5  cm}+2(\mathbf{v}_1\cdot \mathbf{v}_2)^2-14 v_2^2(\mathbf{n}\cdot \mathbf{v}_1)^2+15 v_1^2 v_2^2 \Bigr)\Biggr\}+\big( 1 \leftrightarrow 2\big)\ ,
        \end{align}
    \begin{align}\notag
        L_{2 \mathrm{PC}}= & -\frac{q_1 q_2}{2 r} \\ \notag
        &+ \frac{1}{c^2}\Biggl\{\frac{q_1 q_2}{4 r}\Bigl( \mathbf{v}_1 \cdot \mathbf{v}_2+(\mathbf{n}\cdot \mathbf{v}_1)(\mathbf{n}\cdot \mathbf{v}_2)\Bigr) \Biggr\} \\ \notag
        &+\frac{1}{c^4}\Biggl\{\frac{q_1 q_2}{8}\Bigl(2 (\mathbf{n}\cdot \mathbf{v}_2)(\mathbf{a}_1\cdot \mathbf{v}_2)+(\mathbf{n}\cdot \mathbf{a}_1)(\mathbf{n}\cdot \mathbf{v}_2)^2-v_2^2(\mathbf{n}\cdot \mathbf{a}_1) \Bigr) \\ 
        &\hspace{1cm}+\frac{q_1 q_2}{16 r}\Bigr(2v_2^2(\mathbf{n}\cdot \mathbf{v}_1)^2-3(\mathbf{n}\cdot \mathbf{v}_1)^2(\mathbf{n}\cdot \mathbf{v}_2)^2+2 (\mathbf{v}_1\cdot \mathbf{v}_2)^2-v_1^2 v_2^2 \Bigr)\Biggr\}+\big( 1 \leftrightarrow 2\big)\ ,
    \end{align}
    \begin{align}\notag
        L_{\rm mixed}=& \frac{1}{c^2}\Biggl\{\frac{G m_1 q_1 q_2}{r^2}-\frac{G m_2 q_1^2}{2 r^2} \Biggr\} \\ \notag
        &+\frac{1}{c^4} \Biggl\{ \frac{21 G q_1^2 q_2^2}{8 r^3}+\frac{q_1^2 q_2^2}{4 m_2 r}(\mathbf{n}\cdot \mathbf{a}_1)-\frac{q_1^3 q_2^3}{8 m_1 m_2 r^3}+\frac{G^2 m_2^2}{r^3}\left(q_1^3-q_1 q_2\right) -\frac{2 G m_1 q_1 q_2}{r}(\mathbf{n}\cdot \mathbf{a}_1) \\ 
        &\hspace{1 cm}+ \frac{G^2 m_1 m_2 q_1}{8r}(4 q_1-47  q_2)+\frac{G m_1 q_1 q_2}{2 r^2}\Bigl( v_1^2- 2 (\mathbf{n}\cdot \mathbf{v}_1)^2-\mathbf{v}_1\cdot \mathbf{v}_2\Bigr)\Biggr\}+\big( 1 \leftrightarrow 2\big)\ .
    \end{align}
\end{subequations}
\end{widetext}

This Lagrangian agrees with the 1PN results of Ref.~\cite{Julie:2017rpw,Khalil:2018aaj,Patil:2020dme}, and, in the neutral limit, with the 2PN result of Ref.~\cite{Blanchet:2013haa}.
\section{Equations of motion and Hamiltonian}
\label{sec:eom and Hamiltonian}

In this section we continue our characterization of the 2PN conservative dynamics of a charged BBH, deriving explicitly the EoMs of the two component black holes, the conserved total linear momentum and the two-body Hamiltonian in an ADM-type coordinate system whose definition will be specified below. 

In the neutral limit, $(q_1, q_2) \to (0,0)$, our results  reduce to the known expressions for neutral black holes. Specifically, the EoMs  coincide with those derived in Refs.~\cite{Blanchet:2000ub,Blanchet:2013haa}, while the linear momentum and the Hamiltonian agree with those presented in  Ref.~\cite{deAndrade:2000gf}.

We specify that, at this level, no dissipation is taken into account. The dissipative contributions that, as we shall see, enter the 2PN dynamics at 1.5PN order, will be addressed in Sec.~\ref{sec:dissipations}.

\subsection{Equations of motion and momenta} 
\label{subsec:2.1} 

From the generalized (i.e.~acceleration-dependent) Lagrangian of Eq.~\eqref{eq:: LagrangianHarmonicCoordinates12}, the conservative part of the harmonic-coordinate EoMs for body $A=1,2$ are computed through the functional-derivative equation
\begin{equation}
\label{eq: functional-derivative equation}
    \frac{\delta L}{\delta y^i_A }\equiv \frac{\partial L}{\partial y^i_A}-\frac{d}{dt}\Biggl( \frac{\partial L}{\partial v^i_A} \Biggr)+\frac{d^2}{dt^2}\Biggl( \frac{\partial L}{\partial a^i_A}\Biggr) = 0.
\end{equation}
This is solved perturbatively for $a_A^i$ while removing any derivative of the acceleration via the standard order-reduction procedure, i.e.~by means of the lower-order EoMs determined at the previous perturbative steps.~\footnote{The order reduction procedure can be performed here without any consequential change of coordinates. The same is not true at the level of the harmonic Lagrangian, where the order reduction leads out of the harmonic gauge.} 

For $A=1$, our result is~\cite{SupplementalMaterial}
\begin{widetext}
    \begin{align}
    \label{eq: EOM1}
   \notag a_1^i=&-\frac{G m_2 n^i}{r^2}  \Biggl(1-\frac{q_1 q_2}{G m_1 m_2}\Biggr) \\ \notag
   &+\frac{1}{c^2} \Biggl\{\frac{G m_2(v_1^i-v_2^i)}{r^2}\Biggl[(\mathbf{n}\cdot \mathbf{v}_1) \left(4-\frac{q_1 q_2}{G m_1 m_2}\right)\, - 3 (\mathbf{n}\cdot \mathbf{v}_2)\Biggr]\, \\ \notag 
   &
   \hspace{1cm}+ n^i \Biggl[\frac{3 G m_2}{2 r^2}\left(\mathbf{v}_1-\mathbf{v}_2\right)^2\, + \frac{G m_2}{r^2}\left(1- \frac{q_1 q_2}{G m_1 m_2}\right)\left(\frac{3}{2}(\mathbf{n}\cdot \mathbf{v}_2)^2+ 4\mathbf{v}_1 \cdot \mathbf{v}_2-v_1^2+2v_2^2\right) \\ \notag 
   &\hspace{1.5cm}+\frac{1}{r^3}\left(5 G^2 m_1 m_2 +4 G m_2^2+ G q_2^2-7 G q_1 q_2-5 q_1 q_2 \frac{m_2}{m_1}+ \frac{G q_1^2 q_2^2}{m_1 m_2}\right)\Biggr]\Biggr\} \\ \notag
   &+\frac{1}{c^4}\Biggl\{(v_1^i-v_2^i)\Biggl[\frac{\mathbf{n}\cdot \mathbf{v}_2}{r^3}\left(\frac{55G^2 m_1 m_2}{4}- 2G^2 m_2^2+ \frac{4 G m_2 q_1^2}{m_1}-\frac{33 G q_1 q_2}{2}+ \frac{4 G m_2 q_1 q_2}{m_1}+ G q_2^2+ \frac{7 q_1^2 q_2^2}{4 m_1 m_2}\right) \\ \notag
   &\hspace{2.5cm}+ \frac{\mathbf{n}\cdot \mathbf{v}_1}{r^3}\left(\frac{41 G q_1 q_2}{2} - \frac{63G^2 m_1 m_2}{4}- 2 G^2 m_2^2-\frac{5 G m_2 q_1^2}{m_1}- 2 G q_2^+ \frac{G m_2 q_1 q_2}{m_1}-\frac{11 q_1^2 q_2^2}{4 m_1 m_2}\right)\\ \notag
   & \hspace{2.5cm}+\frac{ G m_2}{2 r^2}\Biggl(9(\mathbf{n}\cdot \mathbf{v}_2)^3 +(\mathbf{n}\cdot \mathbf{v}_1)\left(\frac{q_1 q_2}{ G m_1 m_2}v_1^2- 8( \mathbf{v_1}\cdot \mathbf{v}_2)+ 7 v^2- 9 (\mathbf{n}\cdot \mathbf{v}_2)^2\right) \\ \notag
   &\hspace{2.5cm}+(\mathbf{n}\cdot \mathbf{v}_1) \left(1- \frac{q_1 q_2}{G m_1 m_2}\right) \left(v_2^2- 3 (\mathbf{n}\cdot \mathbf{v}_2)^2\right)+2(\mathbf{n}\cdot \mathbf{v}_2)\left(v_1^2+4 (\mathbf{v}_1 \cdot \mathbf{v}_2)-5 v_2^2 \right)\Biggr)\Biggr]\\ \notag
   &\hspace{1cm}+ n^i \Biggl[ \frac{G m_2}{8 r^2}\Biggl(\frac{q_1 q_2}{G m_1 m_2}v_1^4\,-16(\mathbf{v}_1\cdot \mathbf{v}_2)^2+ 2 (\mathbf{v}_1-\mathbf{v}_2)^2 \left(9 (\mathbf{n}\cdot \mathbf{v}_2)- \frac{q_1 q_2}{G m_1 m_2}\right)+ 28 v_2^2 (\mathbf{v}_1 \cdot \mathbf{v}_2)- 13 v_2^4 \\ \notag
   &\hspace{2cm}+\left(1- \frac{q_1 q_2}{G m_1 m_2}\right) \Bigl(4 v_2^2 (\mathbf{v}_1\cdot \mathbf{v}_2)-15 (\mathbf{n}\cdot \mathbf{v}_2)^2+6 (\mathbf{n}\cdot \mathbf{v}_2)\big( 3 v_2^2-6 (\mathbf{v}_1\cdot \mathbf{v}_2)-v_1^2\big)- 3 v_2^4 \Bigr)\Biggr) \\ \notag
   &\hspace{2cm}+ \frac{G m_2}{4 r^3}\Biggl( v_1^2\Bigl(\frac{28 q_1 q_2}{G m_2}-15 G m_1 - \frac{2 q_1^2}{ m_1}+\frac{2  q_1 q_2}{ m_1}-\frac{5 q_1^2 q_2^2}{G m_1 m_2^2}-\frac{4 q_2^2}{m_2} \Bigr)\\ \notag
   &\hspace{3cm}+v_2^2 \Bigl(5 G m_1 + 16 G m_2 +\frac{2 q_1^2}{m_1}-\frac{18 q_1 q_2}{m_1}+8\frac{q_2^2}{m_2}-\frac{q_1 q_2}{G m_1 m_2^2}\Bigr)\\ \notag
   &\hspace{3cm}+(\mathbf{v}_1\cdot \mathbf{v}_2) \Bigl( \frac{36 q_1 q_2}{m_1}-10 G m_1-32 G m_2-\frac{4 q_1^2}{m_1}-\frac{16 q_2^2}{m_2}+ \frac{2 q_1 q_2}{G m_1 m_2}\Bigr)\\ \notag
   &\hspace{3cm}+(\mathbf{n}\cdot \mathbf{v}_2)^2\Bigl(34 G m_1- 24 G m_2+\frac{4 q_1^2}{m_1}-\frac{42 q_1 q_2}{m_2}+ \frac{40 q_1 q_2}{m_1}-\frac{16 q_2^2}{m_2} \Bigr) \\ \notag
   &\hspace{3cm}+(\mathbf{n}\cdot \mathbf{v}_1)^2\Bigl(78 G m_1+8 G m_2+ \frac{16 q_1^2}{m_1}-\frac{106q_1 q_2}{m_2}-\frac{8 q_2^2}{m_2}+\frac{12 q_1^2 q_2^2}{G m_1 m_2^2} \Bigr) \\ \notag
   &\hspace{3cm}+(\mathbf{n}\cdot \mathbf{v}_1)(\mathbf{n}\cdot \mathbf{v}_2)\Bigl( \frac{212q_1 q_2}{m_2}-16 G m_2-\frac{32 q_1^2}{m_1}-156G m_1+\frac{16 q_2^2}{m_2^2}-\frac{24 q_1^2 q_2^2}{G m_1 m_2^2}\Bigr)\Biggr) \\ \notag
   &\hspace{2cm}+\frac{G^2 m_2}{4 r^4}\Biggl(\frac{q_1 q_2 }{G m_1 m_2}\Bigl(10 q_1^2-80 q_1 q_2+\frac{q_1^2 q_2^2}{m_1 m_2^2}(m_1+2 m_2)+6 q_2^2 \Bigr)+ \frac{2 m_2 q_1}{m_1}\left(17 q_2- 12 q_1\right)\\ \notag
   &\hspace{3cm}+ \frac{q_1 q_2}{G m_2^2}\Bigl(101 G m_1 m_2-27 q_1 q_2+10 q_2^2\Bigr)-3 G \Bigl(19 m_1^2+46 m_1 m_2+6 m_2^2\Bigr)\\ 
   &\hspace{3cm}-14 q_1^2+132 q_1 q_2-24 q_2^2\Biggr)\Biggr]\Biggr\} \ .
    \end{align}
\end{widetext}
The EoMs for body 2 are obtained by exchanging the body labels $(1 \leftrightarrow 2)$, bearing in mind that $\mathbf{n} \rightarrow - \mathbf{n} $ under this exchange. 
With the Lagrangian~\eqref{eq:: LagrangianHarmonicCoordinates12} we can also compute, at 2PN accuracy, the momenta
\begin{subequations}
  \begin{align}
      &p_A^i\equiv\, \frac{\delta L}{\delta v_A^i}=\, \frac{\partial L}{\partial v_A^i}\, - \frac{d}{dt}\Biggl(\frac{\partial L}{\partial a_A^i}\Biggr), \\
      &   \mathcal{Q}_A^i\equiv\, \frac{\delta L}{\delta a_A^i}=\, \frac{\partial L}{\partial a_A^i},
  \end{align}  
\end{subequations}
respectively conjugate to the position $y_A^i$ and velocity $v_A^i$. 
Our results for body 1 read
\begin{widetext}
\begin{subequations}
    \begin{align}
    \label{eq: conjiugate linear momentum}
   \notag     p_1^i=& m_1 v_1^i \\ \notag
   &+\frac{1}{c^2} \Biggl\{v_1^i m_1
   \left(\frac{v_1^2}{2}+\frac{3 G m_2}{r}\right)-\frac{v_2^i}{2r}
   \left(7 G m_1 m_2- q_1 q_2\right)-\frac{n^i}{2r}
   \left(G  m_1 m_2-q_1 q_2\right)(\mathbf{n}\cdot \mathbf{v}_2)\Biggr\}  \\ \notag
   &+\frac{1}{c^4} \Biggl\{v_1^i \Biggl[-\frac{5 G^2 m_1^2
   m_2}{4 r^2}-\frac{2 G m_2
   q_1^2}{r^2}+\frac{q_1 q_2}{8r}\left((\mathbf{n}\cdot \mathbf{v}_2)^2-v_2^2\right) -\frac{q_1^2 q_2^2}{2 r^2 m_2}+\frac{3 m_1
   v_1^4}{8} \\ \notag
   &\hspace{1.5 cm}+\frac{G m_1 m_2 }{8 r} \Biggl(28 v_1^2-13 
   (\mathbf{n}\cdot \mathbf{v}_2)^2-32 (\mathbf{v}_1\cdot \mathbf{v}_2)+23
   v_2^2\Biggr) +\frac{7 G^2 m_1 m_2^2}{2 r^2}+\frac{G m_1}{2r^2} \left(10
   q_1 q_2-q_2^2\right)\Biggr] \\ \notag
   &\hspace{0.7 cm}+v_2^i
   \Biggl[\frac{q_1 q_2}{8r}\left(2(\mathbf{n}\cdot \mathbf{v}_1)
   (\mathbf{n}\cdot \mathbf{v}_2)-(\mathbf{n}\cdot \mathbf{v}_2)^2+2(\mathbf{v}_1\cdot \mathbf{v}_2)+v_2^2\right)+\frac{q_1^2 q_2^2}{2 r^2 m_2}+\frac{G m_2}{2r^2}\left(3 q_1^2-2 q_1 q_2\right)-\frac{7 G^2 m_1 m_2^2}{4 r^2} \\ \notag
   &\hspace{1.5cm}+\frac{G m_1 m_2}{8 r}
   \left(18 (\mathbf{v}_1\cdot \mathbf{v}_2)-2 (\mathbf{n}\cdot \mathbf{v}_1) (\mathbf{n}\cdot \mathbf{v}_2)+ (\mathbf{n}\cdot \mathbf{v}_2)^2-16
   v_1^2-23 v_2^2\right) +\frac{G m_1 }{2r^2}\left(3
   q_2^2-10 q_1 q_2\right)\Biggr] \\ \notag
   &\hspace{0.7cm}+n^i \Biggl[\frac{G^2 m_1^2
   m_2}{4 r^2}
   \left(43 (\mathbf{n}\cdot \mathbf{v}_1)-13 (\mathbf{n}\cdot \mathbf{v}_2)\right) +\frac{2 G  m_2 q_1^2}{r^2}(\mathbf{n}\cdot \mathbf{v}_1)+\frac{ q_1^2 q_2^2}{4 r^2 m_2}\left(5 (\mathbf{n}\cdot \mathbf{v}_1)-(\mathbf{n}\cdot \mathbf{v}_2)\right) \\ \notag
   &\hspace{1.5cm}+\frac{ q_1
   q_2}{8 r}\left((\mathbf{n}\cdot \mathbf{v}_1)
   v_2^2-3
   (\mathbf{n}\cdot \mathbf{v}_1) (\mathbf{n}\cdot \mathbf{v}_2)^2-3
   (\mathbf{n}\cdot \mathbf{v}_2)^3-2(\mathbf{n}\cdot \mathbf{v}_2) (\mathbf{v}_1\cdot \mathbf{v}_2)+3 (\mathbf{n}\cdot \mathbf{v}_2) v_2^2\right) \\ \notag
   &\hspace{1.5cm}+\frac{G m_1 m_2}{8 r} \left(3 (\mathbf{n}\cdot \mathbf{v}_1)
   (\mathbf{n}\cdot \mathbf{v}_2)^2+3 (\mathbf{n}\cdot \mathbf{v}_2)^3+14 (\mathbf{n}\cdot \mathbf{v}_2)
   (\mathbf{v}_1\cdot \mathbf{v}_2)-7 (\mathbf{n}\cdot \mathbf{v}_1) v_2^2 -9 (\mathbf{n}\cdot \mathbf{v}_2)
   v_2^2\right) \\ 
   &\hspace{1.5cm}+\frac{G^2 m_1 m_2^2}{2r^2} \left(2(\mathbf{n}\cdot \mathbf{v}_1)-7
   (\mathbf{n}\cdot \mathbf{v}_2)\right)+\frac{G m_1}{r^2} \left(\frac{7 q_1 q_2}{2}\Bigl(-3
   (\mathbf{n}\cdot \mathbf{v}_1)+ (\mathbf{n}\cdot \mathbf{v}_2)\Bigr) -(\mathbf{n}\cdot \mathbf{v}_1)
   q_2^2\right)\Biggr]\Biggr\},\\ \notag
   \label{eq: conjiugate acceleration momentum}
   \mathcal{Q}_1^i=& \frac{1}{c^4} \Biggr\{n^i
   \Biggl[\frac{ q_1 q_2}{8}\Bigl((\mathbf{n}\cdot \mathbf{v}_2)^2+(\mathbf{n}\cdot \mathbf{a}_2)r-v_2^2)\Bigr)+
   \frac{G m_1 m_2}{8}\Bigl(7 v_2^2-(\mathbf{n}\cdot \mathbf{v}_2)^2\Bigr) -\frac{7G m_1 q_1
   q_2}{4 r}\Biggr]\\ 
   &\hspace{1cm}-\frac{3}{8} r q_1
   q_2\,  a_2^i-\frac{v_2^i}{4}(\mathbf{n}\cdot \mathbf{v}_2) \left(7 G 
   m_1 m_2- q_1
   q_2\right)\Biggr\},
    \end{align}
\end{subequations}
\end{widetext}
while the momenta for body 2 are easily obtained by the label exchange $1 \leftrightarrow 2$.
\subsection{Conservative dynamics in ADM-type coordinates}
\label{subsec:adm_coord} 

At 2PN order, the generalized Lagrangian~\eqref{eq:: LagrangianHarmonicCoordinates12} depends linearly on the accelerations of the two bodies. This is a well-known and unavoidable characteristic of the two-body Lagrangian when it is expressed in terms of harmonic coordinates~\cite{Martin:1979ph}. However, given the gauge nature of the dependence on the acceleration, it is possible to identify a class of alternative coordinates such that the 2PN Lagrangian becomes ordinary, i.e.~dependent only on the positions and the velocities of the two black holes, thus allowing for the computation of the corresponding Hamiltonian.

For BBH with no charge, for instance, the generalized harmonic-coordinate Lagrangian can be recast into an ordinary Lagrangian by transforming the harmonic coordinates into the coordinates of the ADM canonical formalism, as first shown in Ref.~\cite{Damour:1985mt} at 2PN and later extended at 3PN in Ref.~\cite{Damour:2000ni}.
In our case, indeed, we cannot use the same transformations, but we can follow the general procedure presented in Ref.~\cite{Damour:2000ni} to find a family of ADM-type coordinates whose corresponding Lagrangian is ordinary, and then fix the charge-independent part so to reproduce the known ADM results in the neutral limit.

Let us consider an infinitesimal contact transformation between the bodies' trajectories in harmonic coordinates, $y_A^i(t)$, and the corresponding ones in a yet unspecified set of ADM-type coordinates, $Y^i_A(t)$. We denote the difference between the two as 
\begin{equation}
\label{eq: contact transformation generic}
    \delta y^i_A(t)\, \equiv\, Y^i_A(t)\, -y^i_A(t).
\end{equation}
%
This transformation can be assumed to begin at the 2PN order, since accelerations in the harmonic Lagrangian first appear at that order.
Moreover, following Ref.~\cite{deAndrade:2000gf}, it can be taken in the form~\footnote{Working at 2PN accuracy, the ``counter'' term $X^i_A$ introduced in Sec.~IIIB of Ref.~\cite{deAndrade:2000gf} is not needed.}
\begin{equation}
\label{eq: contact transformation}
    \delta y^i_A=\, \frac{1}{m_A}\Biggl[\, \mathcal{Q}_A^i\, +\frac{\partial F}{\partial v_A^i}\Biggr]\, + \mathcal{O}\Biggl(\frac{1}{c^6}\Biggr),
\end{equation}
\newline
where $\mathcal{Q}_A^i$ is the conjugate momentum of the acceleration given in Eq.~\eqref{eq: conjiugate acceleration momentum} and $F$ is an arbitrary function of positions and velocities with leading-order contributions at the 2PN level.

We denote with $L^{\rm ADM-type}$ the Lagrangian associated to the coordinates $Y^i_A(t)$.
Following Ref.~\cite{deAndrade:2000gf}, we have, at linear order in $\delta y^i_A$,
\begin{equation}
\label{eq: lagrangiana ADM transformation}
\begin{aligned}
    &L^{\rm ADM-type}[\mathbf{y}_A,\mathbf{v}_A]=\, L[\mathbf{y}_A,\mathbf{v}_A,\mathbf{a}_A]\,\\
    &\quad \,  +\sum_{A}\frac{\delta L}{\delta y^i_A}\delta y^i_A\, +\frac{d F}{dt},
    \end{aligned}
\end{equation}
\newline
where both sides of the equation are expressed in terms of dummy harmonic variables that must be replaced with the ADM-type ones, $(\mathbf{Y}_A,\mathbf{V}_A)$, once $L^{\rm ADM-type}$ is computed. 

The arbitrariness of the function $F$, entering the contact transformation and the Lagrangian, reflects the fact that there exist an infinite number of ADM-type coordinates that eliminate the acceleration dependence in the Lagrangian, as per Eq.~\eqref{eq: lagrangiana ADM transformation}. 

For neutral BBH, the usual strategy to fix the functional expression of $F$ is to require that the Lagrangian~\eqref{eq: lagrangiana ADM transformation} is exactly the ADM Lagrangian computed, via a Legendre transformation from the PN-expanded Hamiltonian of the canonical ADM-formalism. This is precisely the procedure followed by Ref.~\cite{deAndrade:2000gf}. However in the case of a charged BBH system, the ADM Lagrangian (or Hamiltonian) is not known. This has the consequence that we cannot immediately apply the procedure of Ref.~\cite{deAndrade:2000gf}, but we should instead follow a more general prescription.

More in detail, our prescription is the following. We start with the most general 2PN expression for $F$ that only depends on positions and velocities and is given in terms of unfixed coefficients that can depend on masses and charges. We then partially fix these coefficients by asking that, in the limit $(q_1,q_2)\to(0,0)$, the resulting Lagrangian reduces to the ADM one, finally leaving the residual unfixed coefficients explicit. The function $F$ we consider has the general structure
\begin{widetext}
 \begin{align}
         \label{eq: parametrization of F} \notag
    F\, = \frac{1}{c^4}\Big\{&\mathcal{A}_1 (\mathbf{n}\cdot \mathbf{v}_1) v_1^2\, +\mathcal{A}_2 (\mathbf{n}\cdot \mathbf{v}_2) v_1^2\,+\mathcal{A}_3 (\mathbf{n}\cdot \mathbf{v}_1) (\mathbf{v}_1\cdot \mathbf{v}_2)\,+\mathcal{A}_4 (\mathbf{n}\cdot \mathbf{v}_2) (\mathbf{v}_1\cdot \mathbf{v}_2)\,+\mathcal{A}_5 (\mathbf{n}\cdot \mathbf{v}_1) v_2^2\, +\mathcal{A}_6 (\mathbf{n}\cdot \mathbf{v}_2) v_2^2\\ 
    &+\mathcal{B}_1\frac{(\mathbf{n}\cdot \mathbf{v}_1)}{r}\, +\mathcal{B}_2\frac{(\mathbf{n}\cdot \mathbf{v}_2)}{r}\, +\mathcal{C}_1(\mathbf{n}\cdot \mathbf{v}_1)^3+\mathcal{C}_2(\mathbf{n}\cdot \mathbf{v}_1)^2(\mathbf{n}\cdot \mathbf{v}_2)\, +\mathcal{C}_3(\mathbf{n}\cdot \mathbf{v}_1)(\mathbf{n}\cdot \mathbf{v}_2)^2+\mathcal{C}_4(\mathbf{n}\cdot \mathbf{v}_2)^3 \Bigr\},
 \end{align}
 \end{widetext}
where each coefficient is decomposed in the sum of the corresponding mass-dependent ADM result, obtained in the neutral case in Ref.~\cite{deAndrade:2000gf}, and an unfixed coefficient associated to charge-dependent corrections, which we denote as $(A_1, A_2, A_3, A_4, A_5, A_6, B_1, B_2, C_1, C_2, C_3, C_4)$. Explicitly, we have 
\begin{align} 
   \label{eq: ADM parameters}
    &\mathcal{A}_1\, = A_1, \quad 
    \mathcal{A}_2\, = \frac{G m_1 m_2}{4}\, + A_2, \quad 
    \mathcal{A}_3\, = A_3, \cr 
    &\mathcal{A}_4\, =  A_4, \quad 
    \mathcal{A}_5\, =- \frac{G m_1 m_2}{4}\, + A_5, \quad
    \mathcal{A}_6\, = A_6, \cr 
    &\mathcal{B}_1\, =\frac{G^2 m_1 m_2}{4}\Bigl(\, m_2\, +7m_1\Bigr)\, +B_1, \cr
&\mathcal{B}_2\, =-\frac{G^2 m_1 m_2}{4}\Bigl(\, 7m_2\, +m_1\Bigr)\,+ B_2, \cr
&\mathcal{C}_1\, =C_1, \quad
\mathcal{C}_2\, =C_2, \quad
\mathcal{C}_3\, =C_3, \quad 
\mathcal{C}_4\, =C_4.
\end{align}    
With this choice of $F$, Eq.~\eqref{eq: lagrangiana ADM transformation} yields an ADM-type Lagrangian with no dependence on the acceleration and that reduces to the known ADM Lagrangian in the neutral limit. Sticking to the same contribution separation considered in the previous section, our result is organized as
\begin{align}
    \label{eq: Total_ADM-type_Lagrangian} \notag
    &L^{\rm ADM-type}= L^{\rm ADM-type}_{\rm kin} +  L^{\rm ADM-type}_{\rm PN} \\
    &\quad+ L^{\rm ADM-type}_{\rm PC}  + L^{\rm ADM-type}_{\rm mixed} \ .
    \end{align}
The explicit expression for each of these terms is provided, in terms of the coordinates $(\mathbf{Y}_A,\mathbf{V}_A)$, in Eq.~\eqref{eq: ADM-type_Lagrangian_contributions} of App.~\ref{app: center-of-mass frame}. 

We stress that, although applying the order-reduction procedure to the harmonic Lagrangian would eliminate its dependence on accelerations (and prevent the harmonic gauge to be satisfied~\cite{Martin:1979ph}), the resulting Lagrangian would fail to reduce to the ADM form in the neutral limit and therefore it cannot be obtained from the result~\eqref{eq: Total_ADM-type_Lagrangian} with a particular choice of the free coefficients. Nonetheless it is still possible to recover the order-reduced Lagrangian by properly fixing the coefficients of $F$ without the partial fixing of Eq.~\eqref{eq: ADM parameters}.

Let us now move to the computation of the corresponding ADM-type Hamiltonian.
Noting that the ADM-type momentum $P_A^i$ conjugate to the coordinate $Y^i_A$ is given by the defining relation
\begin{equation}
\label{eq: ADM conjugate momentum}
    P_A^i = \frac{\partial L^{\rm ADM-type}}{\partial V_A^i},
\end{equation}
the ADM-type Hamiltonian, which reduces to the standard ADM Hamiltonian of Ref.~\cite{Damour:2000kk} in the neutral limit, follows from the Legendre transformation 
\begin{equation}
\label{eq: Legrendre transformation}
    H^{\rm ADM-type}\, =\, \sum_{A} P_A^i V_A^i\, -L^{\rm ADM-type}.
\end{equation}
Again, we organize it according to the structure 
    \begin{align}\notag
        \label{eq:Total_ADM-type_Hamiltonian} \notag
        &H^{\rm ADM-type}= H^{\rm ADM-type}_{\rm kin} +  H^{\rm ADM-type}_{\rm PN} \\ 
    &\quad+ H^{\rm ADM-type}_{\rm PC} + H^{\rm ADM-type}_{\rm mixed},
    \end{align}
and separately provide the different contributions in Eq.~\eqref{eq: ADM-type_Hamiltonian_contributions} of App.~\ref{app: center-of-mass frame}.

\section{Transformations to the center of mass frame}
\label{sec:CoM_frame}  

Let us now focus on how to translate all these results in the CoM frame. In the following we will employ harmonic coordinates, knowing that the corresponding transformation to the CoM frame in ADM-type coordinates can be simply obtained by first going to the CoM frame in harmonic coordinates, as explained in detail in this section, and then shifting to ADM-type coordinates, by means of the contact transformation~\eqref{eq: contact transformation}.

In the absence of dissipation, the CoM position $G^i$ satisfies the general equation
\begin{equation} \label{eq:: dGdt}
    \frac{d G^i}{dt}\, =\, \mathcal{P}^i,
\end{equation}
where $\mathcal{P}^i$ is the conserved total linear momentum~\cite{deAndrade:2000gf, Blanchet:2003gy}
\begin{equation}\label{eq: conserved total linear momentum}
    \mathcal{P}^i=\, p_1^i + p_2^i = \frac{\delta L}{\delta v_1^i}+\frac{\delta L}{\delta v_2^i}, \qquad \frac{d\mathcal{P}^i}{dt} = 0.
\end{equation}

To explicitly compute $G^i$ up to 2PN we proceed as follows: (i) we start with a general ansatz for $G^i$ that includes all the dimensionally allowed scalar combinations of $(\mathbf{y}_A,\mathbf{v}_A)$ up to  2PN order, each one paired with a dimensionless coefficient that depends only on masses and charges; (ii) we insert this ansatz in Eq.~\eqref{eq:: dGdt}, order reducing via the EoMs all the accelerations, while in the right hand side we use the explicit 2PN result for $ \mathcal{P}^i$ that follows from the Lagrangian~\eqref{eq:: LagrangianHarmonicCoordinates12}; (iii) since by definition the coefficients of our ansatz for $G^i$ cannot depend on positions and velocities, the equation of the previous step is equivalent to a system of algebraic relations [one for each different scalar combination of $(\mathbf{y}_A,\mathbf{v}_A)$] that we can solve for the coefficients.

The result is unique and reads 
\begin{widetext}
        \begin{align}
        \label{eq: Gconservative}\notag
           G^i =& m_1 y_1^i \\ \notag
           &+\frac{1}{c^2} \Bigl(-\frac{G m_1 m_2}{2 r}+\frac{ q_1 q_2}{2 r}+\frac{1}{2} m_1 v_1^2\Bigr) y_1^i  \cr 
            &   + \frac{1}{c^4} \Biggl\{ v_1^i \biggl[-\frac{7G m_1
   m_2}{4} \Bigl( (\mathbf{n}\cdot \mathbf{v}_1)+ (\mathbf{n}\cdot \mathbf{v}_2)\Bigr) +\frac{ q_1 q_2}{4} (\mathbf{n}\cdot \mathbf{v}_1) +\frac{q_1 q_2}{4} (\mathbf{n}\cdot \mathbf{v}_2) \biggr]
    \cr
   &\hspace{0.8cm}+y_1^i\biggl[G^2 \biggl(-\frac{5 m_1^2 m_2}{4 r^2}+\frac{7 m_1 m_2^2}{4
   r^2}\biggr)+\frac{ q_1 q_2}{8 r}\Bigl((\mathbf{n}\cdot \mathbf{v}_1)^2+2(\mathbf{n}\cdot \mathbf{v}_1) (\mathbf{n}\cdot \mathbf{v}_2)-(\mathbf{n}\cdot \mathbf{v}_2)^2+2(\mathbf{v}_1\cdot \mathbf{v}_2)\Bigr)\cr
   &\hspace{1.3cm}+\frac{ q_1^2 q_2^2}{4 r^2 m_1 m_2}\bigl(m_2-m_1\bigr)+\frac{ q_1 q_2 }{8 r}\Bigl(v_2^2-v_1^2\Bigr)+\frac{3}{8} m_1 v_1^4+G m_2 \biggl(-\frac{q_1^2}{2 r^2}-\frac{3 q_1
   q_2}{r^2}\biggr)+\frac{G m_1}{r^2} \Bigl(2 q_1 q_2+q_2^2\Bigr)\\
   &\hspace{1.3cm}+\frac{G m_1 m_2}{8 r}
   \Bigl(-(\mathbf{n}\cdot \mathbf{v}_1)^2-2(\mathbf{n}\cdot \mathbf{v}_1) (\mathbf{n}\cdot \mathbf{v}_2)+(\mathbf{n}\cdot \mathbf{v}_2)^2-14 (\mathbf{v}_1\cdot \mathbf{v}_2)+19 v_1^2-7 v_2^2\Bigr)\biggr]  \Bigg\} + (1\leftrightarrow 2).
        \end{align}
The transformations to the CoM frame, namely the expression of $\mathbf{y}_A$ in terms of the relative variables $\mathbf{x}=\mathbf{y}_1-\mathbf{y}_2$ and $\mathbf{v}=\mathbf{v}_1-\mathbf{v}_2$, can be found by solving iteratively (PN order by PN order) the condition $G^i\, =0$. Our result for body 1 is~\cite{SupplementalMaterial}  
\begin{align}
\label{eq: com trasnf} 
      \notag  y_1^i\, =& \frac{m_2}{m_1+m_2}x^i\,\\ \notag
      &+\frac{1}{c^2} \Biggl[x^i\Bigl(m_1\, -m_2\Bigr)\Biggl( \frac{m_1 m_2 }{(m_1+m_2)^3}v^2-\frac{G m_1 m_2- q_1 q_2}{2(m_1\, +m_2)^2r}\Biggl)\Biggr]\\ \notag
        &+\frac{1}{c^4}\Biggl\{\frac{v^i}{4 (m_1+m_2)^2} (m_1-m_2) \Bigl(q_1 q_2-7 G m_1 m_2\Bigr)(\mathbf{n}\cdot \mathbf{v})\\ \notag
   &\hspace{1 cm}+\frac{x^i}{8r^2}\Biggl[\frac{2 G^2 m_1 m_2
   }{(m_1+m_2)^3} \Bigl(7 m_1^3+5 m_1^2 m_2-5 m_1 m_2^2-7
   m_2^3\Bigr)\\ \notag
   &\hspace{2cm}+\frac{G}{(m_1+m_2)^4} \Biggl(4 m_2^4 q_1
 (q_1+6 q_2)+m_1 m_2^3 \Bigl( m_2 r ((\mathbf{n}\cdot \mathbf{v})^2-19
   v^2) +16  q_1^2+40  q_1 q_2-8  q_2^2\Bigr)\\ \notag
   & \hspace{4.5cm }-m_1^3
   m_2 \Bigl( m_2 r (3 (\mathbf{n}\cdot \mathbf{v})^2-31 v^2)-8  q_1^2+40  q_1
   q_2+16  q_2^2\Bigr)\\ \notag
   &\hspace{4.5cm}-m_1^2 m_2^2 \Bigl( m_2 r (-3 (\mathbf{n}\cdot \mathbf{v})^2+31
   v^2)-20 q_1^2+20 q_2^2\Bigr)\\ \notag
   &\hspace{4.5cm }-m_1^4 \Bigl( m_2 r
 ((\mathbf{n}\cdot \mathbf{v})^2-19 v^2)+24 q_1 q_2+4 q_2^2\Bigr)\Biggr) \\ \notag
   &\hspace{2cm}+\frac{(m_1-m_2)}{m_1 m_2 (m_1+m_2)^5} 
   \Biggl( m_1 m_2^3 q_1 q_2 \Bigl( m_2 r
   ((\mathbf{n}\cdot \mathbf{v})^2-v^2) +4  q_1 q_2\Bigr)+2 m_2^4  q_1^2 q_2^2\\ \notag
   &\hspace{5.1 cm}+m_1^3 m_2 \Bigl(4  q_1^2
   q_2^2-2 r^2
   (4 m_2^2 r(\mathbf{a}\cdot \mathbf{v})  (\mathbf{n}\cdot \mathbf{v})+3 v^2) +5  m_2 q_1 q_2 r
   ((\mathbf{n}\cdot \mathbf{v})^2-3 v^2)\Bigr)\\ \notag
   &\hspace{5.1 cm}+m_1^2 m_2^2 \Bigl(3 r^2 v^4 m_2^2+5  m_2 q_1 q_2 r ((\mathbf{n}\cdot \mathbf{v})^2-3
   v^2) +4  q_1^2 q_2^2\Bigr)\\ 
   &\hspace{5.1 cm}+m_1^4 \Bigl(3 r^2
   v^4 m_2^2+m_2  q_1 q_2 r ((\mathbf{n}\cdot \mathbf{v})^2-v^2) +2 q_1^2
   q_2^2\Bigr) \Biggr)\Biggr]\Biggr\}.
    \end{align}
\end{widetext}
As usual, the same result for body 2 is simply obtained with the exchange $1 \leftrightarrow 2$ on the body labels, noting that now also $\mathbf{v}$ changes sign under this exchange. 

In the neutral limit $(q_1, q_2) \to (0,0)$, the harmonic CoM position and the associated transformations to the CoM frame reduce to the results of Ref.~\cite{Blanchet:2002mb}. 

The transformations above have been used to derive the 2PN CoM-frame expressions for EoMs, Lagrangian, and Hamiltonian in harmonic coordinates. These results are provided in App.~\ref{app: center-of-mass frame} and include also the dissipative contributions that will be analyzed in Sec.~\ref{sec:dissipations}.

\section{Gauge invariant quantities}
\label{sec:guage_inv_quantities} 

With the conservative 2PN dynamics derived in the previous sections, we are now in the position of computing, with 2PN accuracy, the corrections induced by the charge to the most relevant gauge-invariant quantities for a charged BBH. On the one hand, in the simplifying setting of an adiabatic evolution over a sequence of circular orbits, we will compute the binding energy and the fractional advance of the periastron per radial period.~\footnote{Concerning the binding energy, the assumption of circular orbits makes disappear any dissipative contribution associated to Schott terms, which instead, for more general motions, would appear at 1.5PN (i.e.~the PN order of the leading dissipative effects in the dynamics of charged BBHs). We also notice that 'pseudo-Schott' terms like those recently computed at 4PN in Ref.~\cite{Trestini:2025nzr} for neutral BBHs, which would persist for circular orbits, do not appear before the 3PN order for charged BBHs.} On the other hand, without imposing any restriction on the orbital motion, we will derive the associated conservative scattering angle,~\footnote{In general, the scattering angle also includes dissipative contributions (see, e.g., Refs.~\cite{Bini:2021jmj,Bini:2025rng}); however, these are specific to hyperbolic motions and thus cannot be directly connected (i.e.~via analytic continuation) to the bound case.} which is known to encode the full conservative information of the underlying bound dynamics.

\subsection{Binding energy and periastron advance}
\label{subsec:Eb_and_K}

To compute the binding energy for quasi-circular adiabatic orbits we can use the ADM-type Hamiltonian of Eq.~\eqref{eq:Total_ADM-type_Hamiltonian}. More precisely, it is advantageous to use its expression in the CoM frame, $H^{\rm ADM-type}_{\rm CoM}$, which follows from the procedure presented in the previous section; this is given explicitly in Eq.~\eqref{eq: H in COM ADM}.

In terms of the circular limit of this Hamiltonian, the ($\mu$-rescaled) binding energy is conveniently given by the formula
\begin{equation}
    \hat{E}_b(x_q) = \frac{H^{\rm ADM-type}_{\rm CoM, circ}(x_q)-M}{\mu},
\end{equation}
where $x_q$ is the charge-flexed frequency parameter~\footnote{We highlight that the conditions $\eta_1=q_1/(\sqrt{G}m_1) <1$ and $\eta_2=q_2/(\sqrt{G}m_2) <1$ do imply $\eta_1 \eta_2<1$, therefore $x_q$ is always real and positive. We also recall that we consider $m_1>m_2$.}
\begin{equation}
    \label{eq: xq}
    x_q \equiv \left[\frac{G M \Omega }{c^3}\bigl( 1 - \eta_1 \eta_2 \bigr) \right]^{2/3},
\end{equation}
$\Omega$ being the orbital frequency. The circular-orbit Hamiltonian $H^{\rm ADM-type}_{\rm CoM, circ}(x_q)$ is obtained by expressing the circular limit of $H^{\rm ADM-type}_{\rm CoM}$, Eq.~\eqref{eq: H in COM ADM}, in terms of $x_q$. 

Considering that the motion we are studying is planar, as it is the case for non-spinning BBHs, the ADM-type CoM momentum $\mathbf{P}$ is readily split in its radial and angular components,  $(P_R,P_\Phi)$, considering the relation $P^2 = P_R^2 + P_\Phi^2/R^2$. The circular limit is then equivalent to the radial momentum condition $P_R \to 0$. To recast the angular momentum $P_\Phi$ and the relative separation $R$ in terms of $x_q$, we proceed as follows. First, we compute, up to 2PN, the circular-orbit expansion of $P_\Phi$ in powers of $1/R$, by solving perturbatively the equation
\begin{equation}
0 = \frac{d P_R}{dt} = -\frac{\partial H^{\rm ADM-type}_{\rm CoM}(P_R\to 0, P_\Phi,R)}{\partial R}.
\end{equation}

Then, we compute the expansion of $R$ in power of $x_q$ (again at 2PN accuracy) by solving perturbatively for $R$ the PN expansion of the definition~\eqref{eq: xq}, in which the orbital frequency $\Omega$ is replaced by
\begin{equation}
    \Omega = \frac{d \Phi}{dt } = \frac{\partial H^{\rm ADM-type}_{\rm CoM}(P_R\to 0, P_\Phi,R)}{\partial P_\Phi},
\end{equation}
and where $P_\Phi$, once the derivative relative to it is taken, is expanded in powers of $1/R$ with the result of the previous step.

The resulting 2PN binding energy  reads~\footnote{In expressions of this kind the test-mass limit is readily obtained considering $\nu\to0$ (and $X_{12}\equiv \sqrt{1- 4\nu}\to1)$. } 
\begin{widetext}
    \begin{align}
    \label{eq: Eb expression}
    E_b(x_q) =& - \frac{x_q}{2} \cr
    &+ \frac{x_q^2}{24}\big(1- \eta_1 \eta_2\big)^{-2} \Biggl\{ 9+\nu -2 \left(1+X_{12}\right) \eta_1^2-2 (3+\nu ) \eta_1
   \eta_2-2 \left(1-X_{12}\right) \eta_2^2+(1+\nu
   ) \eta_1^2 \eta_2^2\Bigg\}\cr & + \frac{x_q^3}{48}\big(1- \eta_1 \eta_2\big)^{-4} \Bigg\{ 81-57 \nu +\nu ^2-18\eta_1^2 \left(1+X_{12}\right)-2\nu \left(5-X_{12}\right) \eta_1^2-2(81-84\nu+2\nu^2)\eta_1 \eta_2  \cr 
   &\hspace{2cm}-18(1-X_{12})\eta_2^2-2\nu(5+X_{12})\eta_2^2-2(1+X_{12})^2\eta_1^4+4\eta_1^3\eta_2\left[9+11 \nu+(9-\nu)X_{12}\right]\cr 
   &\hspace{2cm}+(86-238\nu+6\nu^2-4X_{12}^2)\eta_1^2 \eta_2^2+4\eta_1 \eta_2^3\left[9+11 \nu-(9-\nu)X_{12}\right]-2(1-X_{12})^2\eta_2^4\cr &\hspace{2cm}
   +2\eta_1^2 \eta_2^4\left(2 \nu- 19 - 9 X_{12}\right)-2\eta_1^3 \eta_2^3 \left(9-76 \nu +2 \nu
   ^2\right)+\left(1-25 \nu +\nu ^2\right) \eta_1^4 \eta_2^4\cr &\hspace{2cm}+2\eta_1^4 \eta_2^2\left[14 \left(1-X_{12}\right)-\nu \left(19-X_{12}\right)\right]+ 2\eta_1^4 \eta_2^2\left(2 \nu- 19 + 9 X_{12}\right)\cr &\hspace{2cm}+2\eta_1^2 \eta_2^4\left[14 \left(1+X_{12}\right)-\nu
   \left(19+X_{12}\right)  \right]\Bigg\}.
\end{align}
\end{widetext}
As expected, there is no leftover dependence on the free coordinate-dependent coefficients that enter the starting Hamiltonian. Moreover, we have explicitly checked that this result reproduces the 1PN binding energy of Ref.~\cite{Khalil:2018aaj} and, in the neutral limit, the known 2PN results for BBHs with no charge given, e.g., in Ref.~\cite{Blanchet:2013haa}.

The periastron advance for quasi-circular orbits can be obtained from the Hamiltonian $H^{\rm ADM-type}_{\rm CoM}$ thanks to the relation~\cite{Hinderer:2013uwa}
\begin{widetext}
    \begin{align}
    &\frac{\Delta \Phi}{2\pi} = K-1, \cr & K = \lim_{P_R\to0} \Bigg[\bigg(\frac{\partial^2 H^{\rm ADM-type}_{\rm CoM}}{\partial R^2}\frac{\partial^2 H^{\rm ADM-type}_{\rm CoM}}{\partial P_R^2}\bigg)^{-1} \frac{\partial H^{\rm ADM-type}_{\rm CoM}}{\partial P_\Phi} \Bigg].
\end{align}

Following the same procedure detailed for $E_b$, one can express $K$ as a function of the variable $x_q$
Explicitly, our result at 2PN accuracy is
    \begin{align}
\label{eq: K expression}\notag
     K =& 1+ \frac{x_q}{4}\big(1-\eta_1 \eta_2\big)^{-2} \Bigg[ 12- \eta_1^2- 12 \eta_1 \eta_2-\eta_2^2+2 \eta_1^2\eta_2^2+ X_{12}\left(\eta_2^2-\eta_1^2\right)\Biggr] \\ \notag 
    &+\frac{x_q^2}{48}\big(1- \eta_1 \eta_2\big)^{-4} \Biggl\{648-1368 \eta_1 \eta_2+ 204 (1+X_{12})\eta_1^3 \eta_2-120 (1-X_{12})\eta_2^2-120(1+X_{12})\eta_1^2-\frac{7}{2}(1+X_{12})^2\eta_1^4\\ \notag
    &\hspace{1.1cm}+(881+7X_{12}^2)\eta_1^2\eta_2^2-70(1+X_{12})\eta_1^4\eta_2^2+204(1-X_{12})\eta_1 \eta_2^3-192\eta_1^3 \eta_2^3-\frac{7}{2}(1-X_{12})^2 \eta_2^4+10 \eta_1^4 \eta_2^4\\ \notag
    &\hspace{1.1cm}-70(1-X_{12})\eta_1^2 \eta_2^4+ \nu \Biggl[-336-20 \eta_1^2+1008 \eta_1 \eta_2 + 112 \eta_1^3 \eta_2- 20 \eta_2^2-1256 \eta_1^2 \eta_2^2-92 \eta_1^4 \eta_2^2\\
    &\hspace{1.1cm}+ 112 \eta_1 \eta_2^3+ 688 \eta_1^3 \eta_2^3-92\eta_1^2 \eta_2^4-104 \eta_1^4 \eta_2^4+ 16 X_{12}\Biggl(\eta_1^2- 2 \eta_1^3 \eta_2- \eta_2^2+ \eta_1^4 \eta_2^2+ 2 \eta_1 \eta_2^3- \eta_1^2 \eta_2^4 \Biggr)\Biggr]\Biggr\}\ .
\end{align}
\end{widetext}
Note that here, too, there are no leftover free coefficients, and, when taking the neutral limit, our result reproduces up to 2PN the zero-spin limit of the result provided in Ref.~\cite{LeTiec:2013uey}.

\subsection{Scattering angle}
\label{subsec:chi}

The conservative scattering angle for a binary system of two charged black holes has been recently derived, up to $\mathcal{O}(G^3)$ 
in the post-Minkowskian (PM) expansion, in Ref.~\cite{Wilson-Gerow:2023syq,Alonzo-Artiles:2026wbe}.
Here, we compute its PN-expansion up to the 2PN order from the 2PN-accurate conservative dynamics derived in the previous sections. In addition to encoding our 2PN conservative results in a gauge-invariant quantity, this computation permits a non-trivial check against the scattering angle of Refs.~\cite{Wilson-Gerow:2023syq,Alonzo-Artiles:2026wbe}, which can be PN expanded considering that the relative boost factor $\sigma 
$ appearing therein is such that $(\sigma^2-1) \sim 1/c^2$.

The scattering angle associated to the Hamiltonian $H^{\rm ADM-type}_{\rm CoM}$ [see Eq.~\eqref{eq: H in COM ADM}] is given by the integral 
\begin{equation}
    \label{eq: chi general formula}
    \chi = -\pi -2 \int_{R_{\rm min}}^{+\infty} dr \, \frac{\partial P_R(E,P_\Phi,R)}{\partial P_\Phi},
\end{equation}
where the function $P_R(E,P_\Phi,R)$ is obtained by solving perturbatively for the radial momentum the energy-conservation equation $H^{\rm ADM-type}_{\rm CoM}=E$, and $R_{\rm min}$ is the smallest real root of $P_R(E,P_\Phi,R)=0$. The integral in Eq.~\eqref{eq: chi general formula} is then rewritten in terms of $\sigma$ through the relation 
\begin{align}
    &E=Mc^2\sqrt{1+2\nu (\sigma -1)} \cr&\quad=1+\frac{\nu }{2}  \big(\sigma ^2-1\big)-\frac{\nu}{8}   (1+\nu ) \big(\sigma
   ^2-1\big)^2 + \dots,
\end{align} 
and the resulting series of diverging integrals is regularized and evaluated with the standard procedure discussed, e.g., in Refs.~\cite{Damour:2017zjx,Placidi:2024yld}.

To simplify the final expression, our result for the scattering angle $\chi$ is written in terms of the momentum at infinity $p_\infty =  \sqrt{\sigma^2-1}$ and the angular momentum $p_\phi = P_\Phi/\mu$. Up to 2PN order, it reads
\begin{widetext}
    \begin{align}
    \label{eq: chi expl}
    \chi &= \frac{2 G M }{p_{\phi } p_{\infty }}\left(1-\eta_1 \eta_2 \right)-\frac{2 G^3
   M^3 }{3 p_{\phi}^3 p_{\infty }^3}\left(1-\eta_1 \eta_2\right)^3 \cr&
   + \frac{1}{c^2} \Bigg\{ \frac{G M p_{\infty } }{p_{\phi }}\left(4-\eta_1 \eta_2\right)
   +\frac{G^2 M^2  \pi  }{p_{\phi}^2}\bigg[3-\frac{\eta_1^2}{4} \left(1+X_{12}\right) -3  \eta_1 \eta_2-\frac{\eta_2^2}{4}
   \left(1-X_{12}\right) +\frac{\eta_1^2 \eta_2^2}{2}\bigg]\cr&\qquad
   +\frac{G^3 M^3 
   }{p_{\phi }^3 p_{\infty }}\bigg[8-\eta_1^2 \left(1+X_{12}\right)-15\eta_1 \eta_2+\eta_1^3 \eta_2 \left(1+X_{12}\right)
    -\eta_2^2\left(1-X_{12}\right) +8  \eta_1^2 \eta_2^2
    +\eta_1 \eta_2^3\left(1-X_{12}\right)-\eta_1^3 \eta_2^3\bigg] \Bigg\} \cr &
    +\frac{1}{c^4}\Bigg\{ \frac{G^2 M^2 \pi  P_{\infty }^2 }{P_{\phi }^2}\Bigg[\frac{3}{4}
   (5-2 \nu )+\frac{\eta_1^2}{8}\left(1+X_{12}\right) (-3+\nu ) +\frac{3}{2} (-1+\nu ) \eta_1
   \eta_2
    +\frac{\eta_2^2}{8}\left(1-X_{12}\right) (-3+\nu ) -\nu \frac{\eta_1^2
   \eta_2^2}{4 }\Bigg] \cr & \hspace{1cm}
   +\frac{G^3 M^3  p_{\infty }}{p_{\phi }^3} \Bigg[-16 (-3+\nu )-\frac{17}{2}\eta_1^2\left(1+X_{12}\right)
    + \frac{\eta_1^2}{2}\left(1-2 \nu+ X_{12}\right)+ \nu \eta_1^2 \left(1+ X_{12}\right)+\frac{\eta_1 \eta_2}{4}  (128 \nu-225 ) \cr &\hspace{2cm}
+\eta_1^3 \eta_2 \left(1+X_{12}\right)(6- \nu) -\frac{3}{2}\eta_1^3 \eta_2 \left(1- 2\nu+ X_{12}\right)
   -\frac{17}{2} \eta_2^2
   \left(1-X_{12}\right)+\frac{\eta_2^2}{2}\left(1- 2\nu - X_{12}\right) \cr &\hspace{2cm} + \nu \eta_2^2 \left(1- X_{12}\right)+(16-24
   \nu ) \eta_1^2 \eta_2^2 
   +\eta_1 \eta_2^3\left(1- X_{12}\right)(6-\nu)-\frac{3}{2}\eta_1 \eta_2^3\left(1-2\nu-X_{12}\right) +\frac{\eta_1^3 \eta_2^3}{4}(-3+16 \nu )\Bigg] \cr & \hspace{1cm}+\frac{ G M p_{\infty }^3 }{4  p_{\phi }}\eta_1 \eta_2 \Bigg\}.
\end{align}
\end{widetext}
Once again, we notice that the dependence on the free coordinate-related 
coefficients has disappeared. Additionally, we find perfect agreement between this result and the 2PN expansion of the scattering angle provided in Ref.~\cite{Wilson-Gerow:2023syq,Alonzo-Artiles:2026wbe}~\footnote{This also means that, in the post-Lorentzian limit ($G \to 0$), our result also agrees with the one of Ref.~\cite{Bern:2023ccb}.}.

As already mentioned in Sec.~\ref{sec:lagrangian}, we can also use the scattering angle 
as a convenient gauge-invariant 
tool to check 
the alternative 2PN Lagrangian for charged BBHs of Ref.~\cite{Gupta:2022spq}. To this end,  
since Ref.~\cite{Gupta:2022spq} only provides the charge-dependent contribution to the full Lagrangian, 
we completed the result of Ref.~\cite{Gupta:2022spq} following two alternative procedures: (i) by adding the 2PN Lagrangian for neutral BBHs, shown e.g.~in Eq.~(4.1) of Ref.~\cite{deAndrade:2000gf} and, alternatively, (ii) by adding the 2PN EFT Lagrangian provided in \cite{Gilmore:2008gq}.
In both cases, computing the PN-expanded scattering angle yields a result that differs from that of Eq.~\eqref{eq: chi expl} (and thus from the PN-expanded result of Refs.~\cite{Wilson-Gerow:2023syq,Alonzo-Artiles:2026wbe}) by a 2PN term proportional to the charges.

It should be noted, however, that not knowing precisely the arbitrary choices behind the computation of the Lagrangian of Ref.~\cite{Gupta:2022spq}, concerning in particular the associated neutral part, we cannot conclusively determine whether the latter is physically incorrect or whether it should be supplemented with a different neutral-BBH Lagrangian. 

\section{Dissipative effects}
\label{sec:dissipations} 

In this section, we account for the dissipative effects in the 2PN dynamics of a charged BBH. Because of the presence of charge, and thus of a gauge vector field $A_\mu$ coupled to the metric $g_{\mu \nu}$, the radiation emitted by a system of this kind presents also an electromagnetic component of dipolar nature, which enters the dynamics one PN order earlier than the leading 2.5PN quadrupolar radiation associated to the emission of GWs~\cite{Julie:2017rpw,Khalil:2018aaj}. 
This means that, to completely characterize the 2PN dynamics of a charged BBH, we have also to compute the leading 1.5PN dissipative effects associated to this dipolar emission \cite{2001JNS....11..321K}.

More specifically, in the following we will show how to compute the 1.5PN dissipative corrections to the EoMs~\eqref{eq: EOM1} and to the CoM-frame transformations~\eqref{eq: com trasnf}.

\subsection{Leading dissipative effects in the equations of motion}
\label{subsec: dissipations in EoMs}

Varying the action~\eqref{eq: total E-M action} with respect to $y^\mu_A$ yields the EoMs of body $A$, in the form
\begin{align}
\label{eq: geodesic equation generic}
   &\frac{d^2 y^\mu_A}{dt^2}\, +\bigl(\Gamma^{\mu}_{\alpha \beta}\bigr)_A v^\alpha_A v^\beta_A - \bigl(\Gamma^{0}_{\alpha \beta}\bigr)_A v^\alpha_A v^\beta_A v^\mu_A \cr&\quad = -\frac{q_A}{m_A}\, \biggl(g^{\mu \rho}F_{\rho \nu} \sqrt{g_{\alpha \beta} v_A^\alpha v_A^\beta}\biggr)_A v^\nu_A ,
\end{align}
where the parentheses $(...)_A$ are used to denote quantities evaluated at the position $\mathbf{y}_A$ of body $A$.

As mentioned above, we know in advance that the leading dissipative effects we are interested in enter the dynamics at the 1.5PN level. The next-to-leading dissipative effects, being 2.5PN corrections (of both dipolar and quadrupolar nature), can be neglected at the global 2PN accuracy we aim to achieve. Moreover, there is no mixing with the conservative contributions, which are limited to integer (i.e.~time-symmetric) PN orders in the EoMs. Accordingly, for our current purpose, we just need to determine the 1.5PN term in the PN expansion of Eq.~\eqref{eq: geodesic equation generic}, for $\mu=i$.

To this end, we consider the following parametrization of the metric~\cite{PhysRevD.43.3273,Blanchet:1995fr}~\footnote{Even though this parametrization is formally valid only up to the 1PN order, it is actually enough to compute the 1.5PN dissipative contributions we seek here. We notice moreover that we have an overall minus sign with respect to Refs.~\cite{PhysRevD.43.3273,Blanchet:1995fr} due to the different signature employed.
}
\begin{align} 
\notag
   g_{00}\, &=\, 1\, -2V\, +2V^2\, + \dots \\ \label{eq: g param}
   g_{0i}\, &=\, 4 V_i\, +\dots \\ 
\notag g_{ij}\, &=\, -\delta_{ij}\Bigl(1\, +2V \Bigr)\, + \dots
\end{align}
given in terms of retarded potentials, $V \sim
\mathcal{O}\bigl(1/c^2\bigr)$ and $V_i \sim
\mathcal{O}\bigl(1/c^3\bigr)$, that can be obtained from the stress-energy tensor of matter (as shown in Ref.~\cite{Blanchet:1995fr}). Taking also into account the PN scaling of the vector potential's components, $A_0 \sim \mathcal{O}(1)$ and $A_i\sim \mathcal{O}(1/c)$, and considering that $\partial_0 = \frac{1}{c} \partial t$, we find that the 1.5PN dissipative contributions to the EoMs for body $A$ is completely determined by the right hand side of Eq.~\eqref{eq: geodesic equation generic}, and reads
\begin{align}\label{eq:dissipative correction a1 potentials}
 &( a_A^i)_{\rm 1.5PN} =\, \frac{q_A}{m_A}\Biggl[\Bigl( \partial_i A_0^{\rm 1.5PN} \Bigr)_A\, -\frac{1}{c}\Bigl(\partial_t A_i^{\rm 1PN}\Bigr)_A \cr &\quad +\frac{v^j_A}{c}\Bigl(\partial_i A_j^{\rm 1PN} -\partial_j A_i^{\rm 1PN}\Bigr)_A \Biggr].
\end{align}

To complete our derivation, we have to compute the 1.5PN component of $A_0$, the 1PN component of $A_i$, and cure, using the Hadamard regularization~\cite{HadamardReg, Blanchet:2000nu}, the self-field divergencies that appear when such components are evaluated at the positions of the bodies. This is done explicitly in App.~\ref{app: explicit calculation dissipative contributions}. For $A=1$, our final result in harmonic coordinates is
    \begin{align}
    \label{eq: dissipative corretion EoM 1}
       &\bigl( a_1^i\bigr)_{\rm 1.5PN}\, =\frac{2G }{3 r^3 c^3}\frac{q_1 (q_1 m_2\, -q_2 m_1) }{m_1}\cr&\quad\times\biggl(1\, -\frac{q_1 q_2}{G m_1 m_2}\biggr)  \Bigl[3 (\mathbf{n}\cdot \mathbf{v})\, n^i\, -v^i\Bigr],
    \end{align}
and the analogue for $A=2$ is obtained by exchanging the body labels, $1 \leftrightarrow2$, with $\mathbf{n} \to -\mathbf{n}$ and $\mathbf{v} \to -\mathbf{v}$.

As a consistency check for Eq.~\eqref{eq: dissipative corretion EoM 1}, we used it to compute the corresponding flux of energy at infinity,  which is determined using a balance-law approach analogous to the one used in Sec.~VI D of Ref.~\cite{Mirshekari:2013vb}. The resulting energy flux exactly reproduces the leading dipolar flux given in Eq.~(B56) of Ref.~\cite{Khalil:2018aaj}.

\subsection{Leading dissipative effects in the transformations to the center of mass frame}
\label{subsec: dissipations in CoM transf}

The presence of a dissipative correction in the EoMs also affects the linear momentum, the CoM position, and, consequently, the coordinate transformation to the CoM frame. To obtain the correction to the latter, we follow the same procedure outlined in Sec.~\ref{sec:CoM_frame}, modifying both the total linear momentum and the CoM position with the addition of extra terms at the 1.5PN order. 

We begin by defining a modified total linear momentum $\mathbf{\Tilde{P}}$,  expressed as the sum of the conserved linear momentum given in Eq.~\eqref{eq: conserved total linear momentum} and a general ansatz at 1.5PN order. To determine the coefficients entering the ansatz, we require the momentum $\mathbf{\Tilde{P}}$ to be conserved along the full radiation-reacted dynamics. We therefore solve the equation
\begin{equation}
    \frac{d \Tilde{P}^i}{dt}\, = \mathcal{O}\Big(1/c^7\Big),
\end{equation}
while order reducing all the accelerations with the full EoMs, i.e.~complete of the dissipative contribution, as per Eq.~\eqref{eq: dissipative corretion EoM 1}.
The resulting modified total linear momentum reads
\begin{align}
    \label{eq: Ptilde}
    &\Tilde{P}^i=\, P^i\, - \frac{2\,  n^i}{3 m_1 m_2 c^3r}\Bigl(m_1 m_2\, -q_1 q_2\Bigr)\cr&\quad\times\Bigl[ m_1 q_2^2 -m_2q_1^2+(m_1-m_2)q_1 q_2\Bigr].
\end{align}
%

Armed with this result, we can proceed along the same lines as in Sec.~\ref{sec:CoM_frame} and compute the dissipative contributions to the CoM position. We thus consider a modified CoM position $\mathbf{\Tilde{G}}$ that differs from Eq.~\eqref{eq: Gconservative} by a general dissipative contribution depending on free coefficients, which we constrain by imposing
\begin{equation}
    \frac{d \Tilde{G}^i}{dt}\, =\, \Tilde{P}^i.
\end{equation}
Our result is
\begin{equation}
\label{eq: Gtilde}
    \Tilde{G}^i=\, G^i\,  +\frac{ v_1^i (q_1+q_2)}{3\,  m_2 \,c^3}\Bigl(m_1 q_2 \, -m_2 q_1\Bigr)\, +(1\leftrightarrow2).
\end{equation}

Finally, the correction~\eqref{eq: Gtilde} to the CoM momentum determines an associated dissipative contribution in the transformations to the CoM frame, which now follow from the condition $\Tilde{G}^i=0$. In particular, we find that the 1.5PN radiation-reaction correction to Eq.~\eqref{eq: com trasnf} is
\begin{equation}
    \label{eq: 1.5PN com transformation}
  \bigl( y_1^i\bigr)_{\rm 1.5PN} = \frac{2v^i\Bigl(m_2 q_1^2 -q_1 q_2(m_2-m_1) -m_1 q_2^2\Bigr) }{3 (m_1+m_2)^2 c^3}.
\end{equation}
%

\section{Conclusions}
\label{sec:conc}

In this work, we extended the PN knowledge of the dynamics of charged BBHs to 2PN accuracy, addressing both conservative and dissipative contributions. 

In particular, using the EFT approach, we derived the 2PN-accurate Lagrangian in harmonic and Lorenz gauge in Einstein-Maxwell theory. Our result reproduces the known 2PN Lagrangian of GR and the 2PC Lagrangian of electromagnetism in the appropriate limits. Furthermore, we find complete agreement with the PN expansion of the PM Einstein-Maxwell results reported in Refs.~\cite{Wilson-Gerow:2023syq,Alonzo-Artiles:2026wbe}. 
From this Lagrangian, we derived the conservative EoM, as well as the corresponding Hamiltonian in an ADM-type coordinate system defined by the condition that, in the neutral limit, it reduces to the usual ADM one. In the absence of explicit ADM results at 2PN level for the charge-dependent sector, these coordinates and all results expressed in terms of them cannot be fully determined. Instead, they depend on a set of arbitrary charge-dependent coefficients that can be uniquely fixed once the 2PN ADM Hamiltonian for charged BBHs has been computed. This will also provide a non-trivial check for the results obtained in this paper.
In addition, we computed the associated 2PN transformations to the CoM frame. 

Concerning the dissipative part of the dynamics, we provided the leading 1.5PN contributions to the EoMs, which is the only relevant contribution for computations with 2PN accuracy, and we incorporated these corrections consistently into the transformations to the CoM frame. 

Finally, we computed three gauge invariant conservative quantities up to the 2PN order: the periastron advance and the binding energy, for quasi-circular orbits, and the scattering angle along unbound orbits. 

This work represents an important first step toward the complete 2PN characterization of motion and radiation of a charged BBH, and it will be completed by a forthcoming paper focused on the derivation of the 2PN-accurate energy flux, as well as  on the analysis of the extremal limit~\cite{Placidi:future}. Such an endeavor will be crucial for studying these systems through GW observations, particularly once our results are incorporated into analytical waveform models, such as those developed within the effective-one-body formalism \cite{Buonanno:2000qq}.

Future extensions of this work include incorporating spin and finite-size effects of the compact objects for a more accurate description of their dynamics. Moreover, this analysis could be extended to account for magnetic- and electric-type electromagnetic multipole moments of the compact objects, building on the results of Refs.~\cite{Henry:2023guc,Henry:2023len}.

\acknowledgments
\paragraph*{Note:}
We sincerely thank Manfred Kraus and Allan Alonzo for making us aware of several typos in the expressions for the 2PN observables, that are now corrected in this version of the manuscript, and agree with the results of Ref.~\cite{Alonzo-Artiles:2026wbe}.

P.M. and M.P. would like to thank Francisco M. Blanco, Giacomo Brunello, Jung-Wook Kim, Manoj K. Mandal, Raj Patil, Angelo Ricciardone, Trevor Scheopner and Jan Steinhoff for helpful suggestions and discussions over the course of the project.
We thank Donato Bini and Jan Steinhoff for valuable comments and feedback on the manuscript.
A.P., E.G., and  M.O. acknowledge financial support from
the Italian Ministry of University and Research (MUR)
through the program “Dipartimenti di Eccellenza 2018-
2022” (Grant SUPER-C). M. O. acknowledges support from the Italian Ministry of University and Research (MUR) via the PRIN 2022ZHYFA2, GRavitational wavEform models for coalescing compAct binaries with eccenTricity (GREAT) and from “Fondo di Ricerca d’Ateneo”
2023 (GraMB) of the University of Perugia. E.G, and M.O. acknowledge support by the “Center of Gravity”, which is
a Center of Excellence funded by the Danish National
Research Foundation under grant No. 184.  
The work of M.P. is supported by the European Union under the Next Generation EU programme. M.P. gratefully acknowledges financial support from Fondazione Ing. Aldo Gini, and the hospitality and support of the Albert Einstein Institute.
N.B., P.M. and M.P. acknowledge the support of the INFN initiatives
{\it Amplitudes} and {\it InDark}.



\appendix
\section{Expanded fundamental action for Feynman rules}
\label{app:Feynman_rules}
We report the expression for the fundamental action, given by Eq.~\eqref{eq: total E-M action}, expressed in term of Kol-Smolkin variables, given in Eqs.~\eqref{eq:metric_decomposition} and~\eqref{eq:potential_decomposition}, and expanded in the PN parameter using the power counting rules introduced in Sec.~\ref{sec:power_counting}.
The resulting expansion is a polynomial expression in the fields, containing formally an infinite number of terms, which specify the interactions present in the theory. Employing the power counting rules we select only the subset of terms which can contribute to quantities at 2PN order, i.e.~$1/c^4$. 

In Eqs.~\eqref{eq:expanded_fundamental_action} we present the explicit expression for the expanded action. 
For brevity we report only the terms which are actually needed to compute the point-particle conservative Lagrangian at 2PN order. The omitted interaction terms can only appear in diagrams which contribute to higher PN order.
These computations have been performed employing a \texttt{Mathematica} code based on the \texttt{EFTofPNG} package~\cite{Levi:2017kzq}.

\begin{widetext}
    \begin{subequations}
    \label{eq:expanded_fundamental_action}
    \begin{align}
    S = S_{\rm bulk, kin} + S_{\rm bulk} + S_{\rm wl}   
    \end{align}
    \begin{align}
    \label{eq:expanded_fundamental_action_bulk_kinetic}
    S_{\rm bulk, kin} =& \int dt d^3x \left( 
    \Big\{  (\partial_{i}A^{i})^{2}
 - c_{d}\,\partial_{i}\phi\,\partial^{i}\phi
 + \frac{1}{2}\,\partial_{i}\phi_{\text{EM}}\,\partial^{i}\phi_{\text{EM}}
 - \partial_{i}A_{j}\,\partial^{j}A^{i}
 + \partial_{j}A_{i}\,\partial^{j}A^{i} \right. \nonumber\\
&\hspace{1.75cm}
 + \frac{1}{2}\Big[ -(\partial_{i}A_{\text{EM}}^{i})^{2}
   + \partial_{i}A_{\text{EM}\,j}\,\partial^{j}A_{\text{EM}}^{i}
   - \partial_{j}A_{\text{EM}\,i}\,\partial^{j}A_{\text{EM}}^{i} \Big] \nonumber\\
&\hspace{1.75cm}
 + \frac{1}{4}\Big[
    \partial_{j}\sigma^{k}{}_{k}\,\partial^{j}\sigma^{i}{}_{i}
   + 4\,\partial_{i}\sigma^{i}{}_{j}\,\partial_{k}\sigma^{j k}
   - 2\big( 2\,\partial_{j}\sigma_{ik}\,\partial^{k}\sigma^{i j}
           + \partial_{k}\sigma_{i j}\,\partial^{k}\sigma^{i j} \big)
   \Big] \Big\} \nonumber\\
&\hspace{1cm} \left.
 + \frac{1}{c^{2}}{\Big\{
     \frac{1}{2}\,\partial_{t}A_{\text{EM}\,i}\,\partial_{t}A_{\text{EM}}^{i} -\,\partial_{t}A_{i}\,\partial_{t}A^{i}
    + \frac{1}{4}\left( -(\partial_{t}\sigma_{i}{}^{i})^{2}
      + 2\,\partial_{t}\sigma_{ij}\,\partial_{t}\sigma^{ij} \right)
    + c_{d}(\partial_{t}\phi)^{2}
    - \frac{1}{2}(\partial_{t}\phi_{\text{EM}})^{2}
 \Big\}}
 \right)\ , 
 \end{align}
 \begin{align}
    \label{eq:expanded_fundamental_action_bulk}
S_{\rm bulk} &= \int dt d^3x \left( \frac{1}{c^{2}}\Bigg\{
 \frac{c_{d}\!\left(-\,\sigma^{i}{}_{i}\,\partial_{j}\phi\,\partial^{j}\phi
     +2\,\sigma_{ij}\,\partial^{i}\phi\,\partial^{j}\phi\right)}{2\,\Lambda}
 +\frac{\sigma^{i}{}_{i}\,\partial_{j}\phi_{\text{EM}}\,\partial^{j}\phi_{\text{EM}}
     -2\,\sigma_{ij}\,\partial^{i}\phi_{\text{EM}}\,\partial^{j}\phi_{\text{EM}}}{4\,\Lambda} \right.
\nonumber\\
&\hspace{2.5cm}
 -\frac{(\partial_{i}\phi_{\text{EM}}\partial^{i}\phi_{\text{EM}})\,\phi}{\Lambda}
 +\frac{c_{d}\!\left((\partial_{i}A^{i})^{2}-\partial_{i}A_{j}\,\partial^{j}A^{i}
     +\partial_{j}A_{i}\,\partial^{j}A^{i}\right)\phi}{\Lambda}
\nonumber\\
&\hspace{2.5cm}
 -\frac{(-2+c_{d})\!\left((\partial_{i}A_{\text{EM}}^{i})^{2}
     -\partial_{i}A_{\text{EM}\,j}\,\partial^{j}A_{\text{EM}}^{i}
     +\partial_{j}A_{\text{EM}\,i}\,\partial^{j}A_{\text{EM}}^{i}\right)\phi}{2\,\Lambda}
\nonumber\\
&\hspace{2.5cm}
 -\frac{1}{\Lambda}\left(
   \partial_{i}A_{\text{EM}}^{\,i} A^{j}\partial_{j}\phi_{\text{EM}}
  -A^{i}\partial_{i}A_{\text{EM}}^{\,j}\,\partial_{j}\phi_{\text{EM}}
  +A^{k}\partial_{j}\phi_{\text{EM}}\,\partial^{j}A_{\text{EM}\,k}
  +\partial_{i}A^{\,i}\partial_{j}A_{\text{EM}}^{\,j}\,\phi_{\text{EM}}
 \right)
\Bigg\}
\nonumber\\
&\hspace{1.5cm}
+\frac{1}{c^{3}}\Bigg\{
 -\frac{2c_{d}\,A^{i}\partial_{i}\phi\,\partial_{t}\phi}{\Lambda}
 +\frac{\big(A^{i}\partial_{i}\phi_{\text{EM}}+\partial_{i}A^{\,i}\,\phi_{\text{EM}}\big)\,\partial_{t}\phi_{\text{EM}}}{\Lambda}
\nonumber\\
&\hspace{2.5cm}
 +\frac{2\,\partial_{i}\phi_{\text{EM}} \partial_{t}A_{\text{EM}}^{\,i} \,\phi
        -c_{d}\,\partial_{i}A_{\text{EM}}^{\,i}\,\phi_{\text{EM}}\,\partial_{t}\phi
        -2\,\partial_{i}A_{\text{EM}}^{\,i}\,\phi\,\partial_{t}\phi_{\text{EM}}}{\Lambda}
\Bigg\}
\nonumber\\
&\hspace{1.5cm} \left.
+\frac{1}{c^{4}}\Bigg\{
 \frac{\partial_{i}\phi_{\text{EM}}\partial^{i}\phi_{\text{EM}}\,\phi^{2}}{\Lambda^{2}}
 -\frac{c_{d}^{2}\,\phi\,(\partial_{t}\phi)^{2}}{\Lambda}
 +\frac{\partial_{t}\phi_{\text{EM}}\big(2c_{d}\,\phi_{\text{EM}}\,\partial_{t}\phi
      +(2+c_{d})\,\phi\,\partial_{t}\phi_{\text{EM}}\big)}{2\,\Lambda}
\Bigg\} \right) \  ,
    \end{align}
    \begin{align}
    \label{eq:expanded_fundamental_action_worldline}
    S_{\rm wl} &= \sum_{A = 1,2} \int dt_A \left( 
     \Big\{-\,\frac{m_A\,\phi}{\Lambda} \;+\; \frac{q_{\text{EM}, A}\,\phi_{\text{EM}}}{\Lambda_{\text{EM}}}  \Big\} \right. \nonumber\\
&\hspace{1.5cm} + \frac{1}{c}\,\Big\{ \frac{m_A\,A^{i} v_{A, i}}{\Lambda} \;+\; \frac{q_{\text{EM}, A}\,A_{\text{EM}}^{i} v_{A, i}}{\Lambda_{\text{EM}}} \Big\} \nonumber\\
&\hspace{1.5cm}
+ \frac{1}{c^{2}}\,\Big\{  \frac{m_A\, v_{A, i} v_{A, j}\,\sigma^{ij}}{2\,\Lambda} \;-\; \frac{(-1+c_{d})\,m_A\,v_A^2\,\phi}{2\,\Lambda} \;-\; \frac{m_A\,\phi^{2}}{2\,\Lambda^{2}} \Big\} \nonumber\\
&\hspace{1.5cm} + \frac{1}{c^{3}}\,\Big\{ \frac{m_A\,A^{i} v_{A, i}\,v_A^2}{2\,\Lambda} \;+\; \frac{m_A\,A^{i} v_{A, i}\,\phi}{\Lambda^{2}} \Big\}
\nonumber\\
&\hspace{1.5cm}
\left.
+ \frac{1}{c^{4}}\,\Big\{ -\,\frac{(-1+2c_{d})\,m_A\,v_A^{4}\,\phi}{8\,\Lambda}
+\frac{(-1+c_{d})^{2}\,m_A\,v_A^2\,\phi^{2}}{4\,\Lambda^{2}}
- \frac{m_A\,\phi^{3}}{6\,\Lambda^{3}} \Big\} \right) \ ,
    \end{align}
    \end{subequations}     
\end{widetext}
where $v_A^2 = \delta_{ij} v_A^i v_A^j$ and the time dependence of $v_A^i = v_A^i(t_A)$ is understood.

The expanded action in Eqs.~\eqref{eq:expanded_fundamental_action} equivalently represents the action for a $d$-dimensional Euclidean interacting quantum field theory for the five potential fields $W$.
In particular, from the expanded bulk action in Eq.~\eqref{eq:expanded_fundamental_action_bulk_kinetic} we obtain the explicit expressions for the propagators of the potential fields $W$, from the bulk action in Eq.~\eqref{eq:expanded_fundamental_action_bulk} we obtain the Feynman rules for the bulk interaction vertices of the theory, and from the worldline action in Eq.~\eqref{eq:expanded_fundamental_action_worldline} we obtain the Feynman rules for the interactions between the potential fields and the worldlines.

Let us notice that the propagators for the potential fields in Eq.~\eqref{eq:expanded_fundamental_action_bulk_kinetic} have a non-homogeneous scaling in the PN parameter, since the terms with two temporal derivatives are suppressed by the factor $1/c^2$. We therefore treat these terms as perturbative propagator insertions, and they encode the retardation corrections of the PN formalism~\cite{Goldberger:2004jt}.

\section{Explicit results}
\label{app: center-of-mass frame}

In this Appendix, we provide the explicit expression of Lagrangian and Hamiltonian in ADM-coordinates, following the procedure detailed in Sec.~\ref{subsec:adm_coord}.  Moreover, we show the CoM-frame expression of the harmonic EoMs, the harmonic Lagrangian and the ADM-type Hamiltonian.

Thus, the explicit expressions of each component of ADM-type Lagrangian in Eq.~\eqref{eq: Total_ADM-type_Lagrangian} read
\begin{widetext}
   \begin{subequations}
\label{eq: ADM-type_Lagrangian_contributions}
    \begin{equation}
        L^{\rm ADM-type}_{\mathrm{kin}} = \frac{m_1 V_1^2}{2} + \frac{1}{c^2} \left\{ \frac{ m_1 V_1^4}{8} \right\} + \frac{1}{c^4} \left\{ \frac{m_1 V_1^6}{16}  \right\} +\big( 1 \leftrightarrow 2\big) \ ,
    \end{equation}
    \begin{align}
    \label{eq: lagrangian Grav 2PN}\notag
            &L^{\rm ADM-type}_{2\mathrm{PN}} = \frac{G m_1 m_2}{2 R} \\ \notag
            &\quad+\frac{1}{c^2} \Biggl\{\frac{G m_1 m_2}{4 R}\Bigl(3 V_1^2-7\mathbf{V}_1\cdot \mathbf{V}_2+3 V_2^2-(\mathbf{N}\cdot \mathbf{V}_1)(\mathbf{N}\cdot \mathbf{V}_2)\Bigr)-\frac{G^2 m_1 m_2}{4 R^2}(m_1+m_2)\Biggr\} \\ \notag
            &\quad+\frac{1}{c^4}\Biggl\{\frac{G^3 m_1 m_2}{8 R^3}(m_1^2+5 m_1 m_2+m_2^2)+ \frac{G^2 m_1 m_2^2}{8 R^2}\Bigl(15(\mathbf{N}\cdot \mathbf{V}_2)^2+16 V_1^2-30 (\mathbf{V}_1\cdot\mathbf{V}_2)+11 V_2^2\Bigr) \\ \notag
            &\quad \hspace{1 cm}+\frac{m_1 m_2}{16 R}\Biggl[3 (\mathbf{N}\cdot \mathbf{V}_1)^2(\mathbf{N}\cdot \mathbf{V}_2)^2-4 (\mathbf{N}\cdot \mathbf{V}_1)(\mathbf{N}\cdot \mathbf{V}_2)V_1^2-10 (\mathbf{N}\cdot \mathbf{V}_2)^2V_1^2+14 V_1^4 \\ 
            & \quad\hspace{1.5cm}+12(\mathbf{N}\cdot \mathbf{V}_1)(\mathbf{N}\cdot \mathbf{V}_2)(\mathbf{V}_1\cdot \mathbf{V}_2)-14 V_1^2(\mathbf{V}_1\cdot \mathbf{V}_2)+2(\mathbf{V}_1\cdot \mathbf{V}_2)+11 V_1^2 V_2^2\Biggr]\Biggr\}+\big( 1 \leftrightarrow 2\big) \ ,
        \end{align}
    \begin{align}
       \label{eq: Lagranian charged}\notag
        &L^{\rm ADM-type}_{\rm charged}= L^{\rm ADM-type}_{\rm 2PC}+L^{\rm ADM-type}_{\rm mixed}= -\frac{q_1 q_2}{ R} \\ \notag& \hspace{1cm}+\frac{1}{c^2}\Biggl\{\frac{G m_2}{2 R^2} \left(2 q_1 q_2-q_1^2\right)+\frac{G m_1}{2 R^2}(2 q_1 q_2- q_2^2)+\frac{q_1 q_2}{2 R}\Bigl(\mathbf{V}_1\cdot\mathbf{V}_2+(\mathbf{N}\cdot \mathbf{V}_1)(\mathbf{N}\cdot \mathbf{V}_2)\Bigr)\Biggr\}\\  \notag
   &\hspace{1cm}+\frac{1}{c^4}
   \Biggl\{\frac{q_1 q_2}{8 R}\Bigl(V_2^2(\mathbf{N}\cdot\mathbf{V}_1)^2+(\mathbf{N}\cdot \mathbf{V}_2)^2 V_1^2+2 (\mathbf{V}_1\cdot \mathbf{V}_2)^2-3(\mathbf{N}\cdot \mathbf{V}_1)^2(\mathbf{N}\cdot \mathbf{V}_2)^2-V_1^2 V_2^2\Bigr)+\frac{q_1^2 q_2^2}{4R^3}\\ \notag
   &\hspace{1.5 cm}+G^2 \Biggl[
   \frac{m_2^2
   }{4 R^3}\Bigl(4 q_1^2-3 q_1 q_2\Bigr)+\frac{m_2 m_1}{4R^3}
   \Bigl(2q_1^2-17 q_1
   q_2+2 q_2^2\Bigr)+\frac{-3 m_1^2q_1 q_2+4 m_1^2q_2^2}{4R^3}\Biggr] \\ \notag
   &\hspace{2.5cm}+\frac{q_1^2 q_2^2}{8 r^2
   m_2}\Bigl(3
   (\mathbf{N} \cdot \mathbf{V_1})^2-V_1^2\Bigr)+\frac{q_1^3 q_2^3}{4 R^3 m_2 m_1}+\frac{q_1^2
   q_2^2}{8m_1R^2}\left(3
   (\mathbf{N} \cdot \mathbf{V_2})^2-V_2^2\right)\\ \notag
   &\hspace{1.5cm} +G
   \Biggl[\frac{m_2q_1 q_2 }{4 R^2} 
   \Bigl(2(\mathbf{N} \cdot \mathbf{V_1}) (\mathbf{N} \cdot \mathbf{V_2})-13
   (\mathbf{N} \cdot \mathbf{V_2})^2-4(\mathbf{V_1} \cdot \mathbf{V_2})+5 V_2^2\Bigr)\\ \notag
   &\hspace{2.5cm}+\frac{m_1 q_1q_2}{R^2}\Biggl(-\frac{13}{4} 
   (\mathbf{N} \cdot \mathbf{V_1})^2+\frac{(\mathbf{N} \cdot \mathbf{V_1}) (\mathbf{N} \cdot \mathbf{V_2})}{2}+\frac{5
   V_1^2}{4}-(\mathbf{V_1} \cdot \mathbf{V_2})\Biggr)\\ \notag
   &\hspace{2.5cm}+\frac{m_2q_1^2 }{4 R^2} \left(4(\mathbf{N} \cdot \mathbf{V_1})^2-2(\mathbf{N} \cdot \mathbf{V_2})^2-4 V_1^2+6
   (\mathbf{V_1} \cdot \mathbf{V_2})-V_2^2\right) \\ \notag
   &\hspace{2.5cm}+\frac{m_1 q_2^2}{4 R^2}\left(4 (\mathbf{N} \cdot \mathbf{V_2})^2-2(\mathbf{N} \cdot \mathbf{V_1})^2-V_1^2+6 (\mathbf{V_1} \cdot \mathbf{V_2})-4 V_2^2\right)\Biggr]
   \\ \notag
   &\hspace{1.2cm}+A_1 \Biggl[\frac{V_1^2}{R}\Bigl(-(\mathbf{N} \cdot \mathbf{V_1})^2
   +(\mathbf{N} \cdot \mathbf{V_1}) (\mathbf{N} \cdot \mathbf{V_2}) +V_1^2-
   (\mathbf{V_1} \cdot \mathbf{V_2})\Bigr) \\ \notag
   &\hspace{2.2 cm}-\frac{G m_2 }{R^2}\left(1-\frac{q_1 q_2}{G m_1 m_2}\right) \Bigl(2 (\mathbf{N} \cdot \mathbf{V_1})^2+V_1^2\Bigr)\Biggr]\\ \notag
   &\hspace{1.2cm}+A_2
   \Biggl[\frac{V_1^2}{R}\Bigl((\mathbf{N} \cdot \mathbf{V_2})^2-(\mathbf{N} \cdot \mathbf{V_1}) (\mathbf{N} \cdot \mathbf{V_2}) 
   +(\mathbf{V_1} \cdot \mathbf{V_2})- V_2^2\Bigr) \\ \notag
   &\hspace{2.2cm}+\frac{G}{R^2}\left(1-\frac{q_1 q_2}{G m_1 m_2}\right)
   \left(m_1 V_1^2-2 m_2(\mathbf{N} \cdot \mathbf{V_1}) (\mathbf{N} \cdot \mathbf{V_2})\right)\Biggr]\\ \notag
   &\hspace{1.2cm}+A_3 \Biggl[\frac{(\mathbf{V_1} \cdot \mathbf{V_2})}{R}\left(V_1^2-(\mathbf{N} \cdot \mathbf{V_1})^2
   +(\mathbf{N} \cdot \mathbf{V_1}) (\mathbf{N} \cdot \mathbf{V_2}) 
   -(\mathbf{V_1} \cdot \mathbf{V_2})\right)\\  \notag
   &\hspace{2.2cm}+\frac{G}{R^2} \left(1-\frac{q_1 q_2}{G m_1 m_2}\right)\Bigl((\mathbf{N} \cdot \mathbf{V_1})^2
   m_1-m_2(\mathbf{N} \cdot \mathbf{V_1}) (\mathbf{N} \cdot \mathbf{V_2})-m_2(\mathbf{V_1} \cdot \mathbf{V_2})\Bigr)\Biggr]\\ \notag
   &\hspace{1.2cm}+A_4
   \Biggl[\frac{V_2^2}{R}\Bigl((\mathbf{V_1} \cdot \mathbf{V_2})^2-(\mathbf{N} \cdot \mathbf{V_1}) (\mathbf{N} \cdot \mathbf{V_2}) (\mathbf{V_1} \cdot \mathbf{V_2})+(\mathbf{N} \cdot \mathbf{V_2})^2
   (\mathbf{V_1} \cdot \mathbf{V_2})-(\mathbf{V_1} \cdot \mathbf{V_2})\Bigr)+\\ \notag
   &\hspace{2.2cm}+\frac{G}{R^2}\left(1-\frac{q_1 q_2}{G m_1 m_2}\right)
   \Bigl(m_1(\mathbf{N} \cdot \mathbf{V_1}) (\mathbf{N} \cdot \mathbf{V_2})+ m_1(\mathbf{V_1} \cdot \mathbf{V_2})-(\mathbf{N} \cdot \mathbf{V_2})^2 m_2\Bigr)\Biggr]\\ \notag
   &\hspace{1.2cm}+A_5 \Biggl[\frac{ V_2^2}{R}\Bigl(V_1^2-(\mathbf{N} \cdot \mathbf{V_1})^2
   +(\mathbf{N} \cdot \mathbf{V_1}) (\mathbf{N} \cdot \mathbf{V_2}) 
   -(\mathbf{V_1} \cdot \mathbf{V_2})\Bigr) \\ \notag
   &\hspace{2.2cm}+\frac{ G }{R^2}\left(1-\frac{q_1 q_2}{G m_1 m_2}\right)\left( 2 m_1 (\mathbf{N} \cdot \mathbf{V_1})
   (\mathbf{N} \cdot \mathbf{V_2}) - m_2V_2^2 \right)
\Biggr]\\ \notag
   &\hspace{1.2cm}+A_6 \Biggl[\frac{V_2^2}{R}\left((\mathbf{N} \cdot \mathbf{V_2})^2-(\mathbf{N} \cdot \mathbf{V_1}) (\mathbf{N} \cdot \mathbf{V_2})
    +(\mathbf{V_1} \cdot \mathbf{V_2})
   -V_2^2\right)+\frac{G m_1 }{R^2}\left(1-\frac{q_1 q_2}{G m_1 m_2}\right)\Bigl(2
   (\mathbf{N} \cdot \mathbf{V_2})^2+V_2^2\Bigr)\Biggr]\\ \notag
   &\hspace{1.2cm}+B_1 \Biggl[\frac{-2 (\mathbf{N} \cdot \mathbf{V_1}) (\mathbf{N} \cdot \mathbf{V_2})+2
   (\mathbf{N} \cdot \mathbf{V_2})^2+(\mathbf{V_1} \cdot \mathbf{V_2})-V_2^2}{R^2}+\frac{G m_1}{R^3}-\frac{q_1
   q_2}{R^3 m_2}\Biggr]\\ \notag
   &\hspace{1.2cm}+B_2 \Biggl[\frac{-2 (\mathbf{N} \cdot \mathbf{V_1})^2+2 (\mathbf{N} \cdot \mathbf{V_1})
   (\mathbf{N} \cdot \mathbf{V_2})+V_1^2-(\mathbf{V_1} \cdot \mathbf{V_2})}{R^2}-\frac{G m_2}{R^3}+\frac{q_1
   q_2}{R^3 m_1}\Biggr]\\ \notag
   &\hspace{1.2cm}+C_1 \Biggl[-\frac{3(\mathbf{N} \cdot \mathbf{V_1})^2}{R^2}\left( (\mathbf{N} \cdot \mathbf{V_1})^2-
   (\mathbf{N} \cdot \mathbf{V_1}) (\mathbf{N} \cdot \mathbf{V_2})- V_1^2+ (\mathbf{V_1} \cdot \mathbf{V_2})\right)-\frac{3 G m_2}{R^2}\left(1-\frac{q_1 q_2}{G m_1 m_2}\right)(\mathbf{N} \cdot \mathbf{V_1})^2 \Biggr]\\ \notag
   &\hspace{1.2cm}+C_2 \Biggl[+\frac{\mathbf{N} \cdot \mathbf{V_1}}{R} \Bigl(-3 (\mathbf{N} \cdot \mathbf{V_1})^2
   (\mathbf{N} \cdot \mathbf{V_2})+3 (\mathbf{N} \cdot \mathbf{V_1}) (\mathbf{N} \cdot \mathbf{V_2})^2+2  (\mathbf{N} \cdot \mathbf{V_2})
   V_1^2+(\mathbf{N} \cdot \mathbf{V_1}) (\mathbf{V_1} \cdot \mathbf{V_2})\\ \notag
   &\hspace{2.2cm}-2  (\mathbf{N} \cdot \mathbf{V_2})
   (\mathbf{V_1} \cdot \mathbf{V_2})-(\mathbf{N} \cdot \mathbf{V_1}) V_2^2\Bigr)+\frac{G}{R^2} \left(1-\frac{q_1 q_2}{G m_1 m_2}\right)\Bigl((\mathbf{N} \cdot \mathbf{V_1})^2
   m_1-2 (\mathbf{N} \cdot \mathbf{V_1}) (\mathbf{N} \cdot \mathbf{V_2}) m_2\Bigr)\Biggr] \\ \notag
   &\hspace{1.2cm}+C_3
   \Biggl[\frac{\mathbf{N} \cdot \mathbf{V_2}}{R}\Bigl(-3 (\mathbf{N} \cdot \mathbf{V_1})^2 (\mathbf{N} \cdot \mathbf{V_2})+3 (\mathbf{N} \cdot \mathbf{V_1})
   (\mathbf{N} \cdot \mathbf{V_2})^2+(\mathbf{N} \cdot \mathbf{V_2}) V_1^2+2 (\mathbf{N} \cdot \mathbf{V_1})
   (\mathbf{V_1} \cdot \mathbf{V_2})\\  \notag
   &\hspace{2.2cm}-(\mathbf{N} \cdot \mathbf{V_2}) (\mathbf{V_1} \cdot \mathbf{V_2})-2 (\mathbf{N} \cdot \mathbf{V_1}) 
   V_2^2\Bigr) +\frac{G}{R^2}\left(1-\frac{q_1 q_2}{G m_1 m_2}\right) \Bigl(2 (\mathbf{N} \cdot \mathbf{V_1}) (\mathbf{N} \cdot \mathbf{V_2})
   m_1-(\mathbf{N} \cdot \mathbf{V_2})^2 m_2\Bigr)\Biggr]\\ \notag
   &\hspace{1.2cm}+C_4 \Biggl[-\frac{3(\mathbf{N} \cdot \mathbf{V_2})^2}{R}\Bigl(
   (\mathbf{N} \cdot \mathbf{V_1}) (\mathbf{N} \cdot \mathbf{V_2})- (\mathbf{N} \cdot \mathbf{V_2})^2-(\mathbf{V_1} \cdot \mathbf{V_2})+
  V_2^2\Bigr)\\ 
  &\hspace{2.2cm}+\frac{3 G m_1}{R^2}\left(1-\frac{q_1 q_2}{G m_1 m_2}\right)  (\mathbf{N} \cdot \mathbf{V_2})^2\Biggr]\Biggr\},
    \end{align}
\end{subequations}
where we notice the presence of the unfixed coefficient~\eqref{eq: ADM parameters} of the function $F$.

Following the Legendre transformation in Eq~\eqref{eq: Legrendre transformation}, we can compute the ADM-type Hamiltonian~\eqref{eq:Total_ADM-type_Hamiltonian} in terms of the ADM-type conjugated momentum $P_A^i$~\eqref{eq: ADM conjugate momentum}, that reads~\cite{SupplementalMaterial} 

\begin{subequations}
\label{eq: ADM-type_Hamiltonian_contributions}
    \begin{equation}
        H^{\rm ADM-type}_{\rm kin}=\, \frac{P_1^2}{2 m_1}+ \frac{1}{c^2}\Biggl\{ -\frac{P_1^4}{8 m_1^3}\Biggr\}+\frac{1}{c^4}\Biggl\{\frac{P_1^6}{16 m_1^5}\Biggr\} +\big( 1 \leftrightarrow 2\big) \ ,
    \end{equation}
    \begin{align}
        \label{eq: ADM_type_Hamiltonian_pureGrav}\notag
            H^{\rm ADM-type}_{\rm 2PN}=& -\frac{G m_1 m_2}{2R} \\ \notag
            & +\frac{1}{c^2} \Biggl\{ \frac{G}{4 R}(\mathbf{N}\cdot \mathbf{P}_1)(\mathbf{N}\cdot \mathbf{P}_2)-\frac{3 G m_2 P_1^2}{2 R m_1}+\frac{7(\mathbf{P}_1\cdot \mathbf{P}_2)}{4R}+\frac{G^2 m_1 m_2 (m_1+m_2)}{2 R^2}\Biggr\} \\ \notag
            & +\frac{1}{c^4}\Biggl\{\frac{5 G^2 m_2^2 P_1^2}{2 m_1 R^2}-\frac{G^3 m_1 m_2}{8 R^3}(m_1^2+5 m_1 m_2+m_2^2)+\frac{5 G m_2 P_1^4}{8 m_1^3 R}\\ \notag
            &\hspace{1cm}+ \frac{G^2(m_1+m_2)}{8 R^2}\left(19 P_1^2-27 (\mathbf{P}_1\cdot \mathbf{P}_2)-6(\mathbf{N}\cdot \mathbf{P}_1)(\mathbf{N}\cdot \mathbf{P}_2)+19 P_2^2\right) \\ \notag
            &\hspace{1cm}+\frac{G}{8 m_1 m_2 R}\Bigl(5 P_1^2(\mathbf{N}\cdot \mathbf{P}_2)^2-22 P_1^2 P_2^2-4 (\mathbf{P}_1\cdot \mathbf{P}_2)^2-24 (\mathbf{P}_1 \cdot \mathbf{P}_2)(\mathbf{N}\cdot \mathbf{P}_1)(\mathbf{N}\cdot \mathbf{P}_2) \\ 
            &\hspace{2.5cm}-6 (\mathbf{N}\cdot \mathbf{P}_1)^2 (\mathbf{N}\cdot \mathbf{P}_2)^2\Biggr\} +\big( 1 \leftrightarrow 2\big) \ ,
        \end{align}
    \begin{align}
        \label{eq: ADM_type_Hamiltonian, charged part}\notag
       &H^{\rm ADM-type}_{\rm charged}= H^{\rm ADM-type}_{\rm 2PC}+H^{\rm ADM-type}_{\rm mixed}= \frac{q_1 q_2}{ R} \\ \notag
       &\hspace{1cm}+\frac{1}{c^2}\Biggl\{\frac{G}{2R^2}(m_2 q_1^2+m_1 q_2^2)-\frac{2G q_1 q_2 (m_1+m_2)}{R^2}-\frac{q_1 q_2}{4 m_1 m_2 R}\Bigl(\mathbf{P}_1 \cdot \mathbf{P_2}+2(\mathbf{N}\cdot \mathbf{P}_1)(\mathbf{N}\cdot \mathbf{P}_2)\Bigr) \Biggr\} \\ \notag
       &\hspace{1cm}+\frac{1}{c^4}\Biggl\{ \frac{G^2 m_1^2}{4 R^3}(3 q_1 q_2-4 q_2^2)-\frac{G^2 m_1 m_2}{4 R^3}(2 q_1^2-17 q_1 q_2+2 q_2^2)+ \frac{G^2 m_2^2}{4 R^3}(3 q_1 q_2-4 q_1^2) \\ \notag
       & \hspace{1.5 cm}+\frac{G m_1 q_2^2}{m_2 R}\Bigl(P_2^2-(\mathbf{N}\cdot \mathbf{P}_2)^2\Bigr)- \frac{3 G q_1^2 q_2^2}{4 R^3}+\frac{G m_2 q_1^2}{m_1^2 R^2}\Bigl( P_1^2-(\mathbf{N}\cdot \mathbf{P}_1)^2\Bigr)-\frac{q_1^3 q_2^3}{4 m_1 m_2 R^3} \\ \notag
       &\hspace{1.5 cm}+\frac{q_1 q_2}{4 m_1 m_2 R}\left( \frac{P_2^2}{m_2}+\frac{P_1^2}{m_1}\right)\left(\frac{q_1 q_2}{ R} +\mathbf{P_1}\cdot \mathbf{P}_2+(\mathbf{N}\cdot \mathbf{P}_1)(\mathbf{N}\cdot \mathbf{P}_2)\right) \\ \notag
       &\hspace{1.5 cm}+\frac{G }{4 m_1 R^2}\Biggl(q_1 q_2\Bigl(10 (\mathbf{P}_1\cdot \mathbf{P}_2)-12 P_1^2+4(\mathbf{N}\cdot \mathbf{P}_1)^2+4 (\mathbf{N}\cdot \mathbf{P}_1)(\mathbf{N}\cdot \mathbf{P}_2)  \Bigr) \\ \notag
       &\hspace{3 cm}+ q_2^2 P_1^2+ 2 q_2^2 (\mathbf{N}\cdot \mathbf{P}_1)^2-6 q_1^2(\mathbf{P}_1 \cdot \mathbf{P}_2) \Biggr) \\ \notag 
       &\hspace{1.5 cm}+\frac{G }{4 m_2 R^2}\Biggl(q_1 q_2\Bigl(10 (\mathbf{P}_1\cdot \mathbf{P}_2)-12 P_2^2+4(\mathbf{N}\cdot \mathbf{P}_2)^2+4 (\mathbf{N}\cdot \mathbf{P}_1)(\mathbf{N}\cdot \mathbf{P}_2)  \Bigr) \\ \notag
       &\hspace{3 cm}+ q_1^2 P_2^2+ 2 q_1^2 (\mathbf{N}\cdot \mathbf{P}_2)^2-6 q_2^2(\mathbf{P}_1 \cdot \mathbf{P}_2) \Biggr) \\ \notag
       &\hspace{1.5cm}+\frac{q_1 q_2}{8 m_1^2 m_2^2 R }\Biggl( P_1^2 P_2^2-2 (\mathbf{P}_1\cdot \mathbf{P}_2)^2-P_2^2(\mathbf{N}\cdot \mathbf{P}_1)^2-P_1^2(\mathbf{N}\cdot \mathbf{P}_2)^2+3 (\mathbf{N}\cdot \mathbf{P}_1)^2 (\mathbf{N}\cdot \mathbf{P}_2)^2\Biggr)\\ \notag
       &\hspace{1.2cm }+\frac{A_1}{m_1 R} \Biggl[+\frac{G m_2
   }{m_1
   R}\left(1-\frac{q_1 q_2}{G m_1 m_2}\right)\Bigl(P_1^2+2
   (\mathbf{N}\cdot \mathbf{P_1})^2\Bigr)-\frac{P_1^2}{m_1^3}\left(P_1^2-
   (\mathbf{N}\cdot \mathbf{P_1})^2\right)\\ \notag
       &\hspace{2.2cm} +\frac{P_1^2}{m_1^2 m_2}\Bigl((\mathbf{P_1}\cdot \mathbf{P_2}) - (\mathbf{N}\cdot \mathbf{P_1})
   (\mathbf{N}\cdot \mathbf{P_2})\Bigr)\Biggr]\\ \notag
   &\hspace{1.2cm}+\frac{A_2}{m_1 R} \Biggl[-\frac{G}{R}\left(1-\frac{ q_1 q_2}{G m_1 m_2
   }\right)\left(P_1^2-2 (\mathbf{N}\cdot \mathbf{P_1})
   (\mathbf{N}\cdot \mathbf{P_2})\right)+\frac{P_1^2}{m_2^2 m_1} \left(P_2^2-
   (\mathbf{N}\cdot \mathbf{P_2})^2\right)\\ \notag
   &\hspace{2.2cm}+\frac{P_1^2}{m_1^2 m_2} \left(
    (\mathbf{N}\cdot \mathbf{P_1}) (\mathbf{N}\cdot \mathbf{P_2})-(\mathbf{P_1}\cdot \mathbf{P_2})\right)\Biggr]\\ \notag
    &\hspace{1.2cm}+\frac{A_3}{m_1
   R}
   \Biggl[\frac{G}{R}\left(1-\frac{
   q_1 q_2}{G m_2
   m_1 }\right)\Bigl((\mathbf{P_1}\cdot \mathbf{P_2})-(\mathbf{N}\cdot \mathbf{P_1})^2+(\mathbf{N}\cdot \mathbf{P_1}) (\mathbf{N}\cdot \mathbf{P_2})\Bigr)\\ \notag
   &\hspace{2.2cm}+\frac{1}{m_1 m_2^2}\left((\mathbf{P_1}\cdot \mathbf{P_2})^2-(\mathbf{P_1}\cdot \mathbf{P_2}) (\mathbf{N}\cdot \mathbf{P_1})
   (\mathbf{N}\cdot \mathbf{P_2})\right)+\frac{1}{m_1^2 m_2}\left(-(\mathbf{P_1}\cdot \mathbf{P_2}) P_1^2+(\mathbf{P_1}\cdot \mathbf{P_2})
   (\mathbf{N}\cdot \mathbf{P_1})^2\right)\Biggr]
   \\ \notag
   &\hspace{1.2cm}+\frac{A_4}{m_2
   R} \Biggl[-\frac{G}{R}\left(1-\frac{
   q_1 q_2}{G m_2
   m_1 }\right)\Bigl((\mathbf{P_1}\cdot \mathbf{P_2})+(\mathbf{N}\cdot \mathbf{P_1})
   (\mathbf{N}\cdot \mathbf{P_2})-(\mathbf{N}\cdot \mathbf{P_2})^2\Bigr)\\ \notag
   &\hspace{2.2cm}+\frac{P_2^2}{m_2^2 m_1}\left((\mathbf{P_1}\cdot \mathbf{P_2}) -(\mathbf{P_1}\cdot \mathbf{P_2})
   (\mathbf{N}\cdot \mathbf{P_2})^2\right) -\frac{1}{m_1^2 m_2}\left((\mathbf{P_1}\cdot \mathbf{P_2})^2-(\mathbf{P_1}\cdot \mathbf{P_2})
   (\mathbf{N}\cdot \mathbf{P_1}) (\mathbf{N}\cdot \mathbf{P_2})\right)\Biggr]\\ \notag
   &\hspace{1.2cm}+\frac{A_5}{m_2
   R}
   \Biggl[\frac{G}{R}\left(1-\frac{
   q_1 q_2}{G m_2
   m_1 }\right)\Bigl(P_2^2-2 (\mathbf{N}\cdot \mathbf{P_1}) (\mathbf{N}\cdot \mathbf{P_2})\Bigr)-\frac{P_2^2}{m_1^2 m_2}\left(P_1^2-
   (\mathbf{N}\cdot \mathbf{P_1})^2\right)\\ \notag
   &\hspace{2.2cm}+\frac{P_2^2}{m_1
   m_2^2}\Bigl((\mathbf{P_1}\cdot \mathbf{P_2})
   - (\mathbf{N}\cdot \mathbf{P_1}) (\mathbf{N}\cdot \mathbf{P_2})\Bigr)\Biggr] \\ \notag
   &\hspace{1.2cm}+\frac{A_6}{m_2^2
   R} \Biggl[-\frac{G m_1}{R}\left(1-\frac{
   q_1 q_2}{G m_2
   m_1 }\right)
   \Bigl(P_2^2+2
   (\mathbf{N}\cdot \mathbf{P_2})^2\Bigr)\\ \notag
   &\hspace{2.2cm}-\frac{P_2^2}{m_1
   m_2}\Bigl((\mathbf{P_1}\cdot \mathbf{P_2})
  - (\mathbf{N}\cdot \mathbf{P_1}) (\mathbf{N}\cdot \mathbf{P_2})\Bigr)+\frac{P_2^2}{m_2^2}\left(P_2^2-
   (\mathbf{N}\cdot \mathbf{P_2})^2\right)\Biggr] \\ \notag
    &\hspace{1.2cm}+\frac{B_1}{R^2} \Biggl[-\frac{G
   m_1}{R}\left(1-\frac{
   q_1 q_2}{G m_2
   m_1 }\right)-\frac{1}{m_1 m_2}\left((\mathbf{P_1}\cdot \mathbf{P_2})-2 (\mathbf{N}\cdot \mathbf{P_1}) (\mathbf{N}\cdot \mathbf{P_2})\right)+\frac{1}{m_2^2}\left(P_2^2-2
   (\mathbf{N}\cdot \mathbf{P_2})^2\right)\Biggr] \\ \notag
   &\hspace{1.2cm}+\frac{B_2}{R^2} \Biggl[+\frac{G m_2}{R}\left(1-\frac{
   q_1 q_2}{G m_2
   m_1 }\right)+\frac{1}{m_1 m_2}\left((\mathbf{P_1}\cdot \mathbf{P_2})-2 (\mathbf{N}\cdot \mathbf{P_1}) (\mathbf{N}\cdot \mathbf{P_2})\right)-\frac{1}{m_1^2}\left(P_1^2-2
   (\mathbf{N}\cdot \mathbf{P_1})^2\right)\Biggr]\\ \notag
   &\hspace{1.2cm}+\frac{3 C_1}{m_1^2
   R}  (\mathbf{N}\cdot \mathbf{P_1})^2\Biggl[\frac{G  m_2
  }{R}\left(1-\frac{
   q_1 q_2}{G m_2
   m_1 }\right)+\frac{1}{m_1 m_2}\left( (\mathbf{P_1}\cdot \mathbf{P_2}) - (\mathbf{N}\cdot \mathbf{P_1})
   (\mathbf{N}\cdot \mathbf{P_2})\right)-\frac{1}{m_1^2}\left(P_1^2 -
   (\mathbf{N}\cdot \mathbf{P_1})^2\right)\Biggr]\\ \notag
   &\hspace{1.2cm}+\frac{C_2}{m_1
   R}
   \Biggl[-\frac{G}{R}\left(1-\frac{
   q_1 q_2}{G m_2
   m_1 }\right)\Bigl((\mathbf{N}\cdot \mathbf{P_1})^2-2 (\mathbf{N}\cdot \mathbf{P_1}) (\mathbf{N}\cdot \mathbf{P_2})\Bigr)\\ \notag
   &\hspace{2.2cm}+\frac{1}{m_2^2 m_1}\left(P_2^2 (\mathbf{N}\cdot \mathbf{P_1})^2+2
   (\mathbf{P_1}\cdot \mathbf{P_2}) (\mathbf{N}\cdot \mathbf{P_1}) (\mathbf{N}\cdot \mathbf{P_2})-3 (\mathbf{N}\cdot \mathbf{P_1})^2
   (\mathbf{N}\cdot \mathbf{P_2})^2\right)\\ \notag
   &\hspace{2.2cm}-\frac{1}{m_1^2 m_2}\left((\mathbf{P_1}\cdot \mathbf{P_2}) (\mathbf{N}\cdot \mathbf{P_1})^2+2
   P_1^2 (\mathbf{N}\cdot \mathbf{P_1}) (\mathbf{N}\cdot \mathbf{P_2})-3 (\mathbf{N}\cdot \mathbf{P_1})^3
   (\mathbf{N}\cdot \mathbf{P_2})\right)\Biggr]\\ \notag
   &\hspace{1.2cm}+\frac{C_3}{m_2
   R} \Biggl[\frac{G}{R}\left(1-\frac{
   q_1 q_2}{G m_2
   m_1 }\right)\Bigl(2 (\mathbf{N}\cdot \mathbf{P_1})
   (\mathbf{N}\cdot \mathbf{P_2})-(\mathbf{N}\cdot \mathbf{P_2})^2\Bigr)\\ \notag
   &\hspace{2.2cm}+\frac{1}{m_2^2 m_1}\left(2 P_2^2 (\mathbf{N}\cdot \mathbf{P_1}) (\mathbf{N}\cdot \mathbf{P_2})+(\mathbf{P_1}\cdot \mathbf{P_2})
   (\mathbf{N}\cdot \mathbf{P_2})^2-3 (\mathbf{N}\cdot \mathbf{P_1}) (\mathbf{N}\cdot \mathbf{P_2})^3\right)\\ \notag
   &\hspace{2.2cm}-\frac{1}{m_1^2 m_2}\left(2
   (\mathbf{P_1}\cdot \mathbf{P_2}) (\mathbf{N}\cdot \mathbf{P_1}) (\mathbf{N}\cdot \mathbf{P_2})+P_1^2 (\mathbf{N}\cdot \mathbf{P_2})^2-3
   (\mathbf{N}\cdot \mathbf{P_1})^2 (\mathbf{N}\cdot \mathbf{P_2})^2\right)\Biggr]\\ \notag
   &\hspace{1.2cm}+\frac{3 C_4}{m_2^2
   R} (\mathbf{N}\cdot \mathbf{P_2})^2\Biggl[\frac{ G m_1
   }{R}\left(1-\frac{
   q_1 q_2}{G m_2
   m_1 }\right)-\frac{1}{m_1 m_2}\left( (\mathbf{P_1}\cdot \mathbf{P_2}) - (\mathbf{N}\cdot \mathbf{P_1})
   (\mathbf{N}\cdot \mathbf{P_2})\right) \\
   &\hspace{2.2cm}+\frac{1}{m_2^2}\left(P_2^2
   - (\mathbf{N}\cdot \mathbf{P_2})^2\right)\Biggr]\Biggr\} \ .
    \end{align}
\end{subequations}

Now, using the CoM-frame transformation of Eq.~\eqref{eq: com trasnf} and the corresponding notation, the relative acceleration $a^i=\, a_1^i-a_2^i$ reads

\begin{align}
\label{eq: acc in COM}\notag
    a^i\, =&\, -\frac{G M}{r^2} n^i \Bigl(1-\eta_1 \eta_2\Bigr) \\ \notag
    &+\frac{1}{c^2}\Biggl\{ n^i \Biggl[\frac{G^2 M^2}{r^2}\Biggl(4-5 \eta_1 \eta_2+\frac{1}{2}\eta_2^2 \left(1-X_{12}\right)+\frac{1}{2}\eta_1^2\left(1+X_{12}\right)+2\nu \left(1- \eta_1 \eta_2\right)^2 \Biggr) \\ \notag
    &\hspace{1 cm}+ \frac{G M}{r^2}\Biggl(\frac{v^2}{2}\left(-1 -\frac{\eta_1\eta_2}{2}+(-3+3\eta_1 \eta_2)\nu\right)+ \frac{3}{2}\Dot{r}^2 (1- \eta_1 \eta_2) \nu\Biggr)\Biggr] + v^i  \Dot{r}\Biggl[ 4-\eta_1 \eta_2+(-2+2\eta_1 \eta_2)\nu\Biggr]\Biggr\} \\ \notag
    &-\frac{1}{c^3}\Biggl\{\frac{2 G^2 M^4 \nu^2}{3 r^4 }\left(\eta_1- \eta_2\right)^2\left(1- \eta_1 \eta_2\right)^2\Biggr\} \\ \notag
    &+\frac{1}{c^4}\Biggl\{n^i \Biggl[ \frac{G^3 M^3}{r^4}\Biggl(\frac{27}{2}\eta_1 \eta_2 -9- 3\left(1-\frac{1}{4}\eta_1 \eta_2\right)\left( \eta_1^2+\eta_2^2-X_{12}(\eta_2^2-\eta_1^2)\right) \\ \notag
    &\hspace{1cm}+ \nu \Biggl(\frac{7}{4}X_{12}(\eta_2^2- \eta_1^2)(1-\eta_1 \eta_2) -\frac{87}{4}- \frac{7}{4}(\eta_1^2+\eta_2^2)+ \frac{203}{4} \eta_1 \eta_2+\frac{1}{4} \eta_1 \eta_2 (\eta_1^2-107 \eta_1 \eta_2+\eta_2^2+3\eta_1^2 \eta_2^2) \Biggr)\\ \notag
    &\hspace{1 cm}+\frac{G^2 M^2}{r^3}\Biggl(2 \nu^2 (\Dot{r}^2- v^2)(1- \eta_1^2 \eta_2^2)^2+X_{12}(\eta_2^2- \eta_1^2)\bigl(\Dot{r}^2(1+\nu)\, -\frac{v^2}{2}(1+3 \nu)\bigr)+ \frac{v^2}{2}(\eta_1^2+\eta_1\eta_2+\eta_2^2) \\ \notag
    &\hspace{1.5 cm}+\Dot{r}^2(2-\eta_1^2-\eta_2^2)+ \nu\Bigl(\Dot{r}^2(25+ 5 \eta_1^2-39 \eta_1 \eta_2+ 5 \eta_2^2+ 4 \eta_1^2 \eta_2^2)+\frac{v^2}{2}(13-12 \eta_1 \eta_2-\eta_1^2\eta_2^2) \Bigr)\Biggr) \\ \notag
    &\hspace{1 cm}+\frac{G M}{r^2}\Biggl(\nu^2 (1- \eta_1 \eta_2)\Bigl(4 v^4+\frac{45}{8}\Dot{r}^4-6 \Dot{r}^2v^2\Bigr)+ 3 \nu \Bigl(1 -\frac{\eta_1 \eta_2}{2}\Bigr)-\frac{15 \nu}{8}\Dot{r}^4 (1- \eta_1 \eta_2)-\frac{v^4}{8} \eta_1 \eta_2 \Biggr)\Biggr] \\ \notag
    &\hspace{0.3 cm}+v^i \Biggl[\frac{G M }{r^2}\Biggl(\frac{1}{2} \Dot{r} v^2 \eta_1 \eta_2+ \frac{\nu \Dot{r}}{2}(-9 \Dot{r}^2+3 v^2( 5-\eta_1 \eta_2)+ \nu^2 \Dot{r}(1- \eta_1 \eta_2)(2 v^2- 3 \Dot{r}^2)\Biggl)\\ \notag 
    &\hspace{1cm}+\frac{G^2 M^2}{r^3}\Biggl(\Dot{r}(1-\nu)X_{12}(\eta_2^2-\eta_1^2)-4 \nu^2 \Dot{r} (1- \eta_1 \eta_2)^2+ \Dot{r}(\eta_1 \eta_2-\eta_1^2- \eta_2^2-2) \\ 
    &\hspace{1.8cm}+ \frac{\Dot{r}}{2}(56 \eta_1 \eta_2 -4 \eta_1^2 -4 \eta_2^2-7 \eta_1^2 \eta_2^2 -41)\Biggr) \Biggr]\Biggr\}.
\end{align}
We have included here also the 1.5PN dissipative contribution to the EoMs and that is discussed in Sec.~\ref{sec:dissipations}.

Starting from Eq.~\eqref{eq:: LagrangianHarmonicCoordinates12}, that depends linearly on the accelerations, by using Eq.~\eqref{eq: com trasnf}, we can obtain the mass-reduced Lagrangian $\mathcal{L}$ in the CoM frame

\begin{align}
\label{eq: CoM lagrangian }\notag
    \frac{\mathcal{L}}{\mu }=&\,\frac{G M}{r}(1-\eta_1 \eta_2)+\frac{v^2}{2}\\ \notag
    &+\frac{1}{c^2} \Biggl\{ \frac{v^4}{8}(1-3 \nu)+ \frac{G M}{2r}\Biggl(3 v^2+ \nu(1-\eta_1 \eta_2)(v^2+ \Dot{r}^2)\Biggr)+\frac{G^2 M^2}{4r^2}\Biggl(4 \eta_1 \eta_2 - \eta_1^2- \eta_2^2-2+(\eta_2^2- \eta_1^2) X_{12} \Biggr) \Biggr\}\\ \notag
    &+ \frac{1}{c^4}\Biggl\{\frac{v^6}{16}(1-7 \nu+ 13 \nu^2) + \frac{G M \nu}{8}\Bigl( (7-\eta_1 \eta_2)(-2(\mathbf{a}\cdot \mathbf{v})\Dot{r}+ (\mathbf{n}\cdot \mathbf{a})v^2)- \Dot{r}^2 ( \mathbf{n}\cdot \mathbf{a})(1- \eta_1 \eta_2)\Bigr)\\ \notag
    &\hspace{1cm}+ \frac{G^3 M^3}{4r^3}\Biggl(X_{12}(\eta_1^2- \eta_2^2)+ 2- 4 \eta_1 \eta_2+ \eta_1^2+ \eta_2^2+ \nu \Bigl(15+ 2 \eta_1^2- 39 \eta_1 \eta_2+ 2 \eta_2^2+ 21 \eta_1^2 \eta_2^2- \eta_1^3 \eta_2^3\Bigr)\Biggr) \\ \notag
    &\hspace{1cm}+ \frac{G^2 M^2}{r^2}\Biggl[\frac{\eta_2^2- \eta_1^2}{8}(2 \Dot{r}- v^2)(1+ \nu)X_{12}+ \frac{v^2}{8}(14- \eta_1^2- \eta_2^2)+ \frac{\Dot{r}}{4}(2- \eta_1^2- \eta_2^2)+ \frac{\nu^2}{2}(v^2+ 3 \Dot{r}^2)(1- \eta_1 \eta_2)^2 \\ \notag
    &\hspace{2.5cm}+ \frac{\nu}{8}\Biggl( v^2 \Bigl(-27- \eta_1^2- \eta_2^2 -7(1- \eta_- \eta_2)^2\Bigr)+ \Dot{r}^2\Bigl(41+ 10 \eta_1^2+ 10 \eta_2^2-2 \eta_1 \eta_2- 3\eta_1^2 \eta_2^2 \Bigr)\Biggr) \Biggr]\\ \notag
    &\hspace{1cm}+ \frac{GM}{ r}\Biggl[ \frac{7 v^4}{8}+ \frac{\nu}{4}\Bigl( -v^4(5+ \eta_1 \eta_2)+ v^2 \Dot{r}^2(1- \eta_1 \eta_2)\Bigr)+ \frac{\nu^2}{8}(1- \eta_1 \eta_2)\Bigl(3 \Dot{r}^4-9 v^4-10 v^2 \Dot{r}^2\Bigr) \Biggr]\\ 
    &\hspace{1cm}- \frac{G^2 M^2 \nu}{2 r}(\mathbf{n}\cdot \mathbf{a})\eta_1 \eta_2(8-\eta_1 \eta_2)\Biggr\} \ .
\end{align}
\end{widetext}

Now, following the same procedure described in Sec.~\ref{subsec:adm_coord}, we can identify the contact transformation from the relative separation vector $x^i= y_1^i\, -y_2^i$ in harmonic coordinates to the one $X^i=\, Y_1^i\, -Y_2^i$ in ADM-type coordinates:
\begin{equation}
    \label{eq: contact transformation COM}
    \delta x^i=\, X^i\, -x^i\, =\, \delta y_1^i\,-\delta y_2^i
\end{equation}
Applying this transformation to the Lagrangian~\eqref{eq: CoM lagrangian }, we find the corresponding ADM-type Lagrangian in the CoM frame, as a function of $(\mathbf{X}, \mathbf{V})$.  Then, by performing a simple Legendre transformation, we find the ADM-type Hamiltonian in the CoM-frame, which reads

\begin{widetext}
\begin{align}
\label{eq: H in COM ADM}\notag
\frac{H^{\rm ADM-type}_{\rm CoM}}{\mu}\, =&- \frac{c^2}{\nu}\, +\frac{P^2}{2}\, -\frac{G}{R}\Bigl(1-\eta_1 \eta_2\Bigr) \\ 
\notag &+\frac{1}{c^2}\Biggl\{\frac{P^4}{8}(3 \nu- 1)+ \frac{G^2}{4 R^2}\Bigl(2+ \eta_1^2- 4 \eta_1 \eta_2+ \eta_2^2+ X_{12}(\eta_1^2- \eta_2^2)\Bigr)- \frac{G M }{2 R}\Bigl( 3P^2+ \nu (1- \eta_1 \eta_2)(P^2+ P_R^2)\Bigr)\Biggr\} \\ \notag
&+\frac{1}{c^4}\Biggl\{\frac{G^3}{4R^3}\Biggl(4\eta_1 \eta_2- \eta_1^2- \eta_2^2- 2+ X_{12}(\eta_2^2- \eta_1^2)+ \nu\Bigl(23 \eta_1 \eta_2-2\eta_1^2-2 \eta_2^2-3 \eta_1^2 \eta_2^2-\eta_1^3 \eta_2^3-15 \Bigr)\ \Biggr)\\ \notag
&\hspace{1cm}+\frac{G^2}{8 R^2}\Biggl(X_{12}(\eta_1^2- \eta_2^2) \Bigl(P^2- 2 P_R^2\Bigr)(1- \nu)+2 P_R^2(\eta_1^2+ \eta_2^2-2)+ P^2(\eta_1^2+ \eta_2^2+22)\\ \notag
&\hspace{2cm}+\nu \Bigl(P^2(58+7 \eta_1^2-46 \eta_1 \eta_2+ 7 \eta_2^2+ 2 \eta_1^2 \eta_2^2)-P_R^2(32+ 10 \eta_1^2+ 10 \eta_2^2+4 \eta_1 \eta_2)\Bigr)\Biggr)\\ \notag
&\hspace{1cm}+ \frac{G}{8 R}\Biggl(5 P^4+ 2 \nu P^2\Bigl( (1- \eta_1 \eta_2)P_R^2 - P^2(11+\eta_1 \eta_2)\Bigr)- \nu^2 (1- \eta_1 \eta_2)\Bigl(3 P^4+2 P^2 P_R^2+ 3P_R^4\Bigr)\Biggr)\\ \notag
&\hspace{1cm}+\frac{P^6}{16}(1-5 \nu+5 \nu^2)+ \frac{G}{4R^3}(1- \eta_1 \eta_2)(1+12 \nu)+ \frac{P_R^2}{2 R^2}- \frac{P^2}{4 R^2}+ \frac{3\nu}{R^2}(2 P_R^2-P^2)\\ \notag
&\hspace{1cm}-\frac{G \nu}{4 R^2}\Bigl(P^2+2 P_R^2)(1- \eta_1 \eta_2)+  \frac{\nu P^2}{4 R}\Bigl(P^2- P_R^2\Bigr)\\ \notag
&\hspace{1cm}+\mathbf{A}\Biggl[\frac{P^2}{M R}\Bigl(P_R^2- P^2 \Bigr)+ \frac{G}{M R^2}(1- \eta_1 \eta_2)\Bigl(2P_R^2+P^2\Bigr)\Biggr]\\ \notag
&\hspace{1cm}+ \mathbf{B}\Biggl[\frac{2P_R^2- P^2}{M^2 R^2}+ \frac{G}{M^2 R^3}(1- \eta_1 \eta_2) \Biggr]\\ 
&\hspace{1cm}+ \mathbf{C}\Biggl[\frac{3G P_R^2}{M R^2}(1- \eta_1 \eta_2)+\frac{3P_R^2}{M R}\Bigl(P_R^2- P^2\Bigr) \Biggr]\Biggr\}
\end{align}
 \end{widetext}
 where $P^2=\, p^2/\mu^2$, $P_R=p_r/\mu$, $R=r/M$, and $\mathbf{A}, \mathbf{B}, \mathbf{C}$ are coefficients associated to the contact transformation, given by combinations of the coefficients appearing in Eq.~\eqref{eq: parametrization of F}, see Sec.~\ref{subsec:adm_coord}. Explicitly they read
\begin{subequations}
    \begin{align}
        &\mathbf{A}=\, \frac{1}{(m_1+m_2)^3}\Big(\, m_2^3 A_1\,  -m_1 m_2^2(A_2+ A_3)\,  \cr&\quad+ m_1^2 m_2 (A_4+ A_5)\, - m_1^3 A_6 \Big)\, , \\ 
        &\mathbf{B}=\, \frac{1}{m_1 +m_2}\left(\,- m_1 B_1\, + m_2 B_2\right)\,, \\ 
        &\mathbf{C}=\, \frac{1}{(m_1+m_2)^3}\Big(\, m_2^3 C_1\, -m_1 m_2^2 C_2\, \cr&\quad+ m_1^2 m_2 C_3\, -m_1^3 C_4\Big). 
    \end{align}
\end{subequations}
\section{Explicit calculations for the dissipative contributions}
\label{app: explicit calculation dissipative contributions}

In this Appendix, we go over the technical details behind the derivation of the dissipative contributions to the EoMs, Eq.~\eqref{eq: dissipative corretion EoM 1}.

Our starting point is Eq.~\eqref{eq:dissipative correction a1 potentials}, which gives the dissipative contribution we want to compute in terms of the vector potential components. More specifically, we need the terms in the PN expansions of $A_0$ and $A_i$ of order, respectively, 1.5PN and 1PN.

By varying the total action~\eqref{eq: total E-M action} with respect to $A_i$ and $A_0$, the associated field equations at 1PN accuracy are found to be~\footnote{We underline that the next PN order in the field equations would be the 2PN, which we do not need for our current purpose of computing 1.5PN dissipative contributions.}
\begin{subequations}
\label{eq: field equation Ai and A0}
    \begin{align}
    \label{eq:field equation Ai}
        \Box A_i\, &=\, -\frac{4 \pi}{c} \rho_e v^i, \\
        \label{eq:field equation A0}
        \Box A_0\, &=\, 4\pi \rho_e\, -\frac{1}{c^2}\Bigl(2V \nabla^2A_0\, +2\partial_iV \partial_i A_0\Bigr),
    \end{align}
\end{subequations}
where $\Box=\, -\partial_0^2\, +\nabla^2$ and $\rho_e$ is the point-particle electric-charge density defined as
\begin{equation}
\label{eq: rho carica}
    \rho_e(\mathbf{x},t)
    = \sum_A q_A\, \delta^3\big(\mathbf{x}\, -\mathbf{y}_A(t)\big).
\end{equation}

Since our objective here is to compute the leading dissipative contributions to $A_0$ and $A_i$, we are well before the point at which the radiation reaction sector of the dynamics develop tail-related components of non-instantaneous type.~\footnote{In our case the first contributions of this kind would appear at the 3PN level, namely when the leading 1.5PN tail effects enter the dipolar component of the energy flux, which is by itself a quantity that starts at the 1.5PN order.} We can therefore proceed with a direct PN approach similar to the one of Ref.~\cite{Anderson:1975}, and focus on the Poisson-like integrals that follow from the PN expansion of the formal retarded solutions to Eqs.~\eqref{eq: field equation Ai and A0}.

Let us begin with Eq.~\eqref{eq:field equation Ai}. Indeed, its retarded solution reads
\begin{equation}
\label{eq: Ai generico}
    A_i(\mathbf{x},t)\, =\frac{1}{c}\int d^3 \mathbf{z} \frac{\rho_e\Bigl(\mathbf{z}, t-\frac{|\mathbf{x}-\mathbf{z}|}{c}\Bigr)}{| \mathbf{x}-\mathbf{z}|}v^i\,.  
\end{equation}
Once this is expanded in powers of $1/c$, the Poisson-like integrals that spring forth are easily evaluated thanks to the Dirac deltas in $\rho_e$, Eq.~\eqref{eq: rho carica}. Hence, we find
\begin{equation}
\label{eq: Ai espanso}
    A_i\, =\frac{1}{c}\Biggl[\frac{q_1 v_1^i}{|\mathbf{x}-\mathbf{y}_1|}\, -\frac{1}{c}\frac{d}{dt} \Bigl(q_1 v_1^i\Bigr)\Biggr]+(1 \leftrightarrow 2) +\mathcal{O}(1/c^3).
\end{equation}
The leading 0.5PN term in Eq.~\eqref{eq: Ai espanso} fully reproduces the leading order term of $A_i$ provided in Ref.~\cite{Khalil:2018aaj}. In addition, we also find the sought for dissipative contribution 
\begin{equation}
\label{eq: 1PN Ai solution}
    A_i^{\rm 1PN} = -\frac{1}{c^2} \big(q_1 a^i_1+q_2 a^i_2\big).
\end{equation}

In preparation for doing the same with Eq.~\eqref{eq:field equation A0}, it is useful to rewrite its source term using the differential identity
\begin{equation}
    \label{eq: identity}
    \nabla^2(\alpha \beta)\, =\, \alpha \nabla^2 \beta, +\beta \nabla^2\alpha\, +2 \partial_i \alpha \partial_i\beta,
\end{equation}
which yields
\begin{equation}
    \label{eq:simplified field equation A0}
    \Box A_0\, =\, 4 \pi \rho_e\,+\frac{1}{c^2}\Bigl[A_0\, \nabla^2 V -V\, \nabla^2A_0 -\nabla^2(V A_0)\Bigr]. 
\end{equation}
\newline
Here, the potentials $A_0$ and $V$ appearing in the source term can be replaced with the leading order solutions to their field equations, found e.g.~in Ref.~\cite{Khalil:2018aaj}. Introducing the notation $\mathbf{r}_A\equiv\mathbf{x}-\mathbf{y}_A$ and $r_A\equiv|\mathbf{r}_A| $, they simply read
\begin{align}
    \label{eq: leading A0}
    &A_0^{\rm LO} = - \bigg(\frac{q_1}{r_1}+\frac{q_2}{r_2}\bigg), \\
    \label{eq: leading V}
    &V^{\rm LO} = \frac{G m_1}{r_1}+\frac{G m_2}{r_2}.
\end{align}

It is straightforward to see that, once these expressions are substituted in the right hand side of Eq.~\eqref{eq:simplified field equation A0}, the terms $A_0\, \nabla^2 V$ and $-V\, \nabla^2A_0\,$ actually cancel each other out. The retarded solution to the 1PN field equation of $A_0$ is thus
\begin{equation}
\label{eq: A0 leading order general}
 A_0(\mathbf{x},t)\, =\, -\int d^3\mathbf{z} \frac{\rho\Bigl(\mathbf{z}, t-\frac{|\mathbf{x}-\mathbf{z}|}{c}\Bigr)}{| \mathbf{x}-\mathbf{z}|},
\end{equation}
where 
\begin{equation}
    \label{eq: rho 1PN}
    \rho(\mathbf{x},t)= \rho_e(\mathbf{x},t)\, - \frac{1}{4 \pi c^2}\nabla^2\big(V^{\rm LO}A_0^{\rm LO}\big). 
\end{equation}

Similarly as before, we consider the PN expansion of Eq.~\eqref{eq: A0 leading order general} and isolate, in the so-obtained series of Poisson-like integrals, the 1.5PN contribution we want to compute. We find
\begin{align}\label{eq: A0 1.5PN retardations}
    &A_0^{\,  \rm 1.5PN}= - \frac{1}{c^3}\int d^3 \mathbf{z}\Biggl[\frac{| \mathbf{x}-\mathbf{z}|^2}{6}\partial_t^3\rho_e \cr&\quad+\frac{1}{4\pi}\partial_t \nabla^2\big(V^{\rm LO}A_0^{\rm LO}\big)\Biggr].
\end{align}
The first integral, whose support is compact, is easily solved:
\begin{equation}
  \int d^3 \mathbf{z}\frac{| \mathbf{x}-\mathbf{z}|^2}{6}\partial_t^3\rho_e\,\, =\, \frac{q_1}{6}\frac{d^3(r_1^2)}{dt^3}\, +(1 \leftrightarrow2).
\end{equation}
By order reducing the right hand side of the latter via the EoMs \eqref{eq: EOM1}, that is enough to use at leading order in this case, we obtain
\begin{widetext}
\begin{equation}
\label{eq: A0 1.5PN piece 1}
   \int d^3 \mathbf{z}\frac{| \mathbf{x}-\mathbf{z}|^2}{6}\partial_t^3\rho_e\,\, =\, - q_1\Biggl(\frac{G m_2}{r^2}-\frac{q_1 q_2}{m_1 r^2}\Biggr)\Biggl[\mathbf{n}\cdot \mathbf{v}_1\, + \frac{1}{3} \mathbf{r}_1 \cdot \Bigl(3\, (\mathbf{n}\cdot \mathbf{v}) \, \mathbf{n}\, -\mathbf{v}\Bigr)\Biggr]\, +(1 \leftrightarrow2),
\end{equation}
where we recall that $\mathbf{v}=\, \mathbf{v}_1\, -\mathbf{v}_2$. 

For the second integral of Eq.~\eqref{eq: A0 1.5PN retardations}, using Eqs.~\eqref{eq: leading A0} and~\eqref{eq: leading V}, we have
\begin{align}
    \label{eq: A0 next leading order general}
 &\frac{1}{4\pi}\frac{d}{dt}\int d^3\mathbf{z}\, \nabla^2\big(V^{\rm LO}A_0^{\rm LO}\big)=\, \frac{G}{4\pi}\frac{d}{dt}\Bigg( m_1 q_1\,_1\partial_i^2 \int \frac{d^3 \mathbf{z}}{|\mathbf{z}-\mathbf{y}_1|^2}\, + m_2 q_2 \,_2\partial_i^2\int \frac{d^3 \mathbf{z}}{|\mathbf{z}-\mathbf{y}_2|^2} \cr
 &\quad + ( m_1 q_2\, +m_2 q_1)\,_1\partial_i \,_2\partial_i \int \frac{d^3 \mathbf{z}}{|\mathbf{z}-\mathbf{y}_1|\, |\mathbf{z}-\mathbf{y}_2|} \Bigg),
\end{align}
where, following Ref.~\cite{Blanchet:1998vx}, we replaced the partial derivatives $\partial_i$ acting on $z^i$ with the partial derivatives $\,_A\partial_i$ acting on $y_A^i$, exploiting that $\partial_i |\mathbf{z}-\mathbf{y}_A| = -\,_A\partial_i |\mathbf{z}-\mathbf{y}_A|$.

With the same strategy adopted for Eq.~(5.4) of Ref.~\cite{Blanchet:1998vx}, the first two terms can be shown to give no contribution while the last one is evaluated to
\begin{equation}
     \,_1\partial_i \,_2\partial_i \int \frac{d^3 \mathbf{z}}{|\mathbf{z}-\mathbf{y}_1|\, |\mathbf{z}-\mathbf{y}_2|}= -2\pi \,_1\partial_i \,_2\partial_i r =\,\frac{4\pi}{r},
\end{equation}
so that 
\begin{equation}
    \label{eq: A0 1.5PN piece 2}
    \frac{1}{4\pi}\frac{d}{dt}\int d^3\mathbf{z}\, \nabla^2\big(V^{\rm LO}A_0^{\rm LO}\big) = G( m_1 q_2\, +m_2 q_1) \frac{d}{dt} \frac{1}{r}.
\end{equation}

By combining Eqs.~\eqref{eq: A0 1.5PN piece 1} and~\eqref{eq: A0 1.5PN piece 2}, explicitly evaluating the time derivative in the latter, we conclude
\begin{equation}
    \label{eq: 1.5PN A0 total}
    A_0^{\, \rm 1.5PN}=\,- \frac{G m_1}{r^2 c^3}\Biggl\{ m_2 q_1 \Biggl(1-\frac{q_1 q_2}{G m_1 m_2 }\Biggr)\Biggl[\mathbf{n}\cdot \mathbf{v}_1\, + \frac{1}{3} \mathbf{r}_1 \cdot \Bigl(3\, (\mathbf{n}\cdot \mathbf{v}) \, \mathbf{n}\, -\mathbf{v}\Bigr)\Biggr]+q_2 (\mathbf{n}\cdot \mathbf{v})\, \Biggr\}+(1 \leftrightarrow2).
\end{equation}
\end{widetext}

Let us finally move to the evaluation of Eq.~\eqref{eq:dissipative correction a1 potentials}. Now that we have both $A_0^{\rm 1.5PN}$ and $A_i^{\rm 1PN}$, we only need to clarify how to overcome the problem of the infinite self-field of point-particles, which brings forth divergencies when quantities are evaluated at the positions of the bodies. 

Since we aim at a PN accuracy for which there is still no need to resort to the dimensional regularization method~\cite{Blanchet:2003gy}, we can simply regularize all the diverging quantities by taking their Hadamard partie finie~\cite{HadamardReg,Blanchet:2000nu}, of which we review the main technical aspects below.

Let us consider a function $F(\mathbf{x})$ that is smooth everywhere in space except at two isolated singularities, $\mathbf{y}_1$ and $\mathbf{y}_2$. As the field point $\mathbf{x}$ approaches one of the singularities, say $\mathbf{y}_1$ (i.e.~$r_1=|\mathbf{x}-\mathbf{y}_1|\to 0$), the function $F$ admits the power-like singular expansion
\begin{equation}
    \label{eq: F exp}
    F(\mathbf{x};\mathbf{y}_1, \mathbf{y}_2)= \sum_{-k_0 \le k\le 0} r_1^k \, f_k(\mathbf{n}_1;\mathbf{y}_1, \mathbf{y}_2)\, + \mathcal{O}(r_1),
\end{equation}
where $\mathbf{n}_1=(\mathbf{x}-\mathbf{y}_1)/r_1$ and $k \in  \mathbb{Z}$. The Hadamard partie finie of $F$ at the singular point $\mathbf{y}_1$, which gives the  regularized value assumed by $F$ when it is evaluated there, is defined by
\begin{equation}
\label{eq: F1 angular average}
    \bigl(F\bigr)_1\equiv \int \frac{d \Omega(\mathbf{n}_1)}{4 \pi}f_0(\mathbf{n}_1;\mathbf{y}_1, \mathbf{y}_2),
\end{equation}
that is by an angular average of the $k=0$ term in the expansion~\eqref{eq: F exp}, taken with respect to the direction $\mathbf{n}_1$ of approach of the singularity $\mathbf{y}_1$.


%
We have now all we need to evaluate Eq.~\eqref{eq:dissipative correction a1 potentials}: we replace the explicit expression of $A_0^{\rm 1.5PN}$ and $A_i^{\rm 1PN}$, respectively Eqs.~\eqref{eq: 1.5PN A0 total} and~\eqref{eq: 1PN Ai solution}, we compute the derivatives while order reducing each time derivative of the velocities via the EoMs, and we ultimately evaluate everything at the position of the bodies by taking the Hadamard partie finie. The result is precisely Eq.~\eqref{eq: dissipative corretion EoM 1} of the main text.

\newpage
\bibliography{bibliography.bib} 

\end{document}